\documentclass[superscriptaddress, showpacs, nofootinbib, amsmath, amssymb, aps, prd, notitlepage]{revtex4-1}
\usepackage{amsmath,amssymb,amsfonts,mathrsfs,bbold}
\usepackage{yfonts}

\usepackage[dvipsnames]{xcolor}

\usepackage{float}
\usepackage{graphics,graphicx}
\usepackage{dcolumn}
\usepackage{bm}
\usepackage{hyperref}
\usepackage{comment}
\usepackage{xcolor}
\usepackage{slashed}
\usepackage{pstricks}
\usepackage{color}

\usepackage{multirow}

\usepackage{slashed}
\usepackage{simpler-wick}
\usepackage[normalem]{ulem}

\usepackage{tikz}
\usetikzlibrary{decorations.pathmorphing, decorations.markings, shapes.geometric, shapes.misc, calc, math}
\tikzstyle{gluon}=[decorate, decoration={coil,aspect=0.8, amplitude=1.5pt,  segment length=3pt}]

\usepackage{stackrel}
\usepackage[most]{tcolorbox}

\allowdisplaybreaks[1]

\def\sec#1{{Sec.~\ref{#1}}}

\def\eq#1{{Eq.~(\ref{#1})}}
\def\fig#1{{Fig.~\ref{#1}}}
\newcommand{\ben}{\begin{eqnarray*}}
\newcommand{\een}{\end{eqnarray*}}
\newcommand{\un}[1]{\underline{#1}}
\newcommand{\amu}{\alpha_\mu}
\newcommand{\pd}{\partial}
\newcommand{\xx}{\underline{x}}
\newcommand{\pp}{\underline{p}}
\newcommand{\ee}{\underline{\varepsilon}}
\newcommand{\ul}[1]{\underline{#1}}

\newcommand{\tr}{\mbox{tr}}

\newcommand{\llangle}{\Big\langle \!\! \Big\langle}
\newcommand{\rrangle}{\Big\rangle \!\! \Big\rangle}

\newcommand{\as}{\alpha_s}

\begin{document}

\title{Helicity Evolution at Small $x$: the Single-Logarithmic Contribution}

\author{Yuri V. Kovchegov} 
         \email[Email: ]{kovchegov.1@osu.edu}
         \affiliation{Department of Physics, The Ohio State
           University, Columbus, OH 43210, USA}
           
\author{Andrey Tarasov}
         \email[Email: ]{tarasov.3@osu.edu}
         \affiliation{Department of Physics, The Ohio State
           University, Columbus, OH 43210, USA}
         \affiliation{Joint BNL-SBU Center for Frontiers in Nuclear Science (CFNS) at Stony Brook University, Stony Brook, NY 11794, USA}

\author{Yossathorn Tawabutr}
         \email[Email: ]{tawabutr.1@osu.edu}
         \affiliation{Department of Physics, The Ohio State
           University, Columbus, OH 43210, USA}

\begin{abstract}
We calculate single-logarithmic corrections to the small-$x$ flavor-singlet helicity evolution equations derived recently \cite{Kovchegov:2015pbl,Kovchegov:2016zex,Kovchegov:2018znm} in the double-logarithmic approximation. The new single-logarithmic part of the evolution kernel sums up powers of $\as \, \ln (1/x)$, which are an important correction to the dominant powers of $\as \, \ln^2 (1/x)$ summed up by the double-logarithmic kernel from \cite{Kovchegov:2015pbl,Kovchegov:2016zex,Kovchegov:2018znm} at small values of Bjorken $x$ and with $\as$ the strong coupling constant. The single-logarithmic terms arise separately from either the longitudinal or transverse momentum integrals. Consequently, the evolution equations we derive employing the light-cone perturbation theory simultaneously include the small-x evolution kernel and the leading-order polarized DGLAP splitting functions. We further enhance the equations by calculating the running coupling corrections to the kernel. \\ 
\end{abstract}

\maketitle
\tableofcontents


\section{\label{sec:intro}Introduction}

The small-$x$ asymptotics of helicity distributions for quarks and gluons have been the subject of intense studies in recent years \cite{Kovchegov:2015pbl, Altinoluk:2014oxa, Hatta:2016aoc, Kovchegov:2016zex, Kovchegov:2016weo, Kovchegov:2017jxc, Kovchegov:2017lsr, Kovchegov:2018znm, Chirilli:2018kkw, Kovchegov:2019rrz, Boussarie:2019icw, Cougoulic:2019aja, Kovchegov:2020hgb, Cougoulic:2020tbc, Altinoluk:2020oyd, Chirilli:2021lif, Adamiak:2021ppq}. The goal of these ongoing efforts is to provide a reliable first-principles theoretical framework for predicting helicity parton distribution functions (helicity PDFs or hPDFs) at small values of Bjorken $x$. Understanding hPDFs at small $x$ is, in turn, needed to constrain the fraction of the proton (or neutron) spin coming from small-$x$ partons, helping to resolve the proton spin puzzle. The latter is the apparent discrepancy between the spin-$1/2$ of the proton and the amount of proton spin carried by its quarks and gluons, as measured in experiments (see \cite{Aidala:2012mv} and references therein).

Spin sum rules \cite{Jaffe:1989jz, Ji:1996ek} (see also \cite{Leader:2013jra} for a review) represent the proton spin as a sum of quark and gluon helicities and orbital angular momenta (OAM). The Jaffe--Manohar sum rule \cite{Jaffe:1989jz} is
\begin{equation}
S_q+L_q+S_G+L_G=\frac{1}{2},
\label{eqn:JM}
\end{equation}
where $S_q$ and $S_G$ are the total spin of the nucleon carried by the quarks and gluons, respectively, whereas $L_q$ and $L_G$ are their OAM. The quark and gluon spin contributions, $S_q$ and $S_G$, can be written as integrals over $x$,
\begin{align}
S_q(Q^2) = \frac{1}{2} \int\limits_0^1 dx \; \Delta\Sigma(x,Q^2), \ \ \  \ \
S_G(Q^2) = \int\limits_0^1 dx \; \Delta G(x,Q^2),
\label{eqn:SqSG}
\end{align}
where
\begin{equation}
\Delta\Sigma(x,Q^2) = \sum_{f=u,d,s,\ldots} \left[\Delta f(x,Q^2) + \Delta\bar{f}(x,Q^2)\right]
\label{eqn:DeltaSigma}
\end{equation}
is the flavor-singlet quark helicity distribution function while $\Delta G$ is the gluon helicity distribution, both also dependent on the momentum scale $Q^2$. The current values of the quark and gluon spin are $S_q (Q^2 = 10\, \mbox{GeV}^2) \approx 0.15 \div 0.20$, integrated over $0.001 < x <1$, and $S_G (Q^2 = 10 \, \mbox{GeV}^2) \approx 0.13 \div 0.26$, integrated over $0.05 < x <1$ (see \cite{Accardi:2012qut,Leader:2013jra,Aschenauer:2013woa,Aschenauer:2015eha,Proceedings:2020eah} for reviews). Their sum comes up short of $1/2$, especially if one takes into account the error bars: this is the proton spin puzzle. The missing spin of the proton can reside in the OAM and/or at smaller values of $x$.  

While the integrals in \eq{eqn:SqSG} go down to $x=0$, the experimental data needed for extractions of helicity PDFs $\Delta\Sigma(x,Q^2)$ and $\Delta G(x,Q^2)$ are limited to $x > x_{min}$ with the value of $x_{min}$ determined by the finite energy reach of a given experiment. This illustrates the need for theoretical input, in order to determine the helicity PDFs for $x < x_{min}$, and, through \eq{eqn:SqSG}, constrain the small-$x$ quark and gluon helicities in the proton. We conclude that quantitative understanding of helicity PDFs at small $x$ is an essential ingredient for resolving the spin puzzle. 

It is expected that the future experiments, in particular those at the Electron-Ion Collider (EIC) \cite{Accardi:2012qut,Boer:2011fh,Proceedings:2020eah,AbdulKhalek:2021gbh}, will help us constrain helicity PDFs at rather small values of $x$, down to $x \approx 5 \times 10^{-4}$ at $Q^2 = 10 \, \mbox{GeV}^2$ \cite{Accardi:2012qut,AbdulKhalek:2021gbh}. The EIC will present a unique opportunity (i) to verify our theoretical predictions for the proton helicity PDFs at small $x$ and (ii) to extrapolate the PDFs obtained from the future analyses of the EIC data further down to an even smaller values of $x$, beyond the EIC data region, to completely constrain the small-$x$ partons contribution to the proton spin. To accomplish these two goals one needs a theory capable of making predictions for hPDFs in the low-$x$ region to be probed at the EIC and of extrapolation down to even lower $x$, doing both in a rigorous and controlled way. 

Small-$x$ helicity evolution equations, capable of {\sl predicting} helicity PDFs at small $x$, have recently been derived in  \cite{Kovchegov:2015pbl, Kovchegov:2016zex, Kovchegov:2016weo, Kovchegov:2017jxc, Kovchegov:2017lsr, Kovchegov:2018znm, Cougoulic:2019aja}. As was recently demonstrated in \cite{Adamiak:2021ppq}, these equations are capable of precisely accomplishing the low-$x$ predictions and extrapolations listed in the above items (i) and (ii), respectively. The equations from \cite{Kovchegov:2015pbl, Kovchegov:2016zex, Kovchegov:2016weo, Kovchegov:2017jxc, Kovchegov:2017lsr, Kovchegov:2018znm, Cougoulic:2019aja} have recently been verified by an independent calculation in \cite{Chirilli:2021lif} using the background field method.\footnote{More precisely, the calculation done in \cite{Chirilli:2021lif} confirmed the results of \cite{Kovchegov:2015pbl, Kovchegov:2016zex, Kovchegov:2016weo, Kovchegov:2017jxc, Kovchegov:2017lsr, Kovchegov:2018znm, Cougoulic:2019aja} in the gluon sector: the adjoint polarized dipole evolution is given by Eq.~(F.45) in \cite{Chirilli:2021lif}, which, if one neglects the quark operators in it, is identical to Eq.~(70) in \cite{Kovchegov:2018znm}. This confirmation in the pure-glue case also implies agreement in the large-$N_c$ limit between the two approaches. In the quark sector, Eq.~(F.45) in \cite{Chirilli:2021lif} is different from Eq.~(96) in \cite{Kovchegov:2018znm} only by the mixing with the ${\cal Q}_1 \sim {\bar \psi} \gamma^+ \psi$ operator on the right-hand side of the former which is absent in the latter. While more work is needed to understand this apparent discrepancy in the quark sector, we are encouraged by the fact that the mixing with ${\cal Q}_1$ disappears after adding to Eq.~(F.45) from \cite{Chirilli:2021lif} its Hermitian conjugate, like it was done in arriving at Eq.~(F.13) in the same reference in order to compare directly with the object studied in \cite{Kovchegov:2015pbl, Kovchegov:2016zex, Kovchegov:2016weo, Kovchegov:2017jxc, Kovchegov:2017lsr, Kovchegov:2018znm, Cougoulic:2019aja}. For the fundamental polarized dipole, the comparison between the two calculations is hampered by the fact that in \cite{Chirilli:2021lif} the separation into the flavor-singlet and non-singlet contributions was done differently from that in \cite{Kovchegov:2015pbl,  Kovchegov:2016zex, Kovchegov:2018znm}, shown in Eqs.~\eqref{Qdef10} and \eqref{QNSdef10} below, though, again, we are encouraged by the agreement with Eq.~(F.13) in \cite{Chirilli:2021lif}.}

The equations \cite{Kovchegov:2015pbl, Kovchegov:2016zex, Kovchegov:2016weo, Kovchegov:2017jxc, Kovchegov:2017lsr, Kovchegov:2018znm, Cougoulic:2019aja} sum up powers of the leading parameter $\as \ln^2(1/x)$ with $\as$ the strong coupling constant (see \cite{Bartels:1995iu,Bartels:1996wc,Blumlein:1995jp,Blumlein:1996hb} for an earlier attempt to calculate helicity PDFs at small $x$ employing the technique developed in \cite{Kirschner:1983di,Kirschner:1994rq,Kirschner:1994vc,Griffiths:1999dj}). We will refer to the resummation of the parameter $\as \ln^2(1/x)$ as the double-logarithmic approximation (DLA). The equations of \cite{Kovchegov:2015pbl, Kovchegov:2016zex, Kovchegov:2016weo, Kovchegov:2017jxc, Kovchegov:2017lsr, Kovchegov:2018znm, Cougoulic:2019aja} close in the large-$N_c$ and in the large-$N_c \& N_f$ limits (with $N_c$ and $N_f$ the numbers of quark colors and flavors, respectively).  Solution of the DLA evolution equations \cite{Kovchegov:2015pbl, Kovchegov:2016zex, Kovchegov:2018znm} performed in \cite{Kovchegov:2016weo, Kovchegov:2017jxc, Kovchegov:2017lsr} led to the following small-$x$ asymptotics of helicity PDFs at large-$N_c$, 
\begin{align}\label{SigmaG_LargeNc}
\Delta \Sigma (x, Q^2) \bigg|_{\mbox{large-}N_c} \sim \left(\frac{1}{x}\right)^{\alpha_h^q},   \ \ \ \ \
       \Delta G (x, Q^2) \bigg|_{\mbox{large-}N_c} \sim \left(\frac{1}{x}\right)^{\alpha_h^G} 
\end{align}
with
\begin{align}\label{alphas}
\alpha_h^q \bigg|_{\mbox{large-}N_c} = \frac{4}{\sqrt{3}} \, \sqrt{\frac{\as \, N_c}{2 \pi}} , \ \ \ \ \ \alpha_h^G \bigg|_{\mbox{large-}N_c} = \frac{13}{4 \sqrt{3}} \, \sqrt{\frac{\as
        \, N_c}{2 \pi}}.
\end{align}
More recently \cite{Kovchegov:2020hgb}, a numerical solution of the large-$N_c \& N_f$ equations gave
\begin{align}\label{DSigmaMain2}
\Delta \Sigma (x, Q^2)\bigg|_{\mbox{large-}N_c \& N_f} \sim \left( \frac{1}{x} \right)^{\alpha_h^q} \, \cos \left[ \omega_q \, \ln \left( \frac{1}{x} \right) + \varphi_q \right]  
\end{align}  
with $\alpha_h^q$ close to that in \eq{alphas}, the oscillation frequency approximated by
\begin{align}\label{omega_fit}
\omega_q  \approx \frac{0.66 (N_f/N_c)}{1 + 0.3795 (N_f/N_c)}  \, \sqrt{\frac{\as \, N_c}{2 \pi}}  
\end{align}
and the initial-conditions dependent phase $\varphi_q$.

The goal of this project is to, ultimately, find the single-logarithmic corrections to the results in Eqs.~\eqref{SigmaG_LargeNc} and \eqref{DSigmaMain2}. That is, we want to resum powers of $\as \, \ln (1/x)$: we will refer to this as the single-logarithmic approximation (SLA). In this paper we derive the SLA correction to the evolution kernel of the DLA equations of \cite{Kovchegov:2015pbl, Kovchegov:2016zex, Kovchegov:2016weo, Kovchegov:2017jxc, Kovchegov:2017lsr, Kovchegov:2018znm, Cougoulic:2019aja}, obtaining helicity evolution equations containing both the DLA and SLA kernels. 

In the case of unpolarized evolution, the leading-order Balitsky--Fadin--Kuraev--Lipatov (BFKL) \cite{Kuraev:1977fs,Balitsky:1978ic}, Balitsky--Kovchegov (BK) \cite{Balitsky:1995ub,Balitsky:1998ya,Kovchegov:1999yj,Kovchegov:1999ua} and Jalilian-Marian--Iancu--McLerran--Weigert--Leonidov--Kovner
(JIMWLK)
\cite{Jalilian-Marian:1997dw,Jalilian-Marian:1997gr,Weigert:2000gi,Iancu:2001ad,Iancu:2000hn,Ferreiro:2001qy} evolution equations sum up powers of the parameter $\as \, \ln (1/x)$. There is no parameter $\as \, \ln^2 (1/x)$ in the unpolarized evolution, and the $\as \, \ln (1/x)$ part of the evolution kernel generates the leading-logarithmic approximation. The  $\ln (1/x)$ in the unpolarized evolution arises from the longitudinal momentum integral in the emitted gluon's phase space. Transverse momentum integrals in the leading-logarithm unpolarized evolution are both ultraviolet (UV) and infrared (IR) finite and do not generate logarithms of energy. In the case of helicity evolution in the DLA \cite{Kovchegov:2015pbl, Kovchegov:2016zex,Kovchegov:2018znm}, one logarithm of $1/x$ arises from the longitudinal momentum integral (in the phase space of the emitted gluon or quark), while another one is generated by the transverse momentum integral which is regulated in the UV by the center-of-mass energy (see also \cite{Kirschner:1994vc}). As we will see below, in the SLA for helicity evolution the single logarithm of $1/x$ can arise from either the longitudinal or transverse momentum integrals. While the kernel generating the longitudinal single logarithms of $1/x$ in helicity evolution was already found in \cite{Kovchegov:2015pbl}, as a by-product of the DLA kernel calculation, the kernel generating the transverse single logarithms of $1/x$ has not been found in earlier literature and will be constructed below. The transverse-momentum integral origin of the $\ln (1/x)$ we observe here is not found in the unpolarized BFKL/BK/JIMWLK evolution. 

The paper is organized as follows. In Sec.~\ref{sec:LLALT} we explain the origins of the longitudinal and transverse single logarithms of $1/x$. We refer to those as SLA$_L$ and SLA$_T$ respectively. As we mentioned above, SLA$_L$ part of helicity evolution kernel was found before \cite{Kovchegov:2015pbl}, so our goal here is to find the SLA$_T$ part of the evolution kernel. The diagrams contributing to the SLA$_T$ splittings are studied in Sec.~\ref{sec:ingredients}, where the corresponding kernels are constructed as well. The evolution equations for the ``polarized Wilson line" operators defined in \cite{Kovchegov:2017lsr,Kovchegov:2018znm} including both the DLA and the complete SLA (=SLA$_L$+SLA$_T$) terms in the kernel are constructed in Sec.~\ref{sec:eqnnorc}. 

There is one subtlety with the SLA$_T$ terms in the kernel. By their definition, these terms generate logarithms of energy via logarithmically UV divergent transverse momentum integrals. Similar UV divergence leads to the running coupling corrections, as calculated in \cite{Balitsky:2006wa,Kovchegov:2006vj,Gardi:2006rp,Kovchegov:2006wf} for the unpolarized small-$x$ evolution. Hence, there is a need to disentangle the two types of UV divergent terms, the running coupling and the SLA$_T$ ones. This is accomplished in Sec.~\ref{sec:rc}, resulting in the DLA+SLA evolution equations with the running coupling corrections included: these are given in Eqs.~\eqref{evol_rc} and \eqref{Gevol_rc}, which are our main formal result. We note the simplicity of some of the running coupling scales in Eqs.~\eqref{evol_rc} and \eqref{Gevol_rc} compared to the running coupling scales found for the unpolarized evolution in \cite{Balitsky:2006wa,Kovchegov:2006vj}: this simplicity ultimately results from the presence of the SLA$_T$ terms in the helicity evolution at hand. 

The DLA+SLA evolution equations \eqref{evol_rc} and \eqref{Gevol_rc} do not close, the operators on their right-hand sides are not just iterations of the operators on their left-hand sides. This situation is similar to the unpolarized BK/JIMWLK and DLA helicity evolutions. In Sec.~\ref{sec:closed_ee} we obtain closed evolution equations in the large-$N_c$ (Eqs.~\eqref{eqn:Nc17} and \eqref{eqn:Nc18}) and large-$N_c \& N_f$ (Eqs.~\eqref{Qevol}, \eqref{Gamma_bar_evol}, \eqref{Gevol_largeNcNf}, and \eqref{Gamma_evol_largeNcNf}) limits. Note that these equations include nonlinear terms mixing helicity evolution and the unpolarized BK evolution (cf. \cite{Kovchegov:2015pbl, Itakura:2003jp}), with the latter containing saturation corrections \cite{Gribov:1984tu,Iancu:2003xm,Weigert:2005us,JalilianMarian:2005jf,Gelis:2010nm,Albacete:2014fwa,Kovchegov:2012mbw}. The impact of saturation corrections on helicity evolution is to be explored in the future work. We conclude in Sec.~\ref{sec:conclusion} by summarizing our main results.



\section{Types of SLA Correction Terms}
\label{sec:LLALT}

Throughout this paper, we consider the evolution of polarized dipole amplitudes defined in \cite{Kovchegov:2015pbl,Kovchegov:2017lsr,Kovchegov:2018znm}, both fundamental and adjoint. In general, the evolution for a polarized dipole amplitude $G$ (to be properly defined below in \eq{Ggdef}) can be written as
\begin{align}
G &= G^{(0)} + \mathcal{K}\otimes G =  G^{(0)} + \as \left(\mathcal{K}_{\text{DLA}}+\mathcal{K}_{\text{SLA}_L}+\mathcal{K}_{\text{SLA}_T}
\right)\otimes G + {\cal O} (\as^2).
\label{eqn:LLALT1}
\end{align}
Here, $G^{(0)}$ is the initial condition and $\mathcal{K}$ is the (integral) evolution kernel. The latter can be separated into several parts, depending on the type of integrals involved. As we will show below, the evolution equations derived in \cite{Kovchegov:2015pbl} include both the DLA and SLA$_L$ kernels. The subsequent applications and analysis of these equations in \cite{Kovchegov:2016zex, Kovchegov:2016weo, Kovchegov:2017jxc, Kovchegov:2017lsr, Kovchegov:2018znm, Cougoulic:2019aja, Kovchegov:2020hgb} was done in the DLA, retaining only the kernel $\mathcal{K}_{\text{DLA}}$, which contains two logarithmic integrals, one with respect to the longitudinal (momentum) variable and the other over the transverse (coordinate) variable. Concretely, a typical term in $\mathcal{K}_{\text{DLA}}$ is of the form
\begin{align}
\as \, \mathcal{K}_{\text{DLA}} &= \frac{\alpha_s}{2\pi}\int \frac{dz'}{z'} \int \frac{dx^2_{21}}{x^2_{21}},
\label{eqn:LLALT2}
\end{align}
where $z'$ is the longitudinal (light-cone minus) momentum fraction carried by the emitted gluon or quark, while $x_{21} \equiv |{\un x}_2 - {\un x}_1|$ is the transverse distance between the emitted ``daughter" parton at ${\un x}_2$ and the ``parent" parton at ${\un x}_1$. (The transverse vectors are denoted by ${\un x} = (x^1, x^2)$ with ${\un x}_{ij} = {\un x}_i - {\un x}_j$ and $x_{ij} = |{\un x}_{ij}|$, while the light-cone coordinates are $x^\pm = (x^0 \pm x^3)/\sqrt{2}$ with the polarized proton flying in the light-cone plus direction.)
We see that the kernel, when applied to the sample initial condition, $G^{(0)}=1$, gives a logarithm from the longitudinal $z'$-integral and another from the transverse $x^2_{21}$-integral. We purposefully do not specify the integration limits, since they vary somewhat between different terms in the equation(s): however, the limits are such that they ultimately bring in logarithms of energy, and, hence, of $1/x$. From this point on, we will refer to $\mathcal{K}_{\text{DLA}}$ as the ``double-logarithmic (DLA) kernel.''

This work aims to derive the complete sub-leading part of $\mathcal{K}$ that contains only one such logarithmic integral, which can either be the longitudinal or the transverse one. We call the former ``single-logarithmic, longitudinal (SLA$_L$) kernel,'' corresponding to $\mathcal{K}_{\text{SLA}_L}$ in \eqref{eqn:LLALT1}. It is typically of the form
\begin{align}
\as \, \mathcal{K}_{\text{SLA}_L} &= \frac{\alpha_s}{2\pi^2}\int \frac{dz'}{z'} \int d^2 x_{2} \;\Delta P_L(\xx_{20},\xx_{21}),
\label{eqn:LLALT3}
\end{align}
where $\Delta P_L(\xx_{20},\xx_{21})$ is a function of $\xx_{20}$ and $\xx_{21}$, with the terms potentially giving the logarithmic transverse integrals subtracted out (to avoid double-counting with the DLA kernel). Applying $\mathcal{K}_{\text{SLA}_L}$ to $G^{(0)}=1$ gives a single logarithm of energy from the longitudinal $z'$-integral. In practice, these terms can be derived by considering all the various splittings of the dipole of interest and taking the $z'\ll z$ limit, where $z$ is the minus momentum fraction of the parent parton. This process has been performed in \cite{Kovchegov:2015pbl}, resulting in the $\as \left( \mathcal{K}_{\text{DLA}}+\mathcal{K}_{\text{SLA}_L} \right)$ kernel, though, as we mentioned, for the subsequent studies \cite{Kovchegov:2016zex, Kovchegov:2016weo, Kovchegov:2017jxc, Kovchegov:2017lsr, Kovchegov:2018znm, Cougoulic:2019aja, Kovchegov:2020hgb} only the DLA term was kept, since the SLA$_L$ kernel is only a part of the full SLA correction, and it would have been inconsistent to keep it and discard the other SLA term.

This other term, the ``single-logarithmic, transverse (SLA$_T$) kernel'' corresponds to $\mathcal{K}_{\text{SLA}_T}$ in \eqref{eqn:LLALT1}. As we will see below, its typical form is
\begin{align}
\as \, \mathcal{K}_{\text{SLA}_T} &= \frac{\alpha_s}{2\pi^2}\int\limits_0^z dz'\;\Delta P_T \left(\frac{z'}{z}\right) \int \frac{dx^2_{32}}{x^2_{32}},
\label{eqn:LLALT4}
\end{align}
where $\Delta P_T (z'/z)$ is a function of $z'/z$. Similar to the above, applying $\mathcal{K}_{\text{SLA}_T}$ to the sample initial condition $G^{(0)}=1$ gives a single logarithm of energy coming from the transverse $x_{32}^2$-integral, where the ``parent" parton splits into two ``daughter" partons at ${\un x}_2$ and ${\un x}_{3}$. The derivation of this SLA$_T$ kernel involves the splitting of dipoles with $z'\sim z$ but $x^2_{32}\ll x^2_{10}$, i.e., only the transverse separation is ordered. This implies that $\Delta P_T (z'/z)$ actually corresponds to polarized Dokshitzer-Gribov-Lipatov-Altarelli-Parisi (DGLAP)  \cite{Gribov:1972ri,Altarelli:1977zs,Dokshitzer:1977sg} splitting functions, as the notation might have suggested, with the logarithmically divergent parts of the splitting functions subtracted out to eliminate double-counting with the DLA kernel. We carry out the derivation of SLA$_T$ kernel in Sections \ref{sec:ingredients} and \ref{sec:eqnnorc}.


\section{Ingredients for SLA$_T$ Calculation}
\label{sec:ingredients}

In this Section, we derive the SLA$_T$ part of the evolution kernel. We start with the light-cone wave functions corresponding to $q \to qG$, $G \to q \bar q$ and $G \to GG$ splittings calculated at the lowest order in $\as$. The derivation is done in the $A^- =0$ light-cone gauge of the minus-moving projectile partons using the light-cone perturbation theory (LCPT) \cite{Lepage:1980fj,Brodsky:1997de}. We assume that all quarks are massless, since mass-dependent terms do not contribute to the SLA evolution. 


\subsection{$q\to qG$ Splitting Kernels}
\label{sec:ingredientsqqG}

\begin{figure}[h]
\includegraphics[width= 0.42 \textwidth]{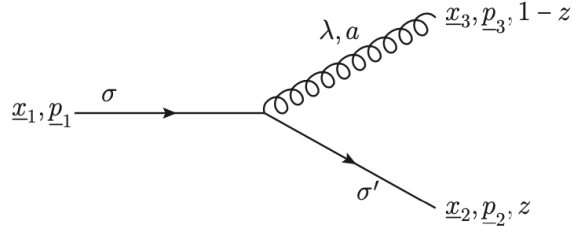} 
\caption{The diagram contributing to the leading-order light-cone wave function for the $q \to qG$ splitting.}
\label{fig:qqG}
\end{figure}

Consider a quark splitting into a quark and a gluon, described by the diagram shown in \fig{fig:qqG}.
By LCPT rules \cite{Lepage:1980fj,Brodsky:1997de} in the convention of \cite{Kovchegov:2012mbw}, the momentum-space $q\to qG$ splitting wave function is of the form
\begin{align}
\tilde{\Psi}^{q\to qG}_{a \sigma\sigma'\lambda}(p_1,p_{2}) &=  - gt^a\delta_{\sigma\sigma'}\sqrt{z}\;\frac{\ee^*_{\lambda}\cdot (\pp_2-z\pp_1 )}{|\pp_2-z\pp_1|^2}\left[1+z+\sigma\lambda(1-z)\right],
\label{eqn:psiqqG1}
\end{align}
where we have used momentum conservation $p_3=p_1-p_2$. Here $\sigma$ and $\sigma'$ are the quark polarizations before and after gluon emission, while $\lambda$ is the polarization of the gluon, which also carries the color $a$, as shown in \fig{fig:qqG}. The SU($N_c$) generators in the fundamental representation are denoted by $t^a$. The transverse momentum vectors are denoted by ${\un p}$, while $z = p_2^-/p_1^-$ is the light-cone momentum fraction of the incoming quark carried by the quark after the splitting. Gluon polarization 4-vector in $A^- =0$ light-cone gauge is $\epsilon_\lambda^\mu (k) = (\tfrac{\ee_{\lambda}\cdot {\un k}}{k^-}, 0, \ee_{\lambda})$ in the $(+, -, \perp)$ component notation with $\ee_{\lambda} = (-1/\sqrt{2}) (\lambda, i)$ \cite{Lepage:1980fj}. 

Performing the transverse Fourier transform with $\xx_{21}$ being the conjugate of $\pp_2$ and $\xx_{31}$ the conjugate of $\pp_1$, yields
\begin{align}
\Psi^{q\to qG}_{a \sigma\sigma'\lambda}(\xx_1,\xx_{2},\xx_3,z) &= \psi^{q\to qG}_{a \sigma\sigma'\lambda}(\xx_{32},z) \, \delta^2\left[z\xx_{21} + (1-z)\xx_{31}\right]
\label{eqn:psiqqG2}
\end{align}
with 
\begin{align}
\psi^{q\to qG}_{a \sigma\sigma'\lambda}(\xx_{32},z) &= \frac{ig}{2\pi} t^a \, \delta_{\sigma\sigma'}\sqrt{z}\left[1+z+\sigma\lambda(1-z)\right]\frac{\ee^*_{\lambda}\cdot\xx_{32}}{x^2_{32}}.
\label{eqn:psiqqG3}
\end{align}

Following the notation similar to \cite{Kovchegov:2015pbl} we define the splitting kernel 
\begin{align}\label{Kdef}
K^{q {\bar q} \to q {\bar q}} = \sum_{\sigma, \lambda} \sigma \, \psi^{q\to qG}_{a \sigma_1 \sigma \lambda}(\xx_{32},z) \left[ \psi^{q\to qG}_{b \sigma_2 \sigma \lambda}(\xx_{32},z) \right]^* + \mbox{virtual} \, \, \, \mbox{term} ,
\end{align}
which is projected on the polarization-dependent channel by the factor of polarization $\sigma$ in the front, coming from the polarized target. (Such a projection, including a sum over $\sigma$, were absent in \cite{Kovchegov:2015pbl}.) The virtual term corresponds to the case where both the $q\to qG$ splitting and the subsequent $qG \to q$ merger reside in the wave function or its complex conjugate. This is similar to, for instance, the derivation of the DGLAP evolution equation \cite{Gribov:1972ri,Altarelli:1977zs,Dokshitzer:1977sg} using LCPT in \cite{Kovchegov:2012mbw}. The virtual corrections can be constructed from the real (first) term in \eq{Kdef} by the unitarity argument \cite{Kovchegov:2012mbw}.

Employing the wave function in Eqs.~\eqref{eqn:psiqqG2} and \eqref{eqn:psiqqG3} we arrive at the following evolution kernels:

\begin{itemize}

\item $q \to qG$ splitting, quark is polarized with momentum fraction $z$: we get
\begin{align}\label{Kqqqq1}
K^{q {\bar q} \to q {\bar q}} = \frac{\as}{2 \pi^2} \, \sigma_1 \, \delta_{\sigma_1 \sigma_2} \, \left[ t^b \otimes t^a \, \int dz \, \frac{1+z^2}{1-z} \, \frac{d^2 x_{32}}{x_{32}^2} - C_F \,  \int dz \, \frac{1+z^2}{1-z} \, \frac{d^2 x_{32}}{x_{32}^2} \right],
\end{align}
where the second term is due to the virtual corrections, while $t^b \otimes t^a$ in the real term denotes the fundamental color matrices whose color indices are not contacted as they will be inserted between the Wilson lines in the evolution equations. All the integration limits will be specified when inserting the kernel in evolution equations. Here $C_F = (N_c^2 -1)/2 N_c$ is the fundamental Casimir operator. The pole at $z=1$ needs to be subtracted out to remove DLA double-counting. In addition, the real emission term, when multiplied by the energy-suppressed helicity-dependent interaction with the polarized target (shock wave) $\sim 1/(z s)$ \cite{Kovchegov:2015pbl, Kovchegov:2016zex, Kovchegov:2018znm} (with $s$ the center-of-mass energy squared for the projectile--target scattering), has a pole at $z=0$, which also needs to be subtracted out to eliminate double-counting with DLA. We are left with
\begin{tcolorbox}[ams align]\label{Kqqqq2}
K^{q {\bar q} \to q {\bar q}} = \frac{\as}{2 \pi^2} \, \sigma_1 \, \delta_{\sigma_1 \sigma_2} \, \left[ - t^b \otimes t^a \, \int dz \, z \ \frac{d^2 x_{32}}{x_{32}^2} + C_F \,  \int dz \, (1+z) \, \frac{d^2 x_{32}}{x_{32}^2} \right].
\end{tcolorbox}

\item $q \to qG$ splitting, gluon is polarized with momentum fraction $z$. Defining the corresponding splitting kernel by
\begin{align}\label{Kdef2}
K^{q {\bar q} \to GG} = \sum_{\sigma, \lambda} \lambda \, \psi^{q\to qG}_{a \sigma_1 \sigma \lambda}(\xx_{32},1-z) \left[ \psi^{q\to qG}_{b \sigma_2 \sigma \lambda}(\xx_{32},1-z) \right]^* ,
\end{align}
we readily obtain
\begin{align}\label{KqqGG1}
K^{q {\bar q} \to GG} = \frac{\as}{2 \pi^2} \, t^b \otimes t^a \, \sigma_1 \, \delta_{\sigma_1 \sigma_2} \, \int dz \, (2-z) \, \frac{d^2 x_{32}}{x_{32}^2}, 
\end{align}
as there is no virtual correction. Subtracting the $z=0$ pole due to the $1/zs$ factor due to interaction with the polarized target to eliminate double-counting with DLA yields 
\begin{tcolorbox}[ams align]\label{KqqGG2}
K^{q {\bar q} \to GG} = - \frac{\as}{2 \pi^2} \, t^b \otimes t^a \, \sigma_1 \, \delta_{\sigma_1 \sigma_2} \, \int dz \, z \ \frac{d^2 x_{32}}{x_{32}^2}. 
\end{tcolorbox}

\end{itemize}


\subsection{$G\to q\bar{q}$ Splitting Kernel}
\label{sec:ingredientsGqq}

\begin{figure}[h]
\includegraphics[width= 0.42 \textwidth]{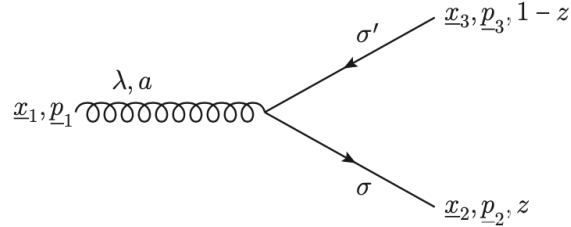} 
\caption{The diagram contributing to the lowest-order light-cone wave function for the $G \to q \bar q$ splitting.}
\label{fig:Gqq}
\end{figure}

Now, consider a gluon splitting into a quark-antiquark pair, described by the diagram in \fig{fig:Gqq}.

Employing the LCPT rules \cite{Lepage:1980fj,Brodsky:1997de,Kovchegov:2012mbw}, the momentum-space $G\to q\bar{q}$ splitting light-cone wave function is
\begin{align}
\tilde{\Psi}^{G\to q \bar q}_{a \lambda\sigma\sigma'}(p_1,p_{2}) &=  gt^a\delta_{\sigma,-\sigma'}\sqrt{z(1-z)}\;\frac{\ee_{\lambda}\cdot(\pp_2-z\pp_1)}{|\pp_2-z\pp_1|^2}\left[1-2z-\sigma\lambda\right],
\label{eqn:psiGqq1}
\end{align}
where we have again used momentum conservation $p_3=p_1-p_2$. Fourier-transforming into the transverse coordinate space we arrive at
\begin{align}
\Psi^{G\to q \bar q}_{a \lambda\sigma\sigma'}(\xx_1,\xx_{2},\xx_3,z) &= \psi_{a\lambda\sigma\sigma'}^{G\to q \bar q}(\xx_{32},z) \ \delta^2\left[z\xx_{21} + (1-z)\xx_{31}\right]
\label{eqn:psiGqq2}
\end{align}
with 
\begin{align}
\psi^{G\to q \bar q}_{a \lambda\sigma\sigma'}(\xx_{32},z) &= - \frac{ig}{2\pi} t^a \delta_{\sigma,-\sigma'}\sqrt{z(1-z)}\left[1-2z-\sigma\lambda\right]\frac{\ee_{\lambda}\cdot\xx_{32}}{x^2_{32}}.
\label{eqn:psiGqq3}
\end{align}

\begin{itemize}

\item $G \to q{\bar q}$ splitting, either quark or anti-quark are polarized with momentum fraction $z$: defining the splitting kernel
\begin{align}\label{Kdef3}
K^{GG \to q {\bar q}} = \sum_{\sigma, \sigma'} \sigma \, \psi^{G\to q \bar q}_{a \sigma \sigma' \lambda_1}(\xx_{32},z) \left[ \psi^{G\to q \bar q}_{b \sigma \sigma' \lambda_2}(\xx_{32},z) \right]^* ,
\end{align}
we get
\begin{align}\label{KGGqq1}
K^{GG \to q {\bar q}} = - \frac{\as}{2 \pi^2} \, t^b \otimes t^a \, \sum_f \, \lambda_1 \, \delta_{\lambda_1 \lambda_2} \, \int dz \, (1 - 2 z) \, \frac{d^2 x_{32}}{x_{32}^2}, 
\end{align}
there is no virtual correction. Subtracting the $z=0$ pole due to $1/zs$ to eliminate double-counting with DLA yields 
\begin{tcolorbox}[ams align]\label{KGGqq2}
K^{GG \to q {\bar q}} = \frac{\as}{\pi^2} \, t^b \otimes t^a \, \sum_f \, \lambda_1 \, \delta_{\lambda_1 \lambda_2} \, \int dz \, z \  \frac{d^2 x_{32}}{x_{32}^2}. 
\end{tcolorbox}

\end{itemize}


\subsection{$G\to GG$ Splitting Kernel}
\label{sec:ingredientsGGG}

\begin{figure}[h]
\includegraphics[width= 0.42 \textwidth]{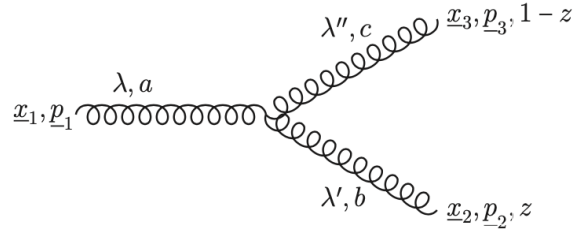} 
\caption{The diagram contributing to the real part of the leading-order light-cone wave function for the $G \to GG$ splitting.}
\label{fig:GGG}
\end{figure}

Finally, consider a gluon splitting into two gluons, described by the diagram in \fig{fig:GGG}. 
The momentum-space $G\to GG$ splitting wave function is 
\begin{align}\label{eqn:psiGGG1}
& \tilde{\Psi}^{G\to GG}_{\lambda\lambda'\lambda''}(p_1,p_{2}) = - 2igf^{abc}  \\ & \times \left[\delta_{\lambda',-\lambda''}z(1-z) \frac{\ee_{\lambda}\cdot(\pp_2-z\pp_1)}{|\pp_2-z\pp_1|^2} + \delta_{\lambda\lambda'}z \frac{\ee_{\lambda''}^*\cdot(\pp_2-z\pp_1)}{|\pp_2-z\pp_1|^2} + \delta_{\lambda\lambda''}(1-z) \frac{\ee^*_{\lambda'}\cdot(\pp_2-z\pp_1)}{|\pp_2-z\pp_1|^2} \right] . \notag
\end{align}
In the transverse coordinate space the wave function becomes
\begin{align}
\Psi^{G\to GG}_{\lambda\lambda'\lambda''}(\xx_1,\xx_{2},\xx_3,z) &= \psi^{G\to GG}_{\lambda\lambda'\lambda''}(\xx_{32},z) \ \delta^2\left[z\xx_{21} + (1-z)\xx_{31}\right]
\label{eqn:psiGGG2}
\end{align}
with 
\begin{align}
\psi^{G\to GG}_{\lambda\lambda'\lambda''}(\xx_{32},z) &= - \frac{g}{\pi}  f^{abc} \left[z(1-z)\delta_{\lambda',-\lambda''}\ee_{\lambda} + z\delta_{\lambda\lambda'}\ee_{\lambda''}^* + (1-z)\delta_{\lambda\lambda''}\ee_{\lambda'}^*\right]\cdot\frac{\xx_{32}}{x^2_{32}}.
\label{eqn:psiGGG3}
\end{align}

\begin{itemize}

\item $G \to GG$ splitting, either gluon is polarized with momentum fraction $z$: similar to the above, we define the splitting kernel as
\begin{align}\label{Kdef4}
K^{GG \to GG} = \sum_{\lambda, \lambda'} \lambda \, \psi^{G\to GG}_{\lambda_1 \lambda \lambda'}(\xx_{32},z) \left[ \psi^{G\to GG}_{\lambda_2 \lambda \lambda'}(\xx_{32},z) \right]^* + \mbox{virtual} \, \, \, \mbox{terms} ,
\end{align}
we obtain
\begin{align}\label{KGGGG1}
K^{GG \to GG} = \frac{\as}{\pi^2} \,  \lambda_1 \, \delta_{\lambda_1 \lambda_2} \, \Bigg[ T^b \otimes T^a \, \int dz \, \frac{2 (1-z)^2 +z}{1-z} \, \frac{d^2 x_{32}}{x_{32}^2}  & - \frac{N_c}{2} \, \int dz \, \left( \frac{z}{1-z} + \frac{1-z}{z} + z (1-z) \right)  \, \frac{d^2 x_{32}}{x_{32}^2} \notag \\ &  - \frac{N_f}{4} \int dz \, [z^2 + (1-z)^2] \, \frac{d^2 x_{32}}{x_{32}^2} \Bigg], 
\end{align}
where the second and third terms are virtual corrections, while $T^a$ are the adjoint color matrices. The poles at $z=1$ and $z=0$ need to be subtracted out to remove DLA double-counting: in the real term the $z=0$ pole is due to $\sim 1/zs$ interaction with the target \cite{Kovchegov:2015pbl, Kovchegov:2016zex, Kovchegov:2018znm}, in the virtual gluon loop term (the second term on the right of \eq{KGGGG1}) the poles are explicitly given in the expression, while the virtual quark loop term (the last term on the right of \eq{KGGGG1}) has neither of the poles. We get
\begin{tcolorbox}[ams align]\label{KGGGG}
K^{GG \to GG} = \frac{\as}{\pi^2} \,  \lambda_1 \, \delta_{\lambda_1 \lambda_2} \, \Bigg[ - T^b \otimes T^a \, \int dz \, 2 z \ \frac{d^2 x_{32}}{x_{32}^2} & + \frac{N_c}{2} \, \int dz \, \left( 2  - z (1-z) \right)  \, \frac{d^2 x_{32}}{x_{32}^2}  \notag \\ & - \frac{N_f}{4} \int dz \, [z^2 + (1-z)^2] \, \frac{d^2 x_{32}}{x_{32}^2} \Bigg]. 
\end{tcolorbox}

\end{itemize}


\section{Evolution Equations}
\label{sec:eqnnorc}

We are now ready to write down the DLA+SLA evolution equations by including both the SLA$_L$ and SLA$_T$ terms into the evolution for the polarized fundamental and adjoint dipole amplitudes. We begin by writing the $g_1$ structure function as \cite{Kovchegov:2015pbl}
\begin{align}\label{Qtog1}
\begin{split}
g_1(x,Q^2) &= - \frac{s}{8\pi^2\alpha_{EM}}\int\limits_{\Lambda^2/s}^1\frac{dz}{z(1-z)}\int\frac{d^2x_0\;d^2x_1}{4\pi} \; \frac{1}{2}\sum_{\sigma,\sigma',\lambda,f} \lambda \left|\Psi_T^{\gamma^*\to q\overline{q}}(\xx_{10},z)\right|^2    \\
&\;\;\;\;\;\times\; \left\langle \mbox{T} \, \mbox{tr} \left[ V_{\ul 0} \, V_{{\un 1}}^{\dagger} (\sigma')\right] + \bar{\mbox{T}} \, \mbox{tr} \left[ V_{{\un 1}}(\sigma') \, V_{\ul 0}^\dagger \right] +  \mbox{T} \, \mbox{tr} \left[ V_{{\un 1}} (\sigma') \, V_{\ul 0}^{\dagger} \right] + \bar{\mbox{T}} \, \mbox{tr} \left[ V_{{\un 1}}^\dagger (\sigma') \, V_{\ul 0} \right] \right\rangle \left(\min\{z,1-z\},z\right) ,
\end{split}
\end{align}
where $\alpha_{EM}$ is the electromagnetic coupling constant. Note that \eq{Qtog1} is valid at both DLA and SLA. Here, $\sigma$, $\sigma'$ and $\lambda$ are helicities of the quark, antiquark and virtual photon, respectively. Also, $\Lambda$ is the infrared cutoff and $s$ is the center-of-mass energy squared of the virtual photon--target scattering. The angle brackets denote averaging over the longitudinally polarized target (proton) wave function, the trace (tr) is over the fundamental indices, and $z$ is the longitudinal momentum fraction of the virtual photon carried by the polarized (anti)quark line. In the $\text{DLA} + \text{SLA}$ evolution, for the reason explained below, one needs to keep track of both $z$ and the smallest of the two longitudinal momentum fractions carried by the quark and the antiquark, $\min\{z,1-z\}$. Time-ordered and anti-time-ordered products are denoted by T and $\bar{\mbox{T}}$, respectively.

In \eq{Qtog1}, the light-cone fundamental Wilson lines are
\begin{align}\label{Vdef}
  V_{\un{x}} [b^-, a^-] = \mathcal{P} \exp \left[ i g \int\limits_{a^-}^{b^-} d x^- \, A^+ (x^+=0, x^-, {\un x}) \right]
\end{align}
with the infinite Wilson lines abbreviated by $V_{\ul 0} = V_{{\ul x}_0} [\infty, -\infty]$. Furthermore, the helicity-dependent Wilson lines are defined as
\begin{align}
V_{{\un 1}}(\sigma) &= V_{{\un 1}} + \sigma V_{{\un 1}}^{\text{pol}},
\label{Qtog2}
\end{align}
where $V_{{\un 1}}^{\text{pol}} = V_{{\un x}_1}^{\text{pol}}$ is the so-called polarized fundamental Wilson line, defined in \cite{Kovchegov:2017lsr, Kovchegov:2018znm} for the DLA and SLA$_L$ evolution as 
\begin{align} 
  \label{eq:Vpol_all} 
  & V^{\text{pol}}_{\un x} = V^{\text{pol} \, g}_{\un x} + V^{\text{pol} \, q}_{\un x} = \frac{i g p_1^+}{s} \,
  \int\limits_{-\infty}^\infty d x^- \, V_{\ul x} [+\infty, x^-] \: 
  F^{12} (x^-, \un{x}) \: V_{\ul x} [x^- , -\infty] \\ & - \frac{g^2 \, p_1^+}{s}
  \int\limits_{-\infty}^\infty d x_1^- \, \int\limits_{x_1^-}^\infty d x_2^- \, V_{\ul x} [+\infty, x_2^-] \:  t^b \, {\psi}_\beta (x_2^-, {\un x})  \, U_{\ul x}^{ba} [ x_2^-,  x_1^-] \left[ \frac{1}{2} \, \gamma^+ \, \gamma^5 \right]_{\alpha\beta} \, {\bar \psi}_\alpha (x_1^-, {\un x}) \, t^a  \: V_{\ul x} [x_1^- , -\infty] \notag 
\end{align}
in terms of the target quark fields $\psi, {\bar \psi}$ and the (sub-eikonal) $F^{12}$ component of the gluon field strength tensor in the fundamental representation. Here $p_1^+$ is the light-cone momentum of the polarized parton in the target giving rise to those sub-eikonal fields and $s$ is the center-of-mass energy squared for the projectile-target scattering. The $g$ and $q$ superscripts in $V^{pol \, g}_{\un x}$ and $V^{pol \, q}_{\un x}$ denote the two terms on the right of \eq{eq:Vpol_all} with the gluon ($F^{12}$) and quark ($\psi, {\bar \psi}$) sub-eikonal operators inserted. 

The remaining object in \eqref{Qtog1} to be discussed is $\Psi_T^{\gamma^*\to q\overline{q}}$, which is the light-cone wave function for the transverse virtual photon splitting into a quark-antiquark pair. Explicitly, it is given by the following expression (see, e.g., \cite{Kovchegov:2012mbw}),
\begin{align}
& \Psi_T^{\gamma^*\to q\overline{q}}(\xx,z) = \frac{eZ_f}{2\pi}\sqrt{z(1-z)}\left[\delta_{\sigma,-\sigma'}\left(1-2z-\sigma\lambda\right)ia_f\frac{\ee_{\lambda}\cdot\xx}{x_{\perp}}K_1\left(x_{\perp}a_f\right) + \delta_{\sigma\sigma'}\frac{m_f}{\sqrt{2}}\left(1+\sigma\lambda\right)K_0\left(x_{\perp}a_f\right)  \right] , \label{Qtog3a} 
\end{align}
where $Q^2$ is the virtuality of the photon and, for each quark flavor $f$, $m_f$ is the quark mass and $Z_f$ denotes the charge of the quark in units of the electron's charge. We also define $a^2_f = z(1-z)Q^2 + m^2_f$.  Plugging Eq.~\eqref{Qtog3a}  into \eqref{Qtog1}, we have
\begin{align}\label{Qtog4}
\begin{split}
g_1(x,Q^2) &=   \frac{s}{4 \pi^3}\int\limits_{\Lambda^2/s}^1dz \int\frac{d^2x_0\;d^2x_1}{4\pi}  \sum_{f}Z_f^2\left[ \left(1- 2 z \right) a_f^2 \left[K_1\left(x_{\perp}a_f\right)\right]^2   - m_f^2 \left[K_0\left(x_{\perp}a_f\right)\right]^2\right] \\
&\;\;\;\;\;\times \;   \left\langle \mbox{T} \, \mbox{tr} \left[ V_{\ul 0} \, V_{{\un 1}}^{\text{pol}\dagger} \right] + \bar{\mbox{T}} \, \mbox{tr} \left[ V_{{\un 1}}^{\text{pol}} \, V_{\ul 0}^\dagger \right] + \mbox{T} \, \mbox{tr} \left[ V_{{\un 1}}^{\text{pol}} \, V_{\ul 0}^\dagger \right] + \bar{\mbox{T}} \, \mbox{tr} \left[ V_{{\un 1}}^{\text{pol} \dagger} \, V_{\ul 0} \right] \right\rangle \left(\min\{z,1-z\},z\right) .
\end{split}
\end{align}

This inspires the following definition of the fundamental polarized dipole amplitude for the flavor-singlet observables \cite{Kovchegov:2016zex, Kovchegov:2018znm}:
\begin{align}\label{Qdef10}
Q_{10} \left(z_{\min},z_{\text{pol}}\right) & = \frac{z_{\text{pol}} \, s}{2 N_c}  \:  \mbox{Re} \:  \left\langle \mbox{T} \, \mbox{tr} \left[ V_{\ul 0} \, V_{{\un 1}}^{\text{pol} \, \dagger} \right] + \mbox{T} \, \mbox{tr} \left[ V_{{\un 1}}^{\text{pol}} \, V_{\ul 0}^\dagger \right] \right\rangle \left(z_{\min},z_{\text{pol}}\right) \\ & \equiv \frac{1}{2 N_c}  \:  \mbox{Re} \:  \llangle \mbox{T} \, \mbox{tr} \left[ V_{\ul 0} \, V_{{\un 1}}^{\text{pol} \, \dagger} \right] + \mbox{T} \, \mbox{tr} \left[ V_{{\un 1}}^{\text{pol}} \, V_{\ul 0}^\dagger \right] \rrangle \left(z_{\min},z_{\text{pol}}\right) ,\notag
\end{align}
leading to
\begin{align}\label{Qtog5}
\begin{split}
g_1(x,Q^2) &=   \frac{N_c}{\pi^3}\int\limits_{\Lambda^2/s}^1\frac{dz}{z} \int\frac{d^2x_0\;d^2x_1}{4\pi}\; Q_{10}(\min\{z,1-z\},z) \\
&\;\;\;\;\;\times \sum_{f}Z_f^2\left[ \left( 1- 2 z \right)  a_f^2 \left[K_1\left(x_{\perp}a_f\right)\right]^2  - m_f^2 \left[K_0\left(x_{\perp}a_f\right)\right]^2\right]   .
\end{split}
\end{align}
As written in \eq{Qdef10}, the amplitude $Q_{10}$ depends on the longitudinal momentum fraction $z_{\text{pol}}$ of the polarized line and on the minimum momentum fraction $z_{\min}$, which we will describe shortly below.  

It is worth noting that the small-$x$ formalism of \cite{Kovchegov:2015pbl, Kovchegov:2016zex, Kovchegov:2017lsr, Kovchegov:2018znm, Cougoulic:2019aja} expresses the $g_1$ structure function along with the quark helicity PDFs and the transverse momentum-dependent PDFs (TMDs) in terms of the dipole amplitude $Q_{10}$. The formalism thus operates with non-zero transverse momenta and a non-zero transverse extent of the dipole: this is typical of the TMD approaches. In this respect the treatment of  \cite{Kovchegov:2015pbl, Kovchegov:2016zex, Kovchegov:2017lsr, Kovchegov:2018znm, Cougoulic:2019aja} is quite similar to the unpolarized small-$x$ evolution \cite{Balitsky:1995ub,Balitsky:1998ya,Kovchegov:1999yj,Kovchegov:1999ua, Jalilian-Marian:1997dw,Jalilian-Marian:1997gr,Weigert:2000gi,Iancu:2001ad,Iancu:2000hn,Ferreiro:2001qy}. However, there is an important difference which arises in the case of helicity, and is not present in the unpolarized case. As one can already see in \eq{eq:Vpol_all}, helicity-dependent interaction in the gluon sector is proportional to the local sub-eikonal operator $F^{12}$ inserted between the light-cone Wilson lines (see \cite{Altinoluk:2020oyd,Chirilli:2021lif,Chirilli:2018kkw} for independent derivations of this result). This is in contrast to the non-local operator $\epsilon^{ij}F^{+i} (x^-) F^{+j} (0^-)$ which describes the gluon helicity in the collinear framework \cite{Jaffe:1989jz} (with $\epsilon^{ij}$ the two-dimensional Levi-Civita symbol). The small-$x$ expression for $g_1$ structure function \eqref{Qtog5} appears to depend on the $F^{12}$ operator, while the large-$Q^2$ expansion for the same $g_1$ structure function leads to the $\epsilon^{ij}F^{+i} (x^-) F^{+j} (0^-)$ operator in the gluon sector (see \cite{Lampe:1998eu} and references therein).  Despite some similarities, there appears to be no obvious way to directly relate these two operators \cite{Kovchegov:2017lsr}. Inserting $F^{12}$ into a dipole Wilson line staple and Fourier-transforming into transverse momentum space gives a gluon helicity TMD operator which has no collinear limit (see, e.g., Eq.~(5.13) in \cite{Chirilli:2021lif}): this TMD is zero when integrated over all transverse momenta. Hence, unlike the Jafffe-Manohar operator $\epsilon^{ij}F^{+i} (x^-) F^{+j} (0^-)$, the $F^{12}$ operator does not lead to DGLAP evolution and does not generate logarithms of $Q^2$, as can be confirmed by a direct calculation \cite{Kovchegov:2018znm}. This difference in operators eventually leads to differences in small-$x$ helicity evolution presented in this work (along with that in \cite{Kovchegov:2015pbl, Kovchegov:2016zex, Kovchegov:2017lsr, Kovchegov:2018znm, Cougoulic:2019aja}) on the one hand and the evolution derived from DGLAP-based approaches \cite{Mertig:1995ny,Moch:2014sna} on the other hand. Curiously, in the quark sector, the helicity operator is ${\bar \psi} \gamma^+ \, \gamma^5 \psi$ in both the small-$x$ and the collinear approaches: no differences appear to arise there. A reconciliation between the two operators for the gluon helicity is the subject of our future work \cite{CKTT}.

The definition \eqref{eq:Vpol_all} has to be extended to include the terms needed for the SLA$_T$ evolution. This is also left for future work \cite{CKTT}. Here, we adopt the diagrammatic approach from \cite{Kovchegov:2015pbl} and think of $V_{{\un 1}}^{\text{pol}}$ as the helicity-dependent part of the (massless) quark scattering amplitude on a longitudinally polarized target (in transverse coordinate space). Hence, all possible diagrams in light-cone perturbation theory that give at least one logarithm of energy must be included in our evolution. The double angle brackets, as defined in the second line of \eq{Qdef10}, are the single angle brackets scaled by $z_{\text{pol}} \, s$ \cite{Kovchegov:2015pbl} to eliminate energy suppression of the sub-eikonal helicity-dependent scattering, with $z_{\text{pol}}$ the longitudinal (minus) momentum fraction of the polarized line (the line at ${\un x}_1$ in \eq{Qdef10}).

For the purpose of this work, it is necessary to keep track of the longitudinal momentum fraction, $z_{\text{pol}}$, of the polarized parton line, in addition to the \emph{minimum} longitudinal momentum fraction, $z_{\min}$, among all the splittings that occurred to the ancestor of the particular dipole. This $z_{\min}$ can be either the softest of all the lines in a correlator (e.g., it may be that $z_{\min} = \min \{ z_1, z_0\}$ for a polarized dipole amplitude $Q_{10}$ with $z_1$ and $z_0$ the minus momentum fractions of lines 1 and 0 respectively) or simply the upper cutoff on longitudinal momentum fractions of subsequent quark and gluon emissions (imposed, for instance, by including a virtual correction in the previous step of the DLA evolution with the momentum fraction $z_2$ such that $z_{\min} = z_2 \ll \min \{ z_1, z_0\}$). In turn, $z_{\text{pol}}$ is the longitudinal momentum fraction of the (only one) polarized line, such that $z_{\text{pol}} = z_1$ for the dipole amplitude $Q_{10}$; by definition of $z_{\min}$ we will always have $z_{\min} \le z_{\text{pol}}$. The DLA and SLA$_L$ helicity evolution \cite{Kovchegov:2015pbl}, along with the standard unpolarized evolution \cite{Balitsky:1995ub,Balitsky:1998ya,Kovchegov:1999yj,Kovchegov:1999ua,Jalilian-Marian:1997dw,Jalilian-Marian:1997gr,Weigert:2000gi,Iancu:2001ad,Iancu:2000hn,Ferreiro:2001qy}, all evolve with $z_{\min}$ due to the logarithmic nature of the longitudinal integrals in their kernels. However, as we have seen in Sec.~\ref{sec:ingredients}, the kernels of SLA$_T$ evolution come in with non-logarithmic $z$-integrals, which need to be evaluated exactly. For this purpose we need to keep the exact $z_{\text{pol}}$ dependence in the arguments of the correlators entering the evolution equations.

\begin{figure}[t]
\begin{center}
\includegraphics[width= 0.8 \textwidth]{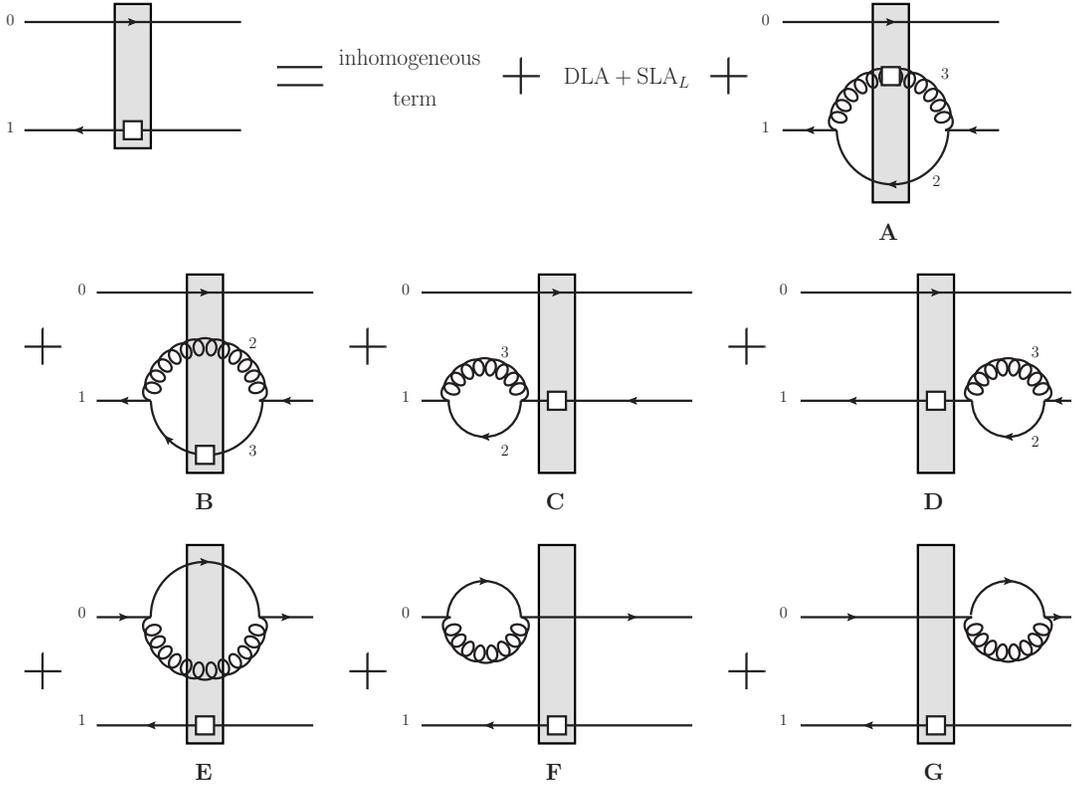}
\caption{Diagrams for SLA$_T$ splittings of a polarized fundamental dipole. For brevity, the inhomogeneous term (the initial condition) and the DLA+SLA$_L$ diagrams are not shown explicitly. }
\label{fig:qkdiagrams}
\end{center}
\end{figure}

The DLA and SLA$_L$ kernels for the fundamental dipole can be deduced directly from \cite{Kovchegov:2015pbl}. To derive the remaining SLA$_T$ terms, we consider all possible splittings in a polarized fundamental dipole with the daughter parton momentum fraction $z'\sim z_{\text{parent}}$, the momentum fraction of the parent (anti)quark line involved in the splitting. Schematically, all such splittings are shown in \fig{fig:qkdiagrams}. (The contributions with $z' \ll z_{\text{parent}}$ for the daughter gluon in all diagrams of \fig{fig:qkdiagrams} and with $z' \ll z_{\text{parent}}$ for the daughter quark in the diagram B are all parts of DLA+SLA$_L$ kernels.)  For brevity, we will only keep the first term in the angle brackets of \eq{Qdef10} both in \fig{fig:qkdiagrams} and in the corresponding equation, and drop the time-ordering sign (T) along with the Re when writing down the latter. As we mentioned above, \fig{fig:qkdiagrams} contains Feynman diagrams, with the solid lines denoting full quark propagators, rather than the Wilson lines from \eq{eq:Vpol_all}. Generalizing \eq{eq:Vpol_all} to include the SLA$_T$ terms is left for future work.

Note that the SLA$_T$ kernel consists of emissions and absorptions by the same parent (anti)quark line, as shown in \fig{fig:qkdiagrams}. The emission by one line and absorption by another generates a logarithm in the transverse position integral only if the daughter parton is far away from the parent dipole, $x_{21} \approx x_{20} \gg x_{10}$ \cite{Kovchegov:2015pbl}. However, the $x^-$-lifetime ordering condition for the emission, necessary for each step of the evolution \cite{Kovchegov:2015pbl, Cougoulic:2019aja}, is $z_{\text{parent}} \, x_{10}^2 \gg z' \, x_{21}^2$, which, for $z' \approx z_{\text{parent}}$, becomes $x_{10}^2 \gg x_{21}^2$, the exact opposite of the transverse logarithmic region condition $x_{21} \gg x_{10}$. Therefore, SLA$_T$ emissions and absorptions have to involve the same parent parton. (This lifetime estimate will be refined shortly, but the conclusion of emissions and absorptions coming from the same parent parton would not change.)

It is shown in Appendix~\ref{sec:unitarity} that diagrams $E$, $F$ and $G$ add up to zero with the SLA$_T$ accuracy, and only the diagrams $A-D$ from \fig{fig:qkdiagrams} contribute. Using the splitting kernels from Sec.~\ref{sec:ingredients} we can write down the following evolution equations for the polarized fundamental dipole amplitude defined in \eqref{Qdef10}:
\begin{align}\label{evol1}
  & \frac{1}{N_c} \, \left\langle \mbox{tr} \left[ V_{\ul{0}} \,
      V_{\ul{1}}^{\text{pol} \, \dagger} \right] \right\rangle \left(z_{\min},z_{\text{pol}}\right) =
  \frac{1}{N_c} \, \left\langle \mbox{tr} \left[ V_{\ul{0}} \,
      V_{\ul{1}}^{\text{pol} \, \dagger} \right] \right\rangle_0 \left(z_{\text{pol}}\right) +
  \frac{\as}{2 \pi^2} \int\limits_{\Lambda^2/s}^{z_{\min}} \frac{d z'}{z_{\text{pol}}} 
  \int\limits_{1/(z' s)} d^2 x_{2} \\ 
  & \;\;\;\; \times \left[ \left(
      \frac{1}{x_{21}^2} \, \theta (x_{10}^2 z_{\min} - x_{21}^2 z') -
      \frac{\ul{x}_{21} \cdot \ul{x}_{20}}{x_{21}^2 \, x_{20}^2} \,
      \theta (x_{10}^2 z_{\min} - \mbox{max} \{ x_{21}^2, x_{20}^2 \} z')
    \right) \right. \frac{2}{N_c} \left\langle \mbox{tr} \left[ t^b \,
      V_{\ul{0}} \, t^a \, V_{\ul{1}}^{\dagger} \right]
    \, U^{\text{pol} \, ba}_{\ul{2}} \right\rangle (z', z') \notag \\ 
    & \;\;\;\;\;\;\;\;\;\; \left. +
    \frac{1}{x_{21}^2} \, \theta (x_{10}^2 z_{\min} - x_{21}^2 z') \,
    \frac{1}{N_c} \left\langle \mbox{tr} \left[ t^b \,
        V_{\ul{0}} \, t^a \, V_{\ul{2}}^{\text{pol} \, \dagger} \right]
      \, U^{ba}_{\ul{1}} \right\rangle (z', z') \right] \notag \\ 
     & + \frac{\as}{\pi^2} \int\limits_{\Lambda^2/s}^{z_{\min}} \frac{d z'}{z'} 
  \int\limits_{1/(z' s)} d^2 x_{2} \, \frac{x_{10}^2}{x_{21}^2 \,
    x_{20}^2} \, \theta (x_{10}^2 z_{\min} - x_{21}^2 z') \notag \\ 
    & \;\;\;\; \times \, \frac{1}{N_c}
  \left[ \left\langle \mbox{tr} \left[ t^b \, V_{\ul{0}} \, t^a
        \, V_{\ul{1}}^{\text{pol} \, \dagger} \right] \, U^{
        ba}_{\ul{2}} \right\rangle (z' , z_{\text{pol}}) - C_F \left\langle \mbox{tr}
      \left[ V_{\ul{0}} \, V_{\ul{1}}^{\text{pol} \, \dagger} \right]
    \right\rangle (z' , z_{\text{pol}}) \right] \notag \\ 
    & \textcolor{blue}{ - \frac{\as}{2 \pi^2} \int\limits_{0}^{z_{\text{pol}}} \frac{d z' \, z'}{z_{\text{pol}}^2} 
  \int\limits_{\frac{z_{\text{pol}}}{z'(z_{\text{pol}}-z')s}} \frac{d^2 x_{32}}{x_{32}^2} \, \theta\left(x^2_{10}z_{\min}z_{\text{pol}} - x^2_{32}z'(z_{\text{pol}}-z')\right) } \notag \\
    & \;\;\;\; \textcolor{blue}{ \times \, \Bigg[ \frac{1}{N_c} \left\langle \mbox{tr} \left[ t^b \,
      V_{\ul{0}} \, t^a \, V_{\ul{x}_1 - \frac{z'}{z_{\text{pol}}} \ul{x}_{32}}^{\dagger} \right]
    \, U^{\text{pol} \, ba}_{\ul{x}_1 + \left( 1 - \frac{z'}{z_{\text{pol}}} \right) \ul{x}_{32}} \right\rangle \left(\min\left\{z_{\min},z', z_{\text{pol}}-z' \right\},z'\right) } \notag \\
    & \;\;\;\;\;\;\;\;\;\; \textcolor{blue}{ +
    \frac{1}{N_c} \left\langle \mbox{tr} \left[ t^b \,
        V_{\ul{0}} \, t^a \, V_{\ul{x}_1 + \left( 1 - \frac{z'}{z_{\text{pol}}} \right) \ul{x}_{32}}^{\text{pol} \, \dagger} \right]
      \, U^{ba}_{\ul{x}_1 - \frac{z'}{z_{\text{pol}}} \ul{x}_{32}} \right\rangle \left(\min\left\{z_{\min},z', z_{\text{pol}}-z' \right\},z'\right) \Bigg]}  \notag \\
      & \textcolor{blue}{ + \frac{\as \, C_F}{2 \pi^2} \int\limits_{0}^{z_{\text{pol}}}  \frac{d z'}{z_{\text{pol}}} \left( 1 + \frac{z'}{z_{\text{pol}}} \right)  \int\limits_{\frac{z_{\text{pol}}}{z'(z_{\text{pol}}-z')s}} \frac{d^2 x_{32}}{x_{32}^2} \, \theta\left(x^2_{10}z_{\min}z_{\text{pol}} - x^2_{32}z'(z_{\text{pol}}-z')\right)  } \notag \\
    & \;\;\;\; \textcolor{blue}{ \times \, \frac{1}{N_c}  \left\langle \mbox{tr}
      \left[ V_{\ul{0}} \, V_{\ul{1}}^{\text{pol} \, \dagger} \right]
    \right\rangle \left(\min\left\{z_{\min},z', z_{\text{pol}}-z' \right\},z_{\text{pol}}\right) , } \notag
\end{align}
where the last five (blue color) lines contain the SLA$_T$ contributions. From this point on, we write all the terms corresponding to SLA$_T$ evolution in blue, while all the DLA+SLA$_L$ terms remain black. Also, $U_{\un x} [b^-, a^-]$ is the adjoint Wilson line, a counterpart of the fundamental one from \eq{Vdef} with $U_{\un 2} = U_{{\un x}_2} [+\infty, -\infty]$, while $U_{\un x}^{\text{pol}}$ is the adjoint polarized Wilson line, defined in \cite{Kovchegov:2018znm} at the DLA+SLA$_L$ level as
\begin{align} 
  \label{eq:UpolFull}
  & (U_{\ul x}^{\text{pol}})^{ab} = (U_{\ul x}^{\text{pol} \, g})^{ab} + (U_{\ul x}^{\text{pol} \, q})^{ab}  =  \frac{2 i \, g \, p_1^+}{s}
  \int\limits_{-\infty}^{+\infty} dx^- \: \left( U_{\ul{x}}[+\infty, x^-] \:
  {\cal F}^{12} (x^+ =0 , x^- , \ul{x}) \: U_{\ul{x}} [x^- , -\infty] \right)^{ab} \\ &  - \frac{g^2 \, p_1^+}{s} \, \int\limits_{-\infty}^\infty d x_1^- \, \int\limits_{x_1^-}^\infty d x_2^- \, U^{aa'}_{\un x} [+\infty, x_2^-] \,  {\bar \psi} (x_2^-, {\un x}) \, t^{a'} \, V_{\un x} [x_2^-, x_1^-] \, \frac{1}{2} \, \gamma^+ \gamma_5 \, t^{b'} \,  \psi (x_1^-, {\un x}) \, U^{b'b}_{\un x} [x_1^-, -\infty] - c.c.  , \notag
\end{align}
with the extension of that definition to include the SLA$_T$ terms left for future work. Here ${\cal F}^{12}$ denotes the adjoint field strength tensor, while $U_{\ul x}^{pol \, g}$ and $U_{\ul x}^{pol \, q}$ again denote the gluon and quark exchange terms, respectively. The inhomogeneous term, given by the initial conditions to the evolution, is denoted by $\langle \ldots \rangle_0$ in \eq{evol1} and everywhere below. It does not depend on $z_{\min}$ \cite{Kovchegov:2016zex}.

Let us note again that here, for brevity, we are only showing the evolution for one of the traces in \eq{Qdef10}, while still keeping only terms corresponding to the evolution of the flavor-singlet amplitude \eqref{Qdef10} on the right-hand side of the equation. 
In particular, the complete version of \eq{evol1} giving evolution for the single-trace operator $\frac{1}{N_c} \, \left\langle \mbox{tr} \left[ V_{\ul{0}} \,
      V_{\ul{1}}^{\text{pol} \, \dagger} \right] \right\rangle \left(z_{\min},z_{\text{pol}}\right)$ would include extra terms we did not show explicitly. Such terms cancel similar terms in the evolution for the second trace in \eq{Qdef10}, $\frac{1}{N_c} \, \left\langle \mbox{tr} \left[  V_{\ul{1}}^{\text{pol}} \,
      V_{\ul{0}}^{\dagger} \right] \right\rangle \left(z_{\min},z_{\text{pol}}\right)$, when the evolution equations for the two traces are added together to construct an evolution equation for $Q_{10} \left(z_{\min},z_{\text{pol}}\right)$. These terms do not contribute to the evolution of the flavor-singlet amplitude $Q_{10} \left(z_{\min},z_{\text{pol}}\right)$ defined in \eq{Qdef10}: we drop them from \eq{evol1} for brevity. However, they do contribute to the evolution of the flavor non-singlet polarized dipole amplitude \cite{Kovchegov:2016zex,Kovchegov:2018znm},
 \begin{align}\label{QNSdef10}
Q^{NS}_{10} \left(z_{\min},z_{\text{pol}}\right) & \equiv \frac{1}{2 N_c}  \:  \mbox{Re} \:  \llangle \mbox{T} \, \mbox{tr} \left[ V_{\ul 0} \, V_{{\un 1}}^{\text{pol} \, \dagger} \right] - \mbox{T} \, \mbox{tr} \left[ V_{{\un 1}}^{\text{pol}} \, V_{\ul 0}^\dagger \right] \rrangle \left(z_{\min},z_{\text{pol}}\right),
\end{align}
since they do not cancel in the evolution equation for the difference of the two traces. Further study of this object and its DLA+SLA$_L$ evolution is presented in \cite{Kovchegov:2016zex, Kovchegov:2018znm, Chirilli:2021lif}.
Below every time we write an evolution equation for one fundamental trace, we will only show the evolution terms surviving in the sum \eqref{Qdef10}. 

The theta function in each term of \eqref{evol1} follows from light-cone lifetime ordering, which is necessary for the splitting functions derived in Sec.~\ref{sec:ingredients}  and for DLA+SLA$_L$ splittings to dominate in each step of evolution. In order to generate a logarithm, the energy denominator in the LCPT rules must be dominated by the quantities related to the latest splitting step. (For virtual corrections derivation of the lifetime ordering condition is more involved \cite{Kovchegov:2016zex}, but results in the same theta-functions.) In the DLA+SLA$_L$ terms of \eq{evol1} the theta-functions are only needed in the DLA limit, providing an unnecessary IR cutoff for the SLA$_L$ terms, in which the transverse integrals are IR- and UV-convergent.

The origin of the theta function in the new SLA$_T$ terms can be illustrated by imposing lifetime ordering in the diagram A from \fig{fig:qkdiagrams} above. Assign to each line in this diagram transverse momentum ${\un k}_i$ and the light-cone momentum fraction $z_i$ with $i=0, 1, 2, 3$ according to the labeling of the lines in the figure. Furthermore, for SLA$_T$ emission we have ${\un k}_2 \approx - {\un k}_3$ and $|{\un k}_2 | \approx |{\un k}_3| \gg |{\un k}_1 |$. In addition, assume that $z_0 \ll z_1$, such that $z_{\min} = z_0$ and $z_{\text{pol}} = z_1$. Then, for the partons 2 and 3 to dominate the light-cone energy denominator we have the following condition
\begin{align}\label{LCordering1}
\frac{{\un k}_0^2}{z_0} \ll \frac{{\un k}_2^2}{z_2} + \frac{{\un k}_3^2}{z_3} \approx {\un k}_2^2 \, \frac{z_1}{z_3 \, (z_1 - z_3)} = {\un k}_2^2 \, \frac{z_{\text{pol}}}{z_3 \, (z_{\text{pol}} - z_3)}, 
\end{align}
where we have used $z_1 = z_2 + z_3$. Since $k_{0\perp} \equiv |{\un k}_0| \approx 1/x_{10}$ and $k_{2\perp} \approx 1/x_{32}$, the condition \eqref{LCordering1} leads to 
\begin{align}\label{LCordering2}
x^2_{10} \, z_{\min} \, z_{\text{pol}} \gg x_{32}^2 \, z_3 \, (z_{\text{pol}} - z_3). 
\end{align}
Identifying $z_3$ in the diagram A from \fig{fig:qkdiagrams} with $z'$ in \eq{evol1}, we reproduce the theta-functions in the SLA$_T$ terms in the latter. Note that, in SLA$_T$ terms we have $z' = z_3 \sim z_{\text{pol}}$, such that the condition \eqref{LCordering2} reduces to $x^2_{10} \, z_{\min} \gg x_{32}^2 \, z_{\text{pol}}$ with the logarithmic accuracy of the associated transverse integrals. This is a more restrictive condition than $x^2_{10} \gg x_{32}^2$ mentioned above only if $z_{\min} \ll z_{\text{pol}}$. Let us point out that even for $z_{\min} \ll z_{\text{pol}}$ the $x^2_{10} \, z_{\min} \gg x_{32}^2 \, z_{\text{pol}}$ condition gives $x_{32}^2 \ll x_{10}^2$, allowing only for emission and absorption from the same parent parton in the SLA$_T$ part of the evolution kernel. Below we will keep the entire condition \eqref{LCordering2} in the theta-functions, without simplifying it, even if by doing this we may somewhat exceed the precision of our calculation. Note that, as one can show, \eq{LCordering2} also applies for the case when $z_0 \gg z_1$, such that $z_{\min} = z_1$.

The lower bound on the transverse position integrals in \eq{evol1} implies a (cutoff) regulator of the UV divergences in the integrals \cite{Mueller:1994rr,Mueller:1994jq,Mueller:1995gb}: these UV divergences are at $x_{21}=0$ in the DLA and SLA$_L$ terms, and at $x_{32}=0$ in the SLA$_T$ terms. Requiring that the lifetime of the partons 2 and 3 from \eqref{LCordering1} is much longer than the width of the shock wave leads to 
\begin{align}\label{LCordering3}
x_{32}^2 \, \frac{z' \, (z_{\text{pol}} - z')}{z_{\text{pol}}} \gg \frac{1}{s},
\end{align}
justifying the lower bound of the $x_{32}$ integral in the SLA$_T$ terms in \eq{evol1}. Since $z' \sim z_{\text{pol}}$ for SLA$_T$ kernels, the condition \eqref{LCordering3} is equivalent to $x_{32}^2 > 1/(z' s)$ condition in the DLA and SLA$_L$ parts of the kernel. Again, we will keep the condition \eqref{LCordering3}, slightly exceeding our calculation's precision. 

It is worth noting that the limits of integration in \eq{evol1} resulting from lifetime ordering are valid up to a multiplicative constant. A constant under the logarithm in the DLA part of the kernel is an SLA-order correction. In Appendix \ref{sec:logconst}, we argue that such a constant under the logarithm can be eliminated by the choice of starting energy/rapidity for the evolution.

The delta-function in \eq{eqn:psiqqG2} relates the positions of partons 1, 2 and 3 in all SLA$_T$ diagrams shown in \fig{fig:qkdiagrams}, requiring that $z_1 {\un x}_1 = z_2 {\un x}_2 + z_3 {\un x}_3$. In arriving at the SLA$_T$ terms in \eq{evol1} we have employed this delta-function to rewrite the positions of the (polarized and regular) Wilson lines in terms of ${\un x}_1$ and ${\un x}_{32}$.

Writing  \cite{Kovchegov:2015pbl}
\begin{align}\label{redef0}
  \left\langle \ldots \right\rangle_\Sigma \left(z_{\min},z_{\text{pol}}\right) = \frac{1}{z_{pol} \, s} \,
  \left\langle \! \left\langle \ldots \right\rangle \!
  \right\rangle_\Sigma \left(z_{\min},z_{\text{pol}}\right),
\end{align} 
and simplifying some of the traces in \eq{evol1} we arrive at (cf. Eq.~(57) in \cite{Kovchegov:2015pbl} )
\begin{align}\label{eqn:qknorc2}
&\frac{1}{N_c}\llangle\text{tr}\left[V_{\underline{0}}V_{\underline{1}}^{\text{pol}\dagger}\right]\rrangle\left(z_{\min},z_{\text{pol}}\right) = \frac{1}{N_c}\llangle\text{tr}\left[V_{\underline{0}}V_{\underline{1}}^{\text{pol }\dagger}\right]\rrangle_0 \left(z_{\text{pol}}\right) + \frac{\alpha_s}{2\pi^2}\int\limits_{\Lambda^2/s}^{z_{\min}}\frac{dz'}{z'} \int\limits_{1/z's}d^2 x_2 \\
&\;\;\;\;\times \bigg[\left(\frac{1}{x^2_{21}}\theta\left(x^2_{10}z_{\min}-x^2_{21}z'\right) - \frac{\xx_{21}\cdot\xx_{20}}{x^2_{21}x^2_{20}}\theta\left(x^2_{10}z_{\min}-\max\left\{x^2_{21},x^2_{20}\right\}z'\right)\right)\frac{2}{N_c}\llangle\text{tr}\left[t^bV_{\underline{0}}t^aV_{\underline{1}}^{\dagger}\right]U_{\underline{2}}^{\text{pol }ba}\rrangle (z',z') \notag \\
&\;\;\;\;\;\;\;\;\;\;+ \frac{1}{x^2_{21}}\theta\left(x^2_{10}z_{\min}-x^2_{21}z'\right) \frac{1}{N_c}\llangle\text{tr}\left[t^bV_{\underline{0}}t^aV_{\underline{2}}^{\text{pol }\dagger}\right]U_{\underline{1}}^{ba}\rrangle (z',z')\bigg] \notag \\
&+ \frac{\alpha_s}{2\pi^2}\int\limits_{\Lambda^2/s}^{z_{\min}}\frac{dz'}{z'} \int\limits_{1/z's} d^2 x_2\;\frac{x^2_{10}}{x^2_{21}x^2_{20}}\;\theta\left(x^2_{10}z_{\min}-x^2_{21}z'\right) \notag \\
&\;\;\;\;\times \frac{1}{N_c}\left[\llangle\text{tr}\left[V_{\underline{0}}V_{\underline{2}}^{\dagger}\right]\text{tr}\left[V_{\underline{2}}V_{\underline{1}}^{\text{pol }\dagger}\right]\rrangle(z',z_{\text{pol}}) - N_c\llangle\text{tr}\left[V_{\underline{0}}V_{\underline{1}}^{\text{pol }\dagger}\right]\rrangle (z',z_{\text{pol}}) \right] \notag \\
&\color{blue} - \frac{\alpha_s}{2\pi^2}\int\limits_0^{z_{\text{pol}}}\frac{dz'}{z_{\text{pol}}} \int\limits_{\frac{z_{\text{pol}}}{z'(z_{\text{pol}}-z')s}}\frac{d^2 x_{32}}{x^2_{32}}\;\theta\left(x^2_{10}z_{\min}z_{\text{pol}} - x^2_{32}z'(z_{\text{pol}}-z')\right) \notag \\
&\color{blue} \;\;\;\;\times \Bigg[\frac{1}{N_c}\llangle\text{tr}\left[t^bV_{\underline{0}}t^aV^{\dagger}_{\xx_1 - \frac{z'}{z_{\text{pol}}}\xx_{32}}\right]U^{\text{pol }ba}_{\xx_1+\left(1-\frac{z'}{z_{\text{pol}}}\right)\xx_{32}}\rrangle \left(\min\left\{z_{\min},z', z_{\text{pol}}-z' \right\},z'\right) \notag \\
&\color{blue} \;\;\;\;\;\;\;\;\;\;+ \frac{1}{N_c}\llangle\text{tr}\left[t^bV_{\underline{0}}t^aV^{\text{pol }\dagger}_{\xx_1+\left(1-\frac{z'}{z_{\text{pol}}}\right)\xx_{32}}\right]U^{ba}_{\xx_1 - \frac{z'}{z_{\text{pol}}}\xx_{32}}\rrangle \left(\min\left\{z_{\min},z', z_{\text{pol}}-z' \right\}, z'\right) \Bigg] \notag \\
&\color{blue} + \frac{\alpha_sC_F}{2\pi^2}\int\limits_0^{z_{\text{pol}}}\frac{dz'}{z_{\text{pol}}}\left(1+\frac{z'}{z_{\text{pol}}}\right)\int\limits_{\frac{z_{\text{pol}}}{z'(z_{\text{pol}}-z')s}}\frac{d^2 x_{32}}{x^2_{32}}\;\theta\left(x^2_{10}z_{\min}z_{\text{pol}} - x^2_{32}z'(z_{\text{pol}}-z')\right) \notag \\ 
&\color{blue} \;\;\;\; \times \frac{1}{N_c}\llangle\text{tr}\left[V_{\underline{0}}V_{\underline{1}}^{\text{pol }\dagger}\right]\rrangle \left(\min\left\{z_{\min},z', z_{\text{pol}}-z'  \right\},z_{\text{pol}}\right). \notag
\end{align}
Equation \eqref{eqn:qknorc2} is almost our final result for the DLA+SLA small-$x$ evolution equation for the fundamental flavor-singlet polarized dipole amplitude. Below we will only further enhance it by specifying the scales of the strong coupling constants in various terms of its kernel. 

\begin{figure}[t]
\begin{center}
\includegraphics[width= 0.67 \textwidth]{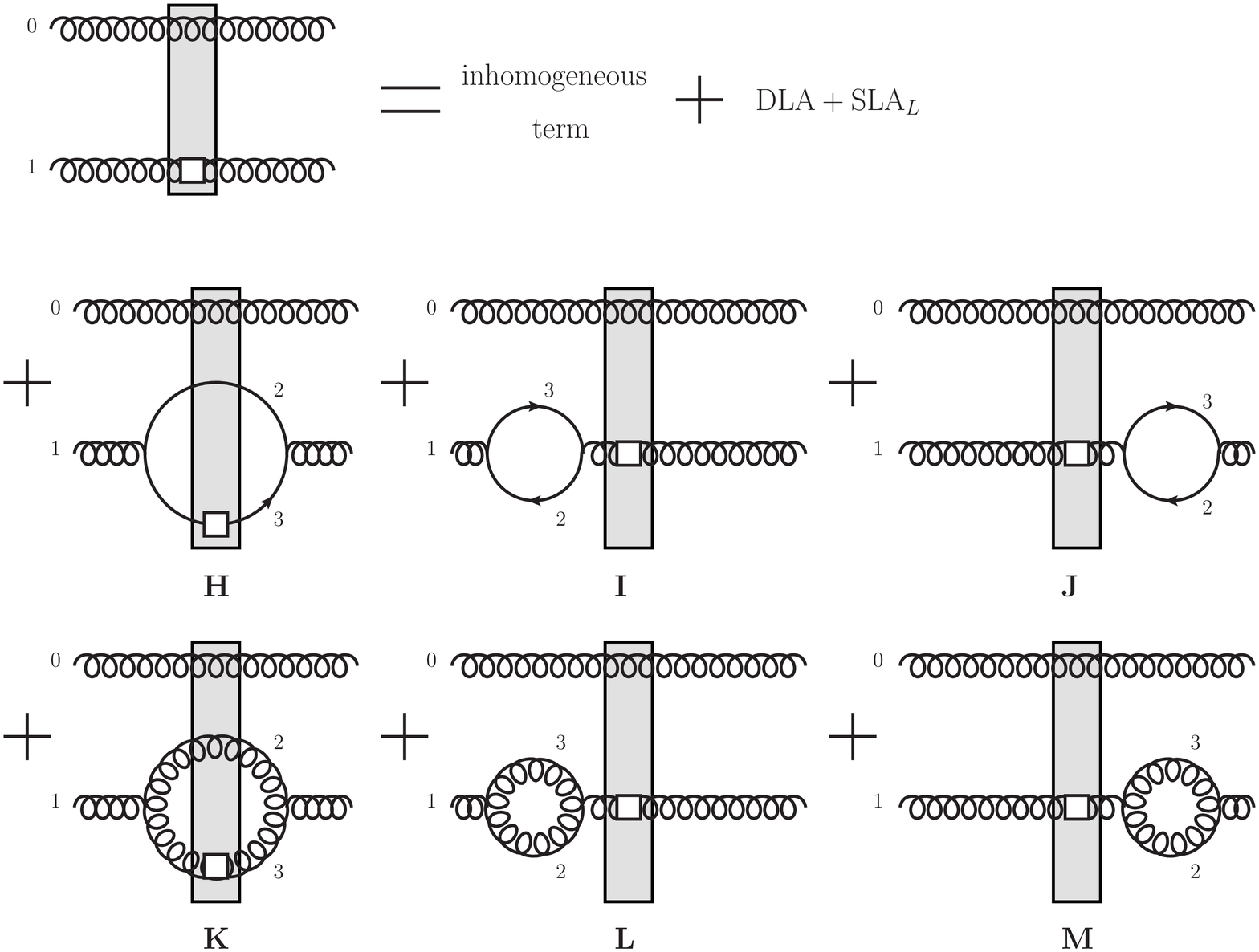} \\~~\\
\includegraphics[width= 0.67 \textwidth]{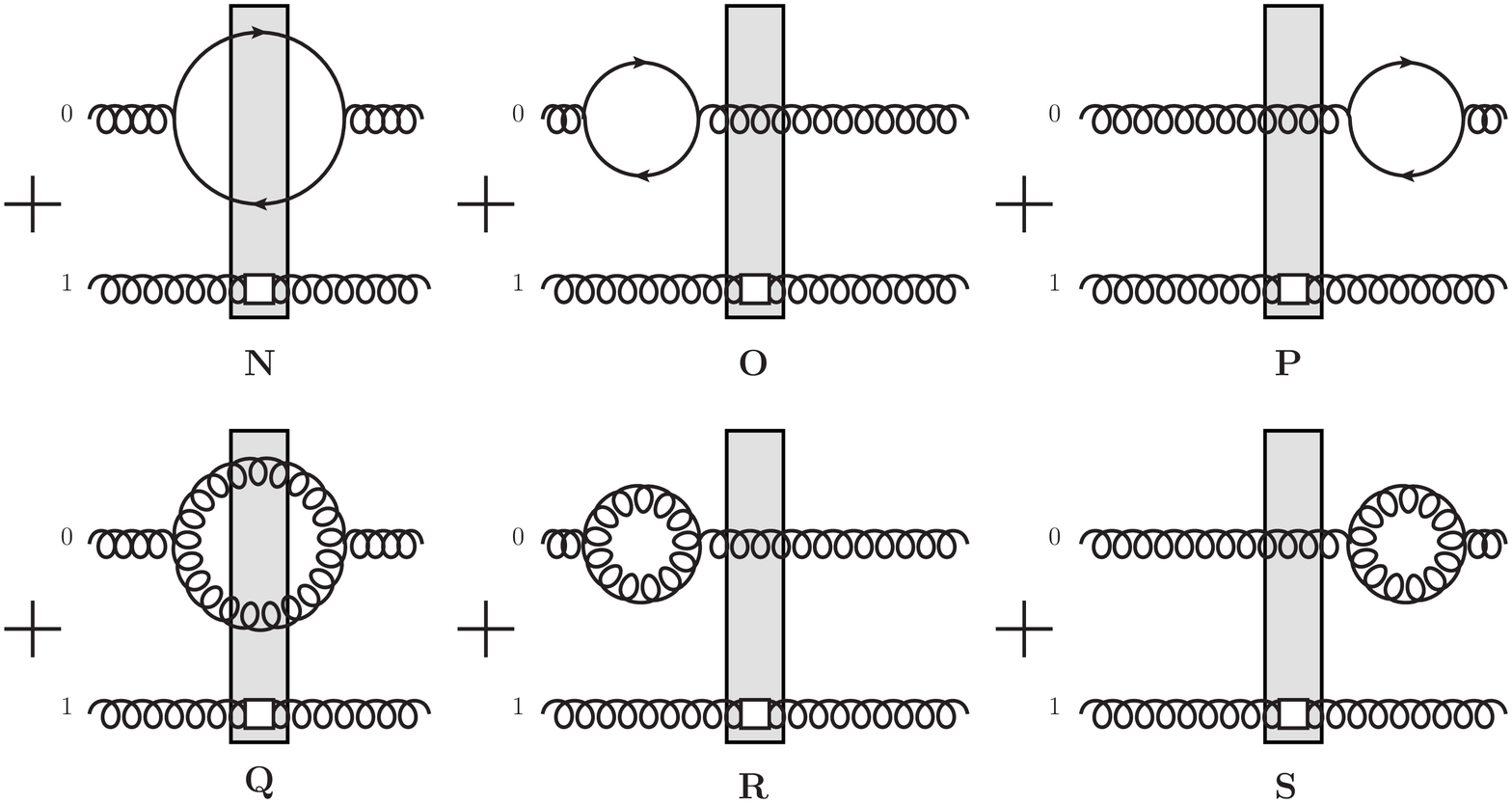}
\end{center}
\caption{Diagrams for SLA$_T$ splittings of a polarized adjoint dipole. For brevity, the initial condition and the DLA+SLA$_L$ diagrams are omitted.}
\label{fig:gldiagrams}
\end{figure}

Before discussing the running of the strong coupling, let us construct the DLA+SLA small-$x$ evolution for the polarized adjoint dipole amplitude.  For the polarized adjoint dipole amplitude, the DLA and SLA$_L$ terms can be deduced from \cite{Kovchegov:2015pbl}, similar to the fundamental dipole case. The SLA$_T$ terms, however, must be derived by employing the $G \to GG$ and $G \to q \bar q$ splitting kernels found in Sec.~\ref{sec:ingredients}; the relevant diagrams are shown in \fig{fig:gldiagrams}. 
We emphasize again that the approach employed here is purely diagrammatic with the gluon lines in the loops being full gluon propagators. Derivation of the SLA$_T$ correction to the polarized adjoint Wilson's line, \eq{eq:UpolFull}, is left for future work. As a result, all possible diagrams that give single-logarithmic integrals in the forward amplitude must be included in \fig{fig:gldiagrams}.

Similar to the case of diagrams $\{E,F,G\}$ from \fig{fig:qkdiagrams}, the diagrams $\{N, O, P\}$ and $\{Q,R,S\}$ from \fig{fig:gldiagrams} separately sum to zero due to unitarity with the SLA$_T$ precision (see Appendix~\ref{sec:unitarity} for details), and only the diagrams $H-M$ in \fig{fig:gldiagrams} contribute. Repeating the same steps as for the fundamental dipole, we arrive at the following evolution equation for the polarized adjoint dipole amplitude:
\begin{align}\label{eqn:glnorc1}
&\frac{1}{N_c^2-1}\llangle\text{tr}\left[U_{\underline{0}}U_{\underline{1}}^{\text{pol }\dagger}\right]\rrangle \left(z_{\min},z_{\text{pol}}\right) = \frac{1}{N_c^2-1}\llangle\text{tr}\left[U_{\underline{0}}U_{\underline{1}}^{\text{pol }\dagger}\right]\rrangle_0 \left(z_{\text{pol}}\right) + \frac{\alpha_s}{2\pi^2}\int\limits_{\Lambda^2/s}^{z_{\min}}\frac{dz'}{z'}\int\limits_{1/z's}d^2 x_2 \\
&\;\;\;\;\times \bigg[\left(\frac{1}{x^2_{21}}\theta\left(x^2_{10}z_{\min}-x^2_{21}z'\right) - \frac{\xx_{21}\cdot\xx_{20}}{x^2_{21}x^2_{20}}\theta\left(x^2_{10}z_{\min}-\max\left\{x^2_{21},x^2_{20}\right\}z'\right)\right)\frac{4}{N_c^2-1}\llangle\text{Tr}\left[T^bU_{\underline{0}}T^aU_{\underline{1}}^{\dagger}\right]U_{\underline{2}}^{\text{pol }ba}\rrangle (z',z') \notag \\
&\;\;\;\;\;\;\;\;- \frac{1}{x^2_{21}}\theta\left(x^2_{10}z_{\min}-x^2_{21}z'\right) \frac{N_f}{N_c^2-1}\llangle\text{tr}\left[t^bV_{\underline{1}}t^aV_{\underline{2}}^{\text{pol }\dagger}\right]U_{\underline{0}}^{ba} + \text{tr}\left[t^bV_{\underline{2}}^{\text{pol}}t^aV_{\underline{1}}^{\dagger}\right]U_{\underline{0}}^{ba}\rrangle (z',z')\bigg] \notag \\
&+ \frac{\alpha_s}{\pi^2}\int\limits_{\Lambda^2/s}^{z_{\min}}\frac{dz'}{z'} \int\limits_{1/z's}d^2 x_2\;\frac{x^2_{10}}{x^2_{21}x^2_{20}}\;\theta\left(x^2_{10}z_{\min}-x^2_{21}z'\right) \notag \\
&\;\;\;\;\times \frac{1}{N_c^2-1}\left[\llangle\text{Tr}\left[T^bU_{\underline{0}}T^aU_{\underline{1}}^{\text{pol }\dagger}\right]U_{\underline{2}}^{ba}\rrangle (z',z_{\text{pol}}) - N_c\llangle\text{Tr}\left[U_{\underline{0}}U_{\underline{1}}^{\text{pol }\dagger}\right]\rrangle(z',z_{\text{pol}}) \right] \notag \\
&\color{blue} - \frac{\alpha_s}{2\pi^2}\int\limits_0^{z_{\text{pol}}}\frac{dz'}{z_{\text{pol}}}\int\limits_{\frac{z_{\text{pol}}}{z'(z_{\text{pol}}-z')s}}\frac{d^2 x_{32}}{x^2_{32}}\;\theta\left(x^2_{10}z_{\min}z_{\text{pol}} - x^2_{32}z'(z_{\text{pol}}-z')\right) \notag \\
&\color{blue} \;\;\;\;\times \frac{4}{N_c^2-1}\llangle\text{Tr}\left[T^bU_{\underline{0}}T^aU^{\dagger}_{\xx_1 - \frac{z'}{z_{\text{pol}}}\xx_{32}}\right]U^{\text{pol }ba}_{\xx_1+\left(1-\frac{z'}{z_{\text{pol}}}\right)\xx_{32}}\rrangle \left(\min\left\{z_{\min},z', z_{\text{pol}}-z' \right\},z'\right) \notag \\
&\color{blue} + \frac{\alpha_s}{\pi^2}\int\limits_0^{z_{\text{pol}}}\frac{dz'}{z_{\text{pol}}} \int\limits_{\frac{z_{\text{pol}}}{z'(z_{\text{pol}}-z')s}}\frac{d^2 x_{32}}{x^2_{32}}\;\theta\left(x^2_{10}z_{\min}z_{\text{pol}} - x^2_{32}z'(z_{\text{pol}}-z')\right) \frac{N_f}{N_c^2-1} \notag \\
&\color{blue} \;\;\;\;\times \llangle\text{tr}\left[t^bV_{\xx_1 - \frac{z'}{z_{\text{pol}}}\xx_{32}}t^aV_{\xx_1+\left(1-\frac{z'}{z_{\text{pol}}}\right)\xx_{32}}^{\text{pol }\dagger}\right]U_{\underline{0}}^{ba} \notag \\
&\color{blue} \hspace*{1.5cm} + \text{tr}\left[t^bV_{\xx_1+\left(1-\frac{z'}{z_{\text{pol}}}\right)\xx_{32}}^{\text{pol}}t^aV_{\xx_1 - \frac{z'}{z_{\text{pol}}}\xx_{32}}^{\dagger}\right]U_{\underline{0}}^{ba}\rrangle \left(\min\left\{z_{\min},z', z_{\text{pol}}-z' \right\},z'\right) \notag \\
&\color{blue} + \frac{\alpha_s}{2\pi^2}\int\limits_0^{z_{\text{pol}}}\frac{dz'}{z_{\text{pol}}}\left[N_c\left(2-\frac{z'}{z_{\text{pol}}}+\frac{z'^2}{z_{\text{pol}}^2}\right) - \frac{N_f}{2}\left(\frac{z'^2}{z_{\text{pol}}^2} + \left(1-\frac{z'}{z_{\text{pol}}}\right)^2\right)\right] \int\limits_{\frac{z_{\text{pol}}}{z'(z_{\text{pol}}-z')s}}\frac{d^2 x_{32}}{x^2_{32}} \notag \\
&\color{blue} \;\;\;\;\times \theta\left(x^2_{10}z_{\min}z_{\text{pol}} - x^2_{32}z'(z_{\text{pol}}-z')\right) \frac{1}{N_c^2-1}\llangle\text{Tr}\left[U_{\underline{0}}U^{\text{pol }\dagger}_{\underline{1}}\right] \rrangle \left(\min\left\{z_{\min},z', z_{\text{pol}}-z' \right\},z_{\text{pol}}\right) . \notag
\end{align}
Here, the SLA$_T$ terms are in the last 7 lines (again, written in blue), and the theta functions follow from the lifetime ordering. In addition, the trace, Tr, is over the adjoint indices. Equation \eqref{eqn:glnorc1} is almost our final result for the DLA+SLA evolution of the polarized adjoint dipole amplitude. Analogous to \eq{eqn:qknorc2}, it only needs to be improved by specifying the scales of the coupling constants in the various terms in the kernel.


\section{Running Coupling Corrections}
 \label{sec:rc}
 
Running coupling corrections for BFKL, BK and JIMWLK evolution equations were derived in \cite{Balitsky:2006wa,Kovchegov:2006vj,Gardi:2006rp,Kovchegov:2006wf} by employing the Brodsky, Lepage, MacKenzie (BLM) prescription \cite{Brodsky:1983gc}: the gluon lines of the leading-order evolution kernel were ``dressed" by quark loops, after which the associated factors of $N_f$ were completed to the full one-loop QCD beta-function via the $N_f \to - 6 \pi \beta_2$ replacement with 
\begin{align}
\beta_2 = \frac{11 N_c - 2 N_f}{12 \pi} .
\end{align}
The quark and antiquark in each loop had comparable longitudinal momentum fractions $z$. The transverse momentum/position integral in each loop generated a logarithm of the renormalization scale $\mu$, which, in the end, was absorbed into the running coupling constant. These features of quark loops seem similar to our SLA$_T$ terms calculated above: by their definition, the SLA$_T$ terms came in with non-logarithmic $z'$-integrals and with logarithmic transverse integrals. The only difference is that the SLA$_T$ terms generate logarithms of the center of mass energy instead of $\mu$. This is due to the lifetime ordering condition \eqref{LCordering3} providing an energy-dependent UV cutoff on the transverse integrals, stemming from the requirement that the $x^-$-lifetime of the loop should be longer than the shock wave width. In principle, the same condition \eqref{LCordering3} should be applied to the quark loops in the running coupling calculation: this would again generate logarithms of energy instead of logarithms of $\mu$. However, as was shown in \cite{Balitsky:2006wa}, the corrections to the BFKL/BK/JIMWLK equations with the quark loops {\sl inside} the shock wave convert all such logarithms of energy into logarithms of $\mu$, effectively replacing the condition \eqref{LCordering3} by $x_{32}^2 \gg 1/\mu^2$. 

It is, therefore, natural to ask the same question in our present helicity evolution calculation: which of the SLA$_T$ terms generate logarithms of $\mu$ instead of logarithms of $s$? This is clearly a relevant question, considering that the $z'$-integral in the last term of \eq{eqn:glnorc1} gives us the beta-function, $(11 N_c - 2 N_f)/6 = 2 \pi \beta_2$ (if we neglect the potential $z'$ dependence in the operator). Ultimately the question is how to take into account the running coupling corrections to our evolution and to eliminate a potential double-counting between those corrections and the SLA$_T$ terms.

The resolution of the problem is in the fact that the SLA$_T$ terms come with the DGLAP-type kernels, as we saw above and as derived in Sec.~\ref{sec:ingredients}. Below, in an almost a toy-model calculation presented in \sec{sec:rcDGLAP}, we show that the QCD beta-function in the DGLAP splitting kernel generates the same running coupling corrections as the BLM prescription \cite{Brodsky:1983gc}. We conclude that, to include the running coupling corrections into the DGLAP-type SLA$_T$ kernels, we should simply run the coupling with the transverse size of the pair of ``daughter"  partons. A more detailed analysis of all the terms in the equations \eqref{eqn:qknorc2} and \eqref{eqn:glnorc1} we conduct below in \sec{sec:rc_terms} gives the running-coupling scales for all the terms in the DLA+SLA kernels, demonstrating that the DGLAP-type running of the coupling in the SLA$_T$ terms even generates the ``triumvirate" structures similar to those found in \cite{Balitsky:2006wa,Kovchegov:2006vj,Kovchegov:2006wf}.


\subsection{Running Coupling in DGLAP} 
\label{sec:rcDGLAP}
 
In our notation the one-loop running of the strong coupling is given by
\begin{align}\label{rc1}
\alpha_s (Q^2) = \frac{\alpha_\mu}{1 + \alpha_\mu \beta_2 \ln (Q^2/\mu^2)} = \frac{1}{\beta_2 \, \ln (Q^2/\Lambda_{\text{QCD}}^2)}
\end{align}
with $\Lambda_{\text{QCD}}$ the QCD confinement scale. Note that the one-loop QCD beta-function is
\begin{align}
\mu^2 \frac{d \alpha_\mu}{d \mu^2} = - \beta_2 \, \alpha_\mu^2 + {\cal O} (\amu^3). 
\end{align}

Imagine that DGLAP evolution is only driven by the $\beta$-function corrections, such that the evolution for the gluon distribution with the $x$-dependence suppressed is
\begin{align}\label{toy_DGLAP}
\mu^2 \frac{d G(\mu^2)}{d \mu^2} = \alpha_\mu \beta_2 G(\mu^2). 
\end{align}
The solution of the toy-model equation \eqref{toy_DGLAP} is
\begin{align}\label{Grun}
G(\mu^2) = G (\mu_0^2) \, \exp \left\{ \int\limits_{\mu_0^2}^{\mu^2} \frac{d \mu^{\prime \, 2}}{\mu^{\prime \, 2}} \alpha_{\mu'} \beta_2 \right\} = \frac{\as (\mu_0^2)}{\as (\mu^2)} \, G (\mu_0^2),
\end{align}
where we have used $\alpha_\mu = \alpha_s (\mu^2)$ along with \eq{rc1}. 

\begin{figure}[ht]
\begin{center}
\includegraphics[width= 0.4 \textwidth]{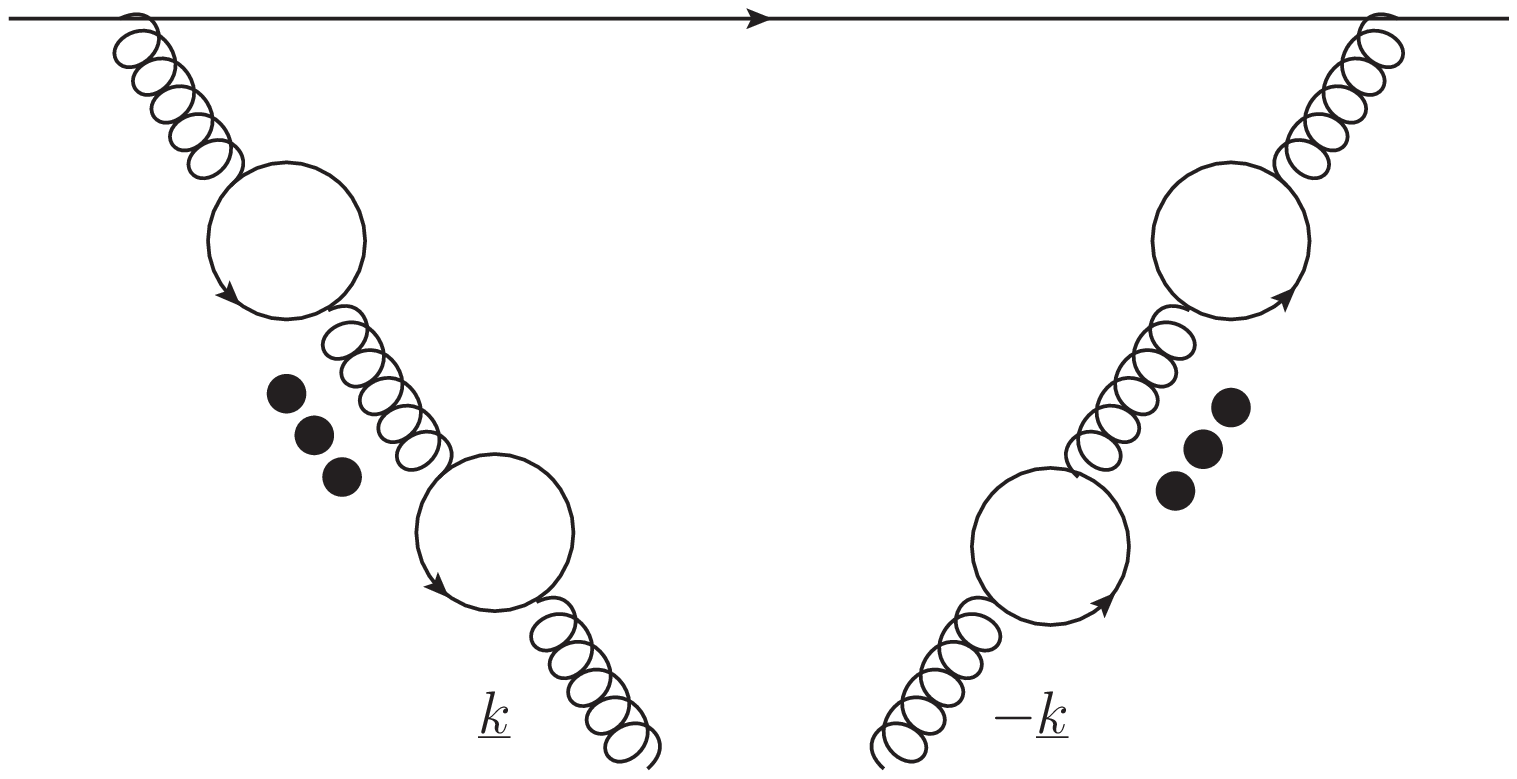} 
\caption{Running coupling corrections to the leading-order gluon distribution in a single quark.}
\label{bubbles}
\end{center}
\end{figure}

Let us next verify that this result is equivalent to a BLM-type calculation of the gluon distribution. Imagine a lowest-order gluon distribution in a single quark. If we put chains of quark loops on the gluon lines, as shown in \fig{bubbles}, after the BLM-prescribed $N_f \to - 6 \pi \beta_2$ replacement we would get
\begin{align}\label{Grun1}
G (\mu^2) - G (\mu_0^2) \sim \int\limits^{\mu^2}_{\mu_0^2} \frac{d k_\perp^2}{k_\perp^2} \, \frac{\amu}{[1 + \amu \beta_2 \ln (k_\perp^2/\mu^2) ]^2} = \int\limits^{\mu^2}_{\mu_0^2} \frac{d k_\perp^2}{k_\perp^2} \, \frac{[\as (k_\perp^2)]^2}{\amu} =\frac{\ln \frac{\mu^2}{\mu_0^2}}{\beta_2 \, \ln \frac{\mu_0^2}{\Lambda_{\text{QCD}}^2}}.
\end{align}
This is in agreement with \eq{Grun} above, which gives
\begin{align}\label{Grun2}
G (\mu^2) - G (\mu_0^2) = \frac{\ln \frac{\mu^2}{\mu_0^2}}{\ln \frac{\mu_0^2}{\Lambda_{\text{QCD}}^2}} \, G (\mu_0^2).
\end{align} 
For the equations \eqref{Grun1} and \eqref{Grun2} to be identical, we need 
\begin{align}
G (\mu_0^2) \sim \frac{1}{\beta_2}.
\end{align}
From \eq{Grun1} we obtain the gluon distribution without the quark loops ``dressing",  by assuming that $\mu_0^2$ is some initial scale at which the quark loop corrections are not yet included, such that 
\begin{align}\label{G0}
G (\mu_0^2) \sim \int\limits_{\Lambda_{\text{QCD}}^2}^{\mu_0^2} \frac{d k_\perp^2}{k_\perp^2} \, \alpha_{\mu_0} = \as (\mu_0^2) \, \ln \frac{\mu_0^2}{\Lambda_{\text{QCD}}^2} = \frac{1}{\beta_2},
\end{align}
as desired. Hence the BLM running coupling scale-setting procedure \cite{Brodsky:1983gc} agrees with the running coupling in DGLAP evolution as long as we run the coupling in the latter with the evolution parameter $\mu^2$.

We conclude that no double-counting between running coupling and SLA$_T$ corrections would happen if we run the coupling with the transverse size of the loop in SLA$_T$. Indeed our toy DGLAP evolution \eqref{toy_DGLAP} can be rewritten in the integral form as
\begin{align}\label{toy_DGLAP2}
G(\mu^2) = G (\mu_0^2) + \int\limits_{\mu_0^2}^{\mu^2} \frac{d \mu^{\prime \, 2}}{\mu^{\prime \, 2}} \, \as (\mu^{\prime \, 2}) \beta_2 \, G(\mu^{\prime \, 2}) 
\end{align}
with the coupling running with the evolution scale $\mu^{\prime \, 2}$. (See \cite{Dokshitzer:1993pf} for more on the coupling scale-setting in DGLAP evolution.)


\subsection{Running Coupling in Different DLA, SLA$_L$ and SLA$_T$ terms} 
\label{sec:rc_terms}

Let us now go through all the terms in our DLA+SLA evolution equations \eqref{eqn:qknorc2} and \eqref{eqn:glnorc1} one-by-one, and deduce how the coupling runs in those if we use the BLM prescription \cite{Brodsky:1983gc}, or, equivalently, as we have just argued, the DGLAP-type prescription. In the terminology below, the `hard' and `soft' refer to the longitudinal momentum fraction of the parton. We also show only one or several representative diagrams in each class. 

\begin{itemize}

\item {\bf Unpolarized soft gluon emissions} (BK emissions): the running coupling for those terms was calculated in \cite{Balitsky:2006wa,Kovchegov:2006vj}, resulting in the running-coupling BK (rcBK) equation. Our evolution equations do not generate quark bubbles on the unpolarized gluon lines, hence there will be no double-counting if we simply use the running coupling results of \cite{Balitsky:2006wa,Kovchegov:2006vj} for the unpolarized soft gluon emission part of our kernel: 
\begin{align}
\mbox{DLA + SLA$_L$} \supset \parbox{8cm}{\includegraphics[width=8cm]{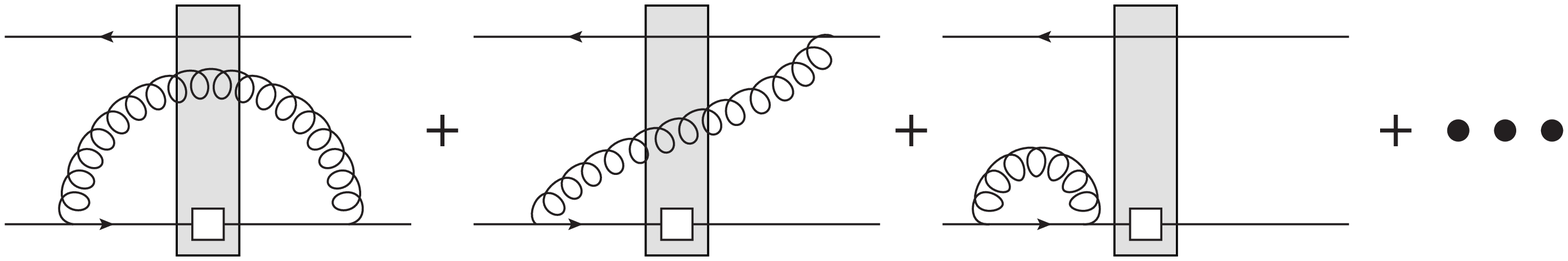}} \Rightarrow \as \ \mbox{from} \ \mbox{rcBK}.
\label{eqn:rcbk}
\end{align}

\item {\bf Hard unpolarized gluon emissions.} For those terms the running of the coupling is clear, since they have only one relevant scale --- the size of the ``daughter" dipole: 
\begin{subequations}
\begin{align}
\mbox{DLA + SLA$_L$} \supset \parbox{2.5cm}{\includegraphics[width=2.5cm]{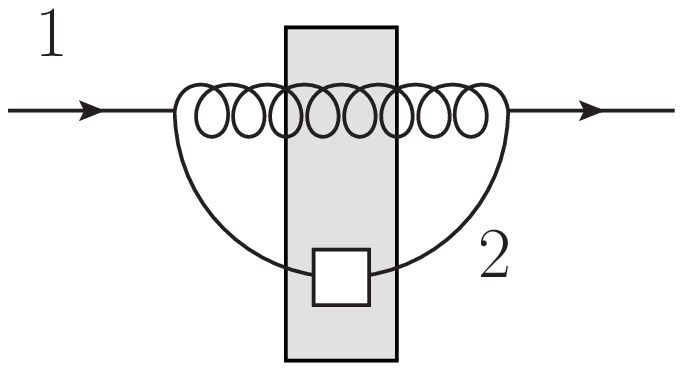}} \Rightarrow \as \left( \frac{1}{x_{21}^2} \right) , \\
\mbox{SLA$_T$} \supset \parbox{2.5cm}{\includegraphics[width=2.5cm]{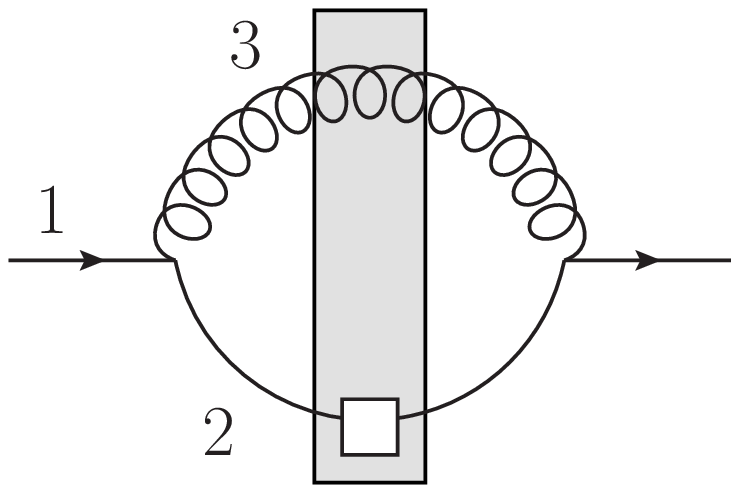}} \Rightarrow \as \left( \frac{1}{x_{32}^2} \right) , \\
\mbox{SLA$_T$} \supset \parbox{2.5cm}{\includegraphics[width=2.5cm]{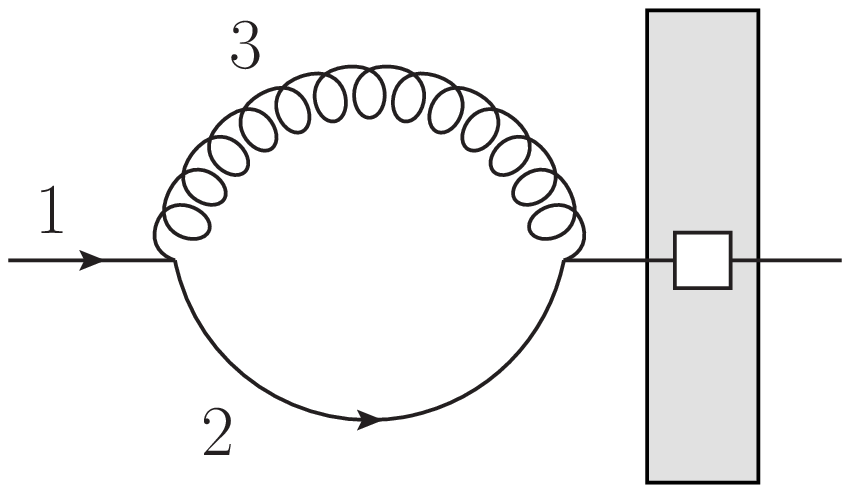}} \Rightarrow \as \left( \frac{1}{x_{32}^2} \right) .
\end{align}
\end{subequations}

\item {\bf Polarized gluon emissions (soft or hard).} The diagrams in question are
\begin{subequations}
\begin{align}
\mbox{DLA} & \supset \parbox{2.5cm}{\includegraphics[width=2.5cm]{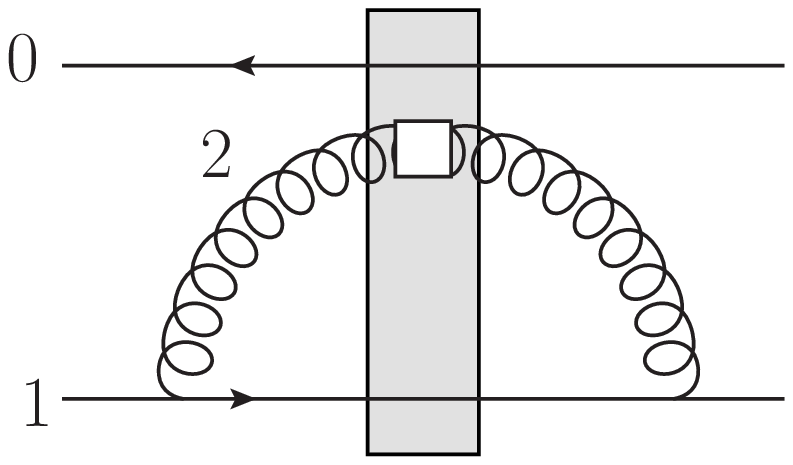}} \Rightarrow \as \left( \frac{1}{x_{21}^2} \right) , \label{pol1} \\
\mbox{DLA+SLA$_L$} & \supset \parbox{2.5cm}{\includegraphics[width=2.5cm]{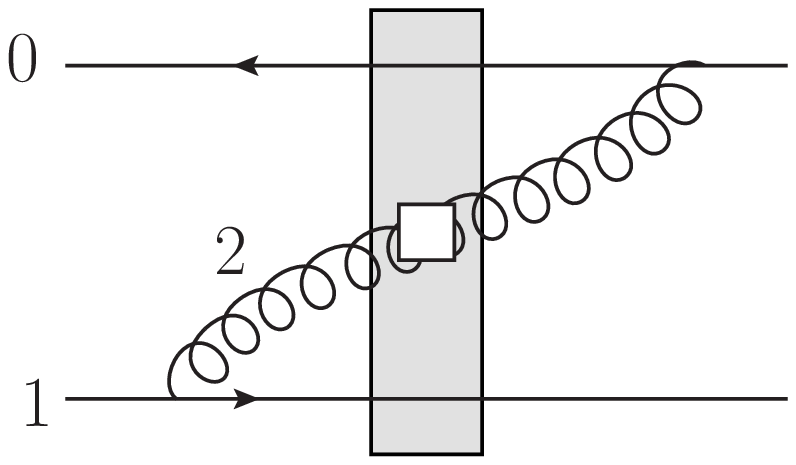}} \Rightarrow \as \left( \min \left\{ \frac{1}{x_{21}^2} , \frac{1}{x_{20}^2} \right\} \right) , \label{pol2} \\
\mbox{SLA$_T$} & \supset \parbox{3cm}{\includegraphics[width=3cm]{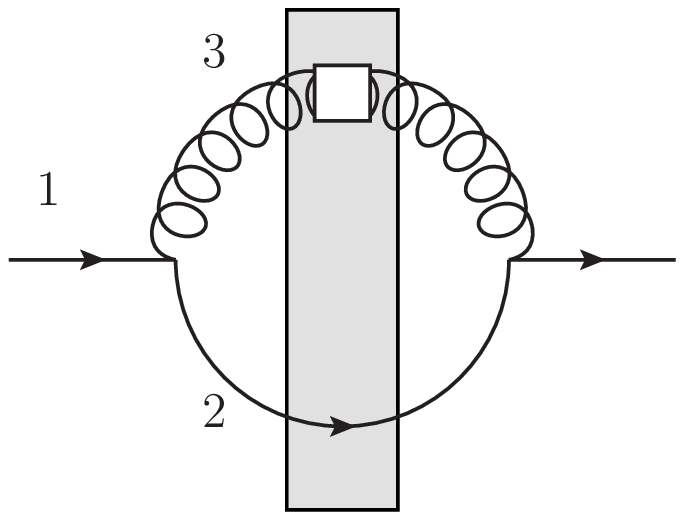}} \Rightarrow \as \left( \frac{1}{x_{32}^2} \right) . \label{pol3} 
\end{align}
\end{subequations}
The scale of the coupling in Eqs.~\eqref{pol1} and \eqref{pol3} is again given by the only relevant scale in the emission, the size of the ``daughter" dipoles 21 and 32, respectively. The origin of the running coupling scales in Eqs.~\eqref{pol1} and \eqref{pol2} is explained in Appendix~\ref{A}. 

\item {\bf Remaining DLA and SLA$_T$ loops.} The diagrams are
\begin{subequations}
\begin{align}
\mbox{DLA} & \supset \parbox{2.5cm}{\includegraphics[width=2.5cm]{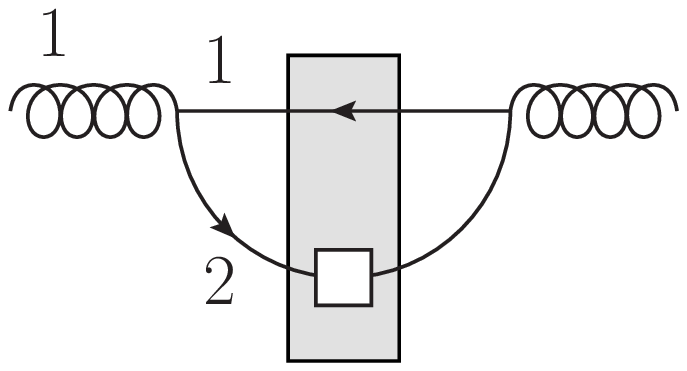}} \Rightarrow \as \left( \frac{1}{x_{21}^2} \right) , \\
\mbox{SLA$_T$} & \supset \parbox{7cm}{\includegraphics[width=7cm]{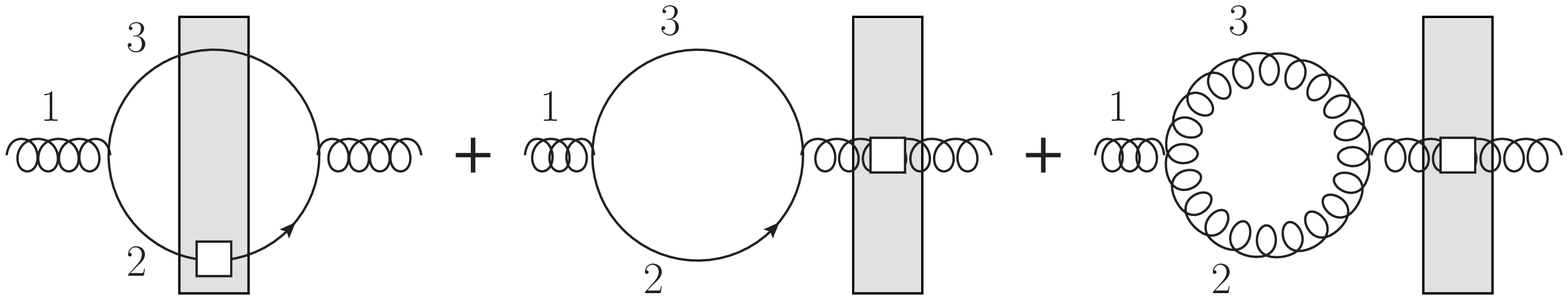}} \Rightarrow \as \left( \frac{1}{x_{32}^2} \right). \label{SLATloop}
\end{align}
\end{subequations}
Here again the coupling runs with the only relevant transverse scale. Note also that for SLA$_T$ terms $x_{32} \sim x_{21} \sim x_{31}$, such that all the distances involved in the above diagrams are equivalent with the approximation precision.

\end{itemize}

The running of the coupling in the diagrams from Eqs.~\eqref{pol1} and \eqref{pol2} is a bit more involved and is explained in detail in Appendix~\ref{A}, where we demonstrate that it agrees with direct diagrammatic calculations. Here, without going into details of those calculations, we note that the diagram in \eq{pol2} is DLA only if $x_{21} \approx x_{20} \gg x_{10}$, in which case there is, again, only one relevant scale determining the running of the coupling, $\as (1/x_{21}^2) \approx \as (1/x_{20}^2)$. In the remaining SLA$_L$ part of the diagram \eqref{pol2} phase space, the difference between $x_{21}$ and $x_{20}$ cannot be parametrically large, by the definition of the SLA$_L$ contribution as not generating logarithms from the transverse integral. Hence, in that part of the phase space the difference between using, say, $\as (1/x_{21}^2)$ or $\as (1/x_{20}^2)$, is outside the precision of our approximation. Therefore, we could, for instance, use $\as (1/x_{21}^2)$ for the part of our kernel coming from the diagram in \eq{pol2}, in both the DLA and SLA$_L$ contributions. However, in the spirit of fixing the scale of the coupling to maximally improve the precision of the approximation \cite{Brodsky:1983gc}, as applied in  \cite{Balitsky:2006wa,Kovchegov:2006vj} to the unpolarized small-$x$ evolution, we performed the calculation described in Appendix~\ref{A} which resulted in the coupling shown in \eq{pol2} (and in \eq{pol1}).


\subsection{DLA+SLA Helicity Evolution with Running Coupling} 

Employing the running coupling prescriptions derived in Sec.~\ref{sec:rc_terms}, we arrive at the following DLA+SLA equations for the small-$x$ helicity evolution. For the fundamental polarized dipole amplitude, \eq{eqn:qknorc2} with the running coupling corrections becomes

\begin{tcolorbox}[ams align]\label{evol_rc}
  & \frac{1}{N_c} \, \llangle \mbox{tr} \left[ V_{\ul{0}} \,
      V_{\ul{1}}^{\text{pol} \, \dagger} \right] \rrangle (z_{\min}, z_{\text{pol}}) =
  \frac{1}{N_c} \, \llangle \mbox{tr} \left[ V_{\ul{0}} \,
      V_{\ul{1}}^{\text{pol} \, \dagger} \right] \rrangle_0 (z_{\text{pol}}) +
  \frac{1}{2 \pi^2} \int\limits_{\Lambda^2/s}^{z_{\min}} \frac{d z'}{z'} 
  \int\limits_{1/(z' s)} d^2 x_{2} \\ & \;\;\;\;\times \left[ \left(
      \frac{\as (1/x_{21}^2)}{x_{21}^2} \, \theta (x_{10}^2 z_{\min} - x_{21}^2 z') - \as (\mbox{min} \{ 1/x_{21}^2, 1/x_{20}^2 \} ) \, 
      \frac{\ul{x}_{21} \cdot \ul{x}_{20}}{x_{21}^2 \, x_{20}^2} \,
      \theta (x_{10}^2 z_{\min} - \mbox{max} \{ x_{21}^2, x_{20}^2 \} z')
    \right) \right. \notag \\ & \;\;\;\;\;\;\;\;\;\;\;\;\times \, \frac{2}{N_c} \llangle \mbox{tr} \left[ t^b \,
      V_{\ul{0}} \, t^a \, V_{\ul{1}}^{\dagger} \right]
    \, U^{\text{pol} \, ba}_{\ul{2}} \rrangle (z' ,  z') \notag \\ & \left. \;\;\;\;\;\;\;\;+
    \frac{\as (1/x_{21}^2)}{x_{21}^2} \, \theta (x_{10}^2 z_{\min} - x_{21}^2 z') \,
    \frac{1}{N_c} \llangle \mbox{tr} \left[ t^b \,
        V_{\ul{0}} \, t^a \, V_{\ul{2}}^{\text{pol} \, \dagger} \right]
      \, U^{ba}_{\ul{1}} \rrangle (z', z') \right] \notag \\ &
  + \frac{1}{2 \pi^2} \int\limits_{\Lambda^2/s}^{z_{\min}} \frac{d z'}{z'} 
  \int\limits_{1/(z' s)} d^2 x_{2} \, K_{\text{rcBK}} ({\un x}_{0}, {\un x}_{1}; {\un x}_{2}) \, \theta (x_{10}^2 z_{\min} - x_{21}^2 z') \notag \\ & \;\;\;\;\times \, \frac{1}{N_c}
  \left[ \llangle
          \mbox{tr} \left[ V_{\ul{0}} \, V^{\dagger}_{\ul{2}} \right] \, \mbox{tr} \left[
            V_{\ul{2}} \, V^{\text{pol} \, \dagger}_{\ul{1}} \right]
        \rrangle (z', z_{\text{pol}}) - N_c 
        \llangle \mbox{tr} \left[ V_{\ul{0}} \,
            V_{\ul{1}}^{\text{pol} \, \dagger} \right] \rrangle (z', z_{\text{pol}}) \right] \notag \\ 
    & \textcolor{blue}{ - \frac{1}{2 \pi^2} \int\limits_{0}^{z_{\text{pol}}} \frac{d z'}{z_{\text{pol}}} 
  \int\limits_{\frac{z_{\text{pol}}}{z'  (z_{\text{pol}} - z') s}} \frac{d^2 x_{32}}{x_{32}^2} \, \theta (x_{10}^2 z_{\min} z_{\text{pol}}  - x_{32}^2 z' (z_{\text{pol}} - z'))  \ \as \left( \frac{1}{x_{32}^2} \right) } \notag \\ & \textcolor{blue}{  \;\;\;\;\times  \, \left[ \frac{1}{N_c} \llangle \mbox{tr} \left[ t^b \,
      V_{\ul{0}} \, t^a \, V_{\ul{x}_1 - \frac{z'}{z_{\text{pol}}} \ul{x}_{32}}^{\dagger} \right]
    \, U^{\text{pol} \, ba}_{\ul{x}_1 + \left( 1 - \frac{z'}{z_{\text{pol}}} \right) \ul{x}_{32}} \rrangle ( \min \{ z_{\min}, z' , z_{\text{pol}}-z' \}  , z') \right.} \notag \\
    & \textcolor{blue}{ \;\;\;\;\;\;\;\;\left. +  \,
    \frac{1}{N_c} \llangle \mbox{tr} \left[ t^b \,
        V_{\ul{0}} \, t^a \, V_{\ul{x}_1 + \left( 1 - \frac{z'}{z_{\text{pol}}} \right) \ul{x}_{32}}^{\text{pol} \, \dagger} \right]
      \, U^{ba}_{\ul{x}_1 - \frac{z'}{z_{\text{pol}}} \ul{x}_{32}} \rrangle (\min \{ z_{\min}, z' , z_{\text{pol}}-z' \} , z') \right] } \notag \\
    & \textcolor{blue}{+ \frac{C_F}{2 \pi^2} \int\limits_{0}^{z_{\text{pol}}}  \frac{d z'}{z_{\text{pol}}} \left( 1 + \frac{z'}{z_{\text{pol}}} \right) \int\limits_{\frac{z_{\text{pol}}}{z'  (z_{\text{pol}} - z') s}} \!\!\!\! \frac{d^2 x_{32}}{x_{32}^2} \, \theta (x_{10}^2 z_{\min} z_{\text{pol}}  - x_{32}^2 z' (z_{\text{pol}} - z'))  \ \as \left( \frac{1}{x_{32}^2} \right) }  \notag \\
      & \textcolor{blue}{ \;\;\;\;\times   \frac{1}{N_c}  \llangle \mbox{tr}
      \left[ V_{\ul{0}} \, V_{\ul{1}}^{\text{pol} \, \dagger} \right]
    \rrangle (\min \{ z_{\min}, z', z_{\text{pol}}-z' \}, z_{\text{pol}}) . } \notag
\end{tcolorbox}

Above we have employed the rcBK kernel $K_{\text{rcBK}} ({\un x}_{0}, {\un x}_{1}; {\un x}_{2})$, which can be taken in the Balitsky prescription \cite{Balitsky:2006wa} 
\begin{align}
K_{\text{rcBK}}^{\text{Bal}} ({\un x}_{0}, {\un x}_{1}; {\un x}_{2}) = \as (1/x_{10}^2) \left[ \frac{x_{10}^2}{x_{21}^2 x_{20}^2} + \frac{1}{x_{21}^2} \left( \frac{\as (1/x_{21}^2)}{\as (1/x_{20}^2)} -1 \right) + \frac{1}{x_{20}^2} \left( \frac{\as (1/x_{20}^2)}{\as (1/x_{21}^2)} -1 \right) \right]
\end{align}
or in the Kovchegov-Weigert one \cite{Kovchegov:2006vj}
\begin{align}
K_{\text{rcBK}}^{\text{KW}} ({\un x}_{0}, {\un x}_{1}; {\un x}_{2}) = \frac{\as (1/x_{21}^2)}{x_{21}^2} - 2 \frac{\as (1/x_{21}^2) \as (1/x_{20}^2)}{\as (1/R^2)} \frac{{\un x}_{21} \cdot {\un x}_{20}}{x_{21}^2 x_{20}^2} + \frac{\as (1/x_{20}^2)}{x_{20}^2}
\end{align}
with
\begin{align}
R^2 = x_{20} \, x_{21}  \left( \frac{x_{21}}{x_{20}} \right)^{\frac{x_{20}^2 + x_{21}^2}{x_{20}^2 - x_{21}^2} - 2 \frac{x_{21}^2 x_{20}^2}{{\un x}_{21} \cdot {\un x}_{20}} \frac{1}{x_{20}^2 - x_{21}^2} }.
\end{align}
The Balitsky prescription appears to more accurately represent all the quark loop corrections \cite{Albacete:2007yr}.

Similarly, the DLA+SLA evolution equation \eqref{eqn:glnorc1} for the adjoint polarized dipole amplitude becomes, after including the running coupling corrections,
\begin{tcolorbox}[ams align]\label{Gevol_rc}
  & \frac{1}{N_c^2 -1} \llangle \mbox{Tr} \left[
      U_{\ul{0}} \, U_{\ul{1}}^{\text{pol} \, \dagger} \right]
  \rrangle (z_{\min}, z_{\text{pol}}) = \frac{1}{N_c^2 -1} \, \llangle \mbox{Tr}
    \left[ U_{\ul{0}} \, U_{\ul{1}}^{\text{pol} \, \dagger} \right]
  \rrangle_0 (z_{\text{pol}}) + \frac{1}{2 \pi^2} \int\limits_{\Lambda^2/s}^{z_{\min}}
  \frac{d z'}{z'} \int\limits_{1/(z' s)} d^2 x_{2}  \\ & \;\;\;\;\times
  \left[ \left( \frac{\as (1/x_{21}^2)}{x_{21}^2} \, \theta (x_{10}^2 z_{\min} - x_{21}^2
      z') - \as (\mbox{min} \{ 1/x_{21}^2, 1/x_{20}^2 \} ) \, \frac{\ul{x}_{21} \cdot \ul{x}_{20}}{x_{21}^2 \, x_{20}^2}
      \, \theta (x_{10}^2 z_{\min} - \mbox{max} \{ x_{21}^2, x_{20}^2 \} z')
    \right) \right. \notag \\ & \;\;\;\;\;\;\;\;\;\;\;\;\times \frac{4}{N_c^2 -1} \llangle
    \mbox{Tr} \left[ T^b \, U_{\ul{0}} \, T^a \, U_{\ul{1}}^{\dagger} \right] \, U^{\text{pol} \, ba}_{\ul{2}} \rrangle
  (z', z') \notag \\ & \;\;\;\;\;\;\;\;\left. - \frac{\as (1/x_{21}^2)}{x_{21}^2} \, \theta (x_{10}^2 z_{\min} -
    x_{21}^2 z') \, \frac{N_f}{N_c^2-1} \llangle \mbox{tr} \left[
        t^b \, V_{\ul{1}} \, t^a \, V_{\ul{2}}^{\text{pol} \, \dagger}
      \right] \, U^{ba}_{\ul{0}} + \mbox{tr} \left[ t^b \,
        V^{\text{pol}}_{\ul{2}} \, t^a \, V_{\ul{1}}^{\dagger} \right]
      \, U^{ba}_{\ul{0}} \rrangle (z', z') \right] \notag \\ &
  + \frac{1}{\pi^2} \int\limits_{\Lambda^2/s}^{z_{\min}} \frac{d z'}{z'} \!
  \int\limits_{1/(z' s)} d^2 x_{2} \, K_{\text{rcBK}} ({\un x}_{0}, {\un x}_{1}; {\un x}_{2}) \, \theta (x_{10}^2 z_{\min} - x_{21}^2 z')  \notag \\ 
& \;\;\;\;\times  \frac{1}{N_c^2 -1} \left[ \llangle \mbox{Tr} \left[ T^b \,
        U_{\ul{0}} \, T^a \, U_{\ul{1}}^{\text{pol} \, \dagger} \right]
      \, U^{ba}_{\ul{2}} \rrangle (z', z_{\text{pol}} ) - N_c \llangle
      \mbox{Tr} \left[ U_{\ul{0}} \, U_{\ul{1}}^{\text{pol} \, \dagger}
      \right] \rrangle (z', z_{\text{pol}} ) \right] \notag \\
      & \textcolor{blue}{  - \frac{1}{2 \pi^2} \int\limits_{0}^{z_{\text{pol}}}
  \frac{d z'}{z_{\text{pol}}} \! \int\limits_{\frac{z_{\text{pol}}}{z'  (z_{\text{pol}} - z') s}} \frac{d^2 x_{32}}{x_{32}^2} \, \theta (x_{10}^2 z_{\min} z_{\text{pol}}  - x_{32}^2 z' (z_{\text{pol}} - z'))  \ \as \left( \frac{1}{x_{32}^2} \right) } \notag \\ 
  & \textcolor{blue}{ \;\;\;\;\times \, \frac{4}{N_c^2 -1} \llangle
    \mbox{Tr} \left[ T^b \, U_{\ul{0}} \, T^a \, U_{\ul{x}_1 - \frac{z'}{z_{\text{pol}}} \ul{x}_{32}}^{\dagger} \right] \, U^{\text{pol} \, ba}_{\ul{x}_1 + \left( 1 - \frac{z'}{z_{\text{pol}}} \right) \ul{x}_{32}} \rrangle
  (\min \{ z_{\min}, z' , z_{\text{pol}}-z' \}, z')} \notag \\ 
  & \textcolor{blue}{  + \frac{1}{\pi^2} \int\limits_{0}^{z_{\text{pol}}}
  \frac{d z'}{z_{\text{pol}}} \, \int\limits_{\frac{z_{\text{pol}}}{z'  (z_{\text{pol}} - z') s}} \frac{d^2 x_{32}}{x_{32}^2} \, \theta (x_{10}^2 z_{\min} z_{\text{pol}}  - x_{32}^2 z' (z_{\text{pol}} - z'))  \ \as \left( \frac{1}{x_{32}^2} \right) \frac{N_f}{N_c^2-1} } \notag \\ 
  & \textcolor{blue}{  \;\;\;\;\times  \llangle \mbox{tr} \left[
        t^b \, V_{\ul{x}_1 - \frac{z'}{z_{\text{pol}}} \ul{x}_{32}} \, t^a \, V_{\ul{x}_1  + \left( 1 - \frac{z'}{z_{\text{pol}}} \right) \ul{x}_{32}}^{\text{pol} \, \dagger} \right] \, U^{ba}_{\ul{0}} } \notag \\ 
  & \textcolor{blue}{ \hspace*{1.5cm} + \mbox{tr} \left[ t^b \,
        V^{\text{pol}}_{\ul{x}_1 + \left( 1 - \frac{z'}{z_{\text{pol}}} \right) \ul{x}_{32}} \, t^a \, V_{\ul{x}_1 - \frac{z'}{z_{\text{pol}}} \ul{x}_{32}}^{\dagger} \right]
      \, U^{ba}_{\ul{0}} \rrangle (\min \{ z_{\min}, z' , z_{\text{pol}}-z' \}, z') } \notag \\ 
  & \textcolor{blue}{  + \frac{1}{2 \pi^2} \int\limits_{0}^{z_{\text{pol}}}
  \frac{d z'}{z_{\text{pol}}} \!\!\!\! \int\limits_{\frac{z_{\text{pol}}}{z'  (z_{\text{pol}} - z') s}} \!\!\!\!\! \frac{d^2 x_{32}}{x_{32}^2} \left[ N_c \left( 2 - \frac{z'}{z_{\text{pol}}} + \frac{z'^2}{z_{\text{pol}}^2} \right) - \frac{N_f}{2} \left( \frac{z'^2}{z_{\text{pol}}^2} + \left( 1 - \frac{z'}{z_{\text{pol}}} \right)^2 \right) \right]   } \notag \\ 
  & \textcolor{blue}{  \;\;\;\;\times  \, \theta \left(x_{10}^2 z_{\min} z_{\text{pol}}  - x_{32}^2 z' (z_{\text{pol}} - z')\right)  \as \left( \frac{1}{x_{32}^2} \right) \, \frac{1}{N_c^2 -1} \llangle
      \mbox{Tr} \left[ U_{\ul{0}} \, U_{\ul{1}}^{\text{pol} \, \dagger}
      \right] \rrangle (\min \{ z_{\min}, z' , z_{\text{pol}}-z' \}, z_{\text{pol}}) } \notag  .
\end{tcolorbox}

Equations \eqref{evol_rc} and \eqref{Gevol_rc} are our main results for the DLA+SLA small-$x$ helicity evolution, outside the closed equations obtained in the large-$N_c$ and in the large-$N_c \& N_f$ limits below. Deriving such closed equations from Eqs.~\eqref{evol_rc} and \eqref{Gevol_rc} is our next step. 



\section{Closed Evolution Equations}
\label{sec:closed_ee}

As we see from Sections~\ref{sec:eqnnorc} and \ref{sec:rc}, the DLA+SLA helicity evolution equations do not close. However, the equations close once we take the large-$N_c$ or the large $N_c\& N_f$ limit. We explore each of the two limits in this Section.


\subsection{Large-$N_c$ Limit}
\label{sec:nc}

In this Subsection we consider the 't Hooft's large-$N_c$ limit \cite{tHooft:1973alw}, where $N_c\gg 1$ and $N_c\gg N_f$, while $\as \, N_c$ is constant and, for our perturbative calculation, is assumed to be small. While we do not have explicit expressions for the polarized Wilson lines $U^{\text{pol}}$ and  $V^{\text{pol}}$ at the SLA$_T$ accuracy, we will assume that the large-$N_c$ identities derived for $U^{\text{pol}}$ and  $V^{\text{pol}}$ at the DLA+SLA$_L$ level \cite{Kovchegov:2017lsr, Kovchegov:2018znm} will apply, since they already incorporate the right color and spin factors. 

Further, as was already apparent at the DLA+SLA$_L$ level \cite{Kovchegov:2018znm}, a relation between $U^{\text{pol}}$ and  $V^{\text{pol}}$ (see Eq.~(74) in \cite{Kovchegov:2018znm} and \eq{eq:Wpol4} below) can only be obtained in the large-$N_c$ limit by working in the gluon sector, while neglecting interactions with the quark background fields in the shock wave. Therefore, even at DLA, the large-$N_c$ limit can be constructed by keeping only gluon exchanges with the shock wave \cite{Kovchegov:2015pbl, Kovchegov:2016zex, Kovchegov:2018znm}, that is, by putting $N_f =0$. The rationale for this approximation is that at large-$N_c$ gluons dominate all the dynamics. (One possible exception for this logic arises when the original polarized dipole is made out of a true quark--antiquark pair, as discussed in \cite{Kovchegov:2020hgb}: certainly the original quark/antiquark could emit a hard gluon and become soft, or, equivalently, go into the shock wave. Such limit of the original quark-antiquark polarized dipole, with no other (dynamical) quarks in the problem has not been fully explored yet, though the studies presented in \cite{Kovchegov:2020hgb} appear to indicate that the resulting small-$x$ asymptotics may not be very different from the pure-glue interpretation of the large-$N_c$ limit.)  The apparent need to work in the gluons-only sector is further reinforced by the SLA$_T$ terms, which, as we saw in Section~\ref{sec:ingredients}, are essentially polarized DGLAP splitting functions (with the DLA singularities subtracted). While the real emission parts of the kernels  $K^{q{\bar q} \to q{\bar q}}$, $K^{q{\bar q} \to GG}$, $K^{GG \to q{\bar q}}$, and $K^{GG \to GG}$ are all proportional to each other (for the reasons unclear to us at this point), the virtual terms in $K^{q{\bar q} \to q{\bar q}}$ and $K^{GG \to GG}$ are also proportional to each other (if we carry out the $z$-integrals) but with a different proportionality constant compared to the real terms in these kernels, even in the large-$N_c$ limit (see Eqs.~\eqref{Kqqqq2} and \eqref{KGGGG}). 

Therefore, in what follows we will take the large-$N_c$ limit of the evolution equation \eqref{Gevol_rc} for the adjoint polarized dipole, discarding quark contributions and keeping gluon lines only. We will employ the identities derived in \cite{Kovchegov:2018znm} for the correlators of the Wilson lines in the gluons-only shock wave background. We will thus employ the following definition of the polarized fundamental dipole amplitude (cf. \eq{Qdef10} and Eq.~(75) in \cite{Kovchegov:2018znm}),
\begin{align}\label{Ggdef}
G_{10} (z_{\min},z_{\text{pol}} ) = \frac{1}{2 N_c}\llangle\text{tr}\left[V_{\underline{0}} \left( V_{\underline{1}}^{\text{pol} \, g} \right)^{\dagger}\right]\rrangle \left(z_{\min},z_{\text{pol}}\right) + \frac{1}{2 N_c}\llangle\text{tr}\left[V_{\underline{1}}^{\text{pol} \, g}V_{\underline{0}}^{\dagger}\right]\rrangle \left(z_{\min},z_{\text{pol}}\right) ,
\end{align}
where $g$ in the superscript of the polarized Wilson lines denotes gluons-only shock wave interactions (see \eq{eq:Vpol_all}). Continuing with our abbreviated notation we have omitted the Re sign and the time-ordered product sign T (cf. \eq{Qdef10}): both of these are implied. 

Below we will also encounter the unpolarized dipole amplitude employed in the BK and JIMWLK evolution equations  \cite{Balitsky:1995ub,Balitsky:1998ya,Kovchegov:1999yj,Kovchegov:1999ua,Jalilian-Marian:1997dw,Jalilian-Marian:1997gr,Weigert:2000gi,Iancu:2001ad,Iancu:2000hn,Ferreiro:2001qy}
\begin{align}
S_{10} (z_{\min}) = \frac{1}{2N_c}\left\langle\text{tr}\left[V_{\underline{0}}V_{\underline{1}}^{\dagger}\right]\right\rangle \left(z_{\min} \right) + \frac{1}{2N_c}\left\langle\text{tr}\left[V_{\underline{1}}V_{\underline{0}}^{\dagger}\right]\right\rangle \left(z_{\min} \right) \approx \frac{1}{N_c}\left\langle\text{tr}\left[V_{\underline{0}}V_{\underline{1}}^{\dagger}\right]\right\rangle \left(z_{\min} \right),
\label{eqn:Nc4}
\end{align}
where we have ignored the odderon contribution, which is small at low $x$ compared to the $C$-even ``pomeron" contribution \cite{Kovchegov:2003dm,Hatta:2005as}. Here, $S$ depends on $z_{\min}$ but not $z_{\text{pol}}$ because $S$ is unrelated to helicity and there are no polarized Wilson lines in its definition. The BK/JIMWLK evolution of $S_{10} \left(z_{\min}\right)$ is single-logarithmic in our terminology. This is why, in the prior works on helicity evolution one could put $S=1$ with the DLA accuracy. However, now that we are constructing the DLA+SLA evolution equations, $S_{10} \left(z_{\min}\right)$ again will be non-trivial, obeying the BK/JIMWLK evolution equations, and, therefore, incorporating the saturation dynamics as well. Below we will also employ the initial condition for the evolution of $S_{10} \left(z_{\min}\right)$ denoted by $S_{10}^{(0)}$. Note that this initial condition is independent of energy (and, hence, of $z_{\min}$) \cite{McLerran:1993ni,McLerran:1993ka,McLerran:1994vd,Mueller:1989st,Kovchegov:1996ty}.

Employing the relation between the adjoint and fundamental Wilson lines,
\begin{align}
U_{\underline{x}}^{ba} = 2\text{ tr}\left[t^bV_{\underline{x}}t^aV_{\underline{x}}^{\dagger}\right],
\label{eqn:Nc2}
\end{align}
along with the relation between the polarized adjoint and fundamental Wilson lines derived in \cite{Kovchegov:2018znm},
\begin{align} 
  \label{eq:Wpol4} 
  (U^{\text{pol} \, g}_{\un x})^{ab} =  4 \, \tr \left[ \left( V_{\ul x}^{\text{pol} \, g} \right)^{\dagger} t^a V_{\ul x} t^b \right] + 4 \,  \tr \left[ V^\dagger_{\ul x} t^a V_{\ul x}^{\text{pol} \, g} t^b \right] , 
\end{align}
and applying the Fierz identity twice, we obtain (with the large-$N_c$ accuracy) 
\begin{align}
\frac{1}{N_c^2-1}\llangle\text{Tr}\left[U_{\underline{0}} \, \left( U_{\underline{1}}^{\text{pol} \, g} \right)^{\dagger}\right]\rrangle \left(z_{\min},z_{\text{pol}}\right) = 4 \, G_{10} (z_{\min},z_{\text{pol}}) \, S_{10} (z_{\min})
\label{eqn:Nc5}
\end{align}
for the adjoint polarized dipole amplitude in \eq{Gevol_rc}. 

Similarly, multiple applications of the Fierz identity along with Eqs.~\eqref{eqn:Nc2} and \eqref{eq:Wpol4} yield
\begin{align}\label{eqn:Nc6}
\frac{1}{N_c^2 -1} \llangle\text{Tr}\left[T^bU_{\underline{0}}T^aU_{\underline{1}}^{\dagger}\right] \left( U_{\underline{2}}^{\text{pol} \, g} \right)^{ba}\rrangle\left(z_{\min},z_{\text{pol}}\right) 
= 2 \, N_c \, S_{10} \left(z_{\min}\right) \, \Big[ & \, S_{20} (z_{\min}) \, G_{21} (z_{\min},z_{\text{pol}}) \\
&+ S_{21} (z_{\min}) \, \Gamma^{gen}_{20, 21} (z_{\min},z_{\text{pol}})\Big] , \notag 
\end{align}
again with the large-$N_c$ accuracy (see Appendix~A of \cite{Kovchegov:2016zex} noting that Eq.~(A5) there is missing an overall factor of 2 on its right-hand side, as was pointed out in the more detailed calculation presented in \cite{Kovchegov:2018znm}). In \eq{eqn:Nc6} we have employed the generalized (polarized) dipole amplitude
\begin{align}
\Gamma^{gen}_{10, 32} (z_{\min},z_{\text{pol}}) = G_{10} (z_{\min},z_{\text{pol}} ) \, \theta\left(x_{32}-x_{10}\right) + \Gamma_{10,32} (z_{\min},z_{\text{pol}}) \, \theta (x_{10}-x_{32} ) 
\label{eqn:Nc7}
\end{align}
originally introduced in \cite{Kovchegov:2017lsr}.  In the second term on the right of \eq{eqn:Nc7}, $\Gamma_{10,32} \left(z_{\min},z_{\text{pol}}\right)$ is the so-called neighbor dipole amplitude, which is a polarized dipole amplitude defined by \eq{Ggdef}, but with the upper limit on the lifetime of the subsequent emissions being $z_{\min} \, x^2_{32}$, unrelated to the transverse size $x_{10}$ of the dipole. Such amplitude arises for DLA helicity evolution in a dipole which is a ``neighbor" of the smallest shortest-lived dipole $32$ (see \cite{Kovchegov:2015pbl} for a more detailed discussion of the neighbor dipole amplitude).

Once more, Fierz identity with Eqs.~\eqref{eqn:Nc2} and \eqref{eq:Wpol4} give us
\begin{align}\label{eqn:Nc8}
\frac{1}{N_c^2 -1}  \llangle\text{Tr}\left[T^bU_{\underline{0}}T^a \left( U_{\underline{1}}^{\text{pol} \, g} \right)^{\dagger}\right]U_{\underline{2}}^{ba}\rrangle \left(z_{\min},z_{\text{pol}}\right) = 2 \, N_c \, S_{20} (z_{\min}) \, \Big[& \, S_{10} (z_{\min}) \, G_{21} (z_{\min},z_{\text{pol}} ) \\
& + S_{21} (z_{\min}) \, \Gamma^{gen}_{10,21} (z_{\min},z_{\text{pol}} )\Big] , \notag
\end{align}
also at large $N_c$.

Applying equations \eqref{eqn:Nc5}, \eqref{eqn:Nc6}, and \eqref{eqn:Nc8} to \eqref{Gevol_rc} we get, in the large-$N_c$ limit (with $N_c\gg N_f$), 
\begin{align}\label{eqn:Nc9}
&G_{10} (z_{\min},z_{\text{pol}}) \, S_{10} (z_{\min}) = G^{(0)}_{10} (z_{\text{pol}}) \, S^{(0)}_{10}  + \frac{N_c}{\pi^2} \int\limits_{\Lambda^2/s}^{z_{\min}}\frac{dz'}{z'} \int\limits_{1/(z's)}d^2 x_2 \\
&\;\;\;\;\times \left(\frac{\as(1/x^2_{21})}{x^2_{21}}\;\theta\left(x^2_{10}z_{\min}-x^2_{21}z'\right) - \as (\min\{1/x^2_{21},1/x^2_{20}\})\;\frac{\xx_{21}\cdot\xx_{20}}{x^2_{21}x^2_{20}}\;\theta\left(x^2_{10}z_{\min}-\max\left\{x^2_{21},x^2_{20}\right\}z'\right)\right) \notag \\
&\;\;\;\;\times \left[G_{21} (z',z' ) \, S_{20} (z' )  + \Gamma^{gen}_{20,21} (z',z') \, S_{21} (z' ) \right] \, S_{10} (z') \notag \\
&+ \frac{N_c}{2\pi^2} \int\limits_{\Lambda^2/s}^{z_{\min}}\frac{dz'}{z'} \int\limits_{1/z's} d^2 x_2 \, K_{\text{rcBK}} ({\un x}_{0}, {\un x}_{1}; {\un x}_{2}) \, \theta\left(x^2_{10}z_{\min}-x^2_{21}z'\right) \notag \\
&\;\;\;\;\times  \left[G_{21} (z',z_{\text{pol}} ) \, S_{20} (z') \, S_{10} (z') + \Gamma^{gen}_{10, 21} (z',z_{\text{pol}}) \, S_{20} (z') \, S_{21} (z' )  - 2 \, \Gamma^{gen}_{10, 21} (z',z_{\text{pol}}) \, S_{10} (z')  \right] \notag \\
&\color{blue} - \frac{N_c}{\pi^2}\int\limits_0^{z_{\text{pol}}}\frac{dz'}{z_{\text{pol}}}\int\limits_{\frac{z_{\text{pol}}}{z'  (z_{\text{pol}} - z') s}} \frac{d^2 x_{32}}{x_{32}^2} \;\as\left(\frac{1}{x^2_{32}}\right) \theta (x_{10}^2 z_{\min} z_{\text{pol}}  - x_{32}^2 z' (z_{\text{pol}} - z'))   \, S_{\xx_{1}-\frac{z'}{z_{\text{pol}}} \xx_{32}, \xx_{0} } (\min \left\{ z_{\min}, z' , z_{\text{pol}} - z' \right\} ) \notag \\
&\color{blue} \;\;\;\;\;\;\;\;\;\;\;\;\times \left[G_{\xx_{1}+\left(1-\frac{z'}{z_{\text{pol}}} \right) \xx_{32}, \, \xx_{1}-\frac{z'}{z_{\text{pol}}} \xx_{32}} ( \min \left\{ z_{\min}, z' , z_{\text{pol}} - z' \right\}, z' ) \, S_{\xx_{1}+\left(1-\frac{z'}{z_{\text{pol}}} \right) \xx_{32}, \, \xx_{0}} (\min\left\{z_{\min},z', z_{\text{pol}} - z' \right\}) \right. \notag \\
&\color{blue} \;\;\;\;\;\;\;\;\left. + \Gamma_{\xx_{1}+\left(1-\frac{z'}{z_{\text{pol}}} \right) \xx_{32}, \, \xx_{0}; x_{32}} ( \min\left\{z_{\min},z' , z_{\text{pol}} - z' \right\},z' ) \, S_{\xx_{1}+\left(1-\frac{z'}{z_{\text{pol}}} \right) \xx_{32}, \, \xx_{1}-\frac{z'}{z_{\text{pol}}} \xx_{32}} ( \min \left\{ z_{\min}, z' , z_{\text{pol}} - z' \right\}) \right] \notag \\
&\color{blue} + \frac{N_c}{2\pi^2}\int\limits_0^{z_{\text{pol}}}\frac{dz'}{z_{\text{pol}}} \left(2-\frac{z'}{z_{\text{pol}}}+\frac{z'^2}{z_{\text{pol}}^2}\right) \int\limits_{\frac{z_{\text{pol}}}{z'  (z_{\text{pol}} - z') s}} \frac{d^2 x_{32}}{x_{32}^2} \;\as\left(\frac{1}{x^2_{32}}\right) \theta (x_{10}^2 z_{\min} z_{\text{pol}}  - x_{32}^2 z' (z_{\text{pol}} - z')) \notag \\
&\color{blue} \;\;\;\;\times \Gamma_{10,32} (\min\left\{z_{\min},z' , z_{\text{pol}} - z' \right\},z_{\text{pol}} ) \, S_{10} (\min\left\{z_{\min},z', z_{\text{pol}} - z' \right\} ). \notag
\end{align}
Note that the virtual splitting terms, both in the DLA+SLA$_L$ and SLA$_T$ parts of the evolution kernel, yield the generalized dipole amplitude, $\Gamma^{gen}_{10,21}$, because the lifetime of subsequent emissions may depend on the transverse size of the virtual loop $x_{21}$, which is different from the dipole's transverse size, $x_{10}$.  Now, we realize that in the SLA$_T$ terms the following kinematic constraints apply,
\begin{align}
x_{32}\ll x_{10}\;\;\;\;\;\text{and}\;\;\;\;\;z_{\min} \lesssim z_{\text{pol}} \sim z' \sim z_{\text{pol}} - z'.
\label{eqn:Nc11}
\end{align}
Therefore, in those terms $\Gamma^{gen}$ becomes $\Gamma$, since indeed the lifetime of the subsequent emissions is controlled by the (small) size of the virtual loop, $x_{32}$ in the SLA$_T$ terms. Note also that while either $z'$ or $z_{\text{pol}} - z'$ can be smaller than $z_{\min}$, the regions where this happens are small: for instance, to generate a large $\ln (z_{\min}/z') \sim 1/\as$, one needs $z' < z_{\min} e^{-1/\as}$. Since the integrands in the SLA$_T$ terms are regular at both $z'=0$ and $z'=z_{\text{pol}}$, we conclude that such regions' impact on the coefficient in front of the logarithm in SLA$_T$ terms is very small and can be neglected. Hence, for the SLA$_T$ terms one can assume that $z' \approx z_{\text{pol}} - z' \approx  z_{\text{pol}}  \gtrsim z_{\min}$, and only use $z_{\min}$ instead.

In our calculations we assume that the ``parent" dipole size $x_{10}$ is perturbatively small. We further assume that the variation of all the dipole amplitudes with the dipole impact parameter is slow compared to their dependence on the dipole size \cite{Kovchegov:1999yj,Kovchegov:1999ua}, since the impact parameter varies over the longer non-perturbative distance scales of the order of the diameter of the proton (or nuclear) target. We therefore can employ the conditions from \eqref{eqn:Nc11} to approximate
\begin{align}\label{32approx}
& S_{\xx_{1}-\frac{z'}{z_{\text{pol}}} \xx_{32}, \xx_{0} } (\min \left\{ z_{\min}, z' , z_{\text{pol}} - z' \right\} )  \approx S_{10} (z_{\min}), \ \ \ S_{\xx_{1}+\left(1-\frac{z'}{z_{\text{pol}}} \right) \xx_{32}, \, \xx_{0}} (\min \left\{ z_{\min}, z' , z_{\text{pol}} - z' \right\} ) \approx S_{10} (z_{\min}), \\ & \Gamma_{\xx_{1}+\left(1-\frac{z'}{z_{\text{pol}}} \right) \xx_{32}, \, \xx_{0}; x_{32}} ( \min\left\{z_{\min},z' , z_{\text{pol}} - z' \right\},z' ) \approx \Gamma_{10, 32} ( z_{\min} , z' ), \ \ \ S_{\xx_{1}+\left(1-\frac{z'}{z_{\text{pol}}} \right) \xx_{32}, \, \xx_{1}-\frac{z'}{z_{\text{pol}}} \xx_{32}}  \approx S_{11} = 1 \notag
\end{align}
in \eq{eqn:Nc9}. The last approximation, $S \approx 1$, is due to the fact that deviations of $S$ from 1 are proportional to positive order-one powers of $x_{32}$ \cite{Kovchegov:2012mbw}, rendering the $x_{32}$ integral non-logarithmic, and, thus making the terms non-SLA$_T$ (cf. the cancellations in Appendix~\ref{sec:unitarity}).\footnote{Note that here and in Appendix~\ref{sec:unitarity} we assume that our projectile dipoles are much smaller than the transverse size of the nucleon target, which is the usual perturbative QCD assumption.}  Similar approximation cannot be applied to $G_{\xx_{1}+\left(1-\frac{z'}{z_{\text{pol}}} \right) \xx_{32}, \, \xx_{1}-\frac{z'}{z_{\text{pol}}} \xx_{32}}$, since the $x_{32}$ dependence in this term is either logarithmic or power-law with a perturbatively small power \cite{Kovchegov:2017jxc}, which is essential for getting the right logarithms coming from the $x_{32}$ integration. (We can replace $\min \left\{ z_{\min}, z', z_{\text{pol}} - z' \right\}  \to z_{\min}$ in its first argument though.)

Performing the approximations \eqref{32approx} in \eq{eqn:Nc9} we rewrite the latter as
\begin{align}\label{eqn:Nc10}
&G_{10} (z_{\min},z_{\text{pol}}) \, S_{10} (z_{\min}) = G^{(0)}_{10} (z_{\text{pol}}) \, S^{(0)}_{10} + \frac{N_c}{\pi^2} \int\limits_{\Lambda^2/s}^{z_{\min}}\frac{dz'}{z'} \int\limits_{1/(z's)}d^2 x_2 \\
&\;\;\;\;\times \left(\frac{\as(1/x^2_{21})}{x^2_{21}}\;\theta\left(x^2_{10}z_{\min}-x^2_{21}z'\right) - \as (\min\{1/x^2_{21},1/x^2_{20}\})\;\frac{\xx_{21}\cdot\xx_{20}}{x^2_{21}x^2_{20}}\;\theta\left(x^2_{10}z_{\min}-\max\left\{x^2_{21},x^2_{20}\right\}z'\right)\right) \notag \\
&\;\;\;\;\times \left[G_{21} (z',z' ) \, S_{20} (z' )  + \Gamma^{gen}_{20,21} (z',z') \, S_{21} (z' ) \right] \, S_{10} (z') \notag \\
&+ \frac{N_c}{2\pi^2} \int\limits_{\Lambda^2/s}^{z_{\min}}\frac{dz'}{z'} \int\limits_{1/z's} d^2 x_2 \, K_{\text{rcBK}} ({\un x}_{0}, {\un x}_{1}; {\un x}_{2}) \, \theta\left(x^2_{10}z_{\min}-x^2_{21}z'\right) \notag \\
&\;\;\;\;\times  \left[G_{21} (z',z_{\text{pol}} ) \, S_{20} (z')  - \Gamma^{gen}_{10, 21} (z',z_{\text{pol}})  \right]  \, S_{10} (z') \notag \\
&\color{blue} - \frac{N_c}{\pi^2}\int\limits_0^{z_{\text{pol}}}\frac{dz'}{z_{\text{pol}}}\int\limits_{\frac{z_{\text{pol}}}{z'  (z_{\text{pol}} - z') s}} \frac{d^2 x_{32}}{x_{32}^2} \;\as\left(\frac{1}{x^2_{32}}\right) \theta (x_{10}^2 z_{\min} z_{\text{pol}}  - x_{32}^2 z' (z_{\text{pol}} - z'))   \, S_{10} (z_{\min} ) \notag \\
&\color{blue} \;\;\;\;\;\;\;\;\;\;\;\;\times \left[G_{\xx_{1}+\left(1-\frac{z'}{z_{\text{pol}}} \right) \xx_{32}, \, \xx_{1}-\frac{z'}{z_{\text{pol}}} \xx_{32}} (z_{\min}, z' ) \, S_{10} (z_{\min}) + \Gamma_{10,32} ( z_{\min}, z' ) \right] \notag \\
&\color{blue} + \frac{N_c}{2\pi^2}\int\limits_0^{z_{\text{pol}}}\frac{dz'}{z_{\text{pol}}} \left(2-\frac{z'}{z_{\text{pol}}}+\frac{z'^2}{z_{\text{pol}}^2}\right) \int\limits_{\frac{z_{\text{pol}}}{z'  (z_{\text{pol}} - z') s}} \frac{d^2 x_{32}}{x_{32}^2} \;\as\left(\frac{1}{x^2_{32}}\right) \theta (x_{10}^2 z_{\min} z_{\text{pol}}  - x_{32}^2 z' (z_{\text{pol}} - z')) \notag \\
&\color{blue} \;\;\;\;\times \Gamma_{10,32} (z_{\min},z_{\text{pol}} ) \, S_{10} (z_{\min}) \notag \\
&\color{purple} + \frac{N_c}{2\pi^2} \int\limits_{\Lambda^2/s}^{z_{\min}}\frac{dz'}{z'} \int d^2 x_2 \, K_{\text{rcBK}} ({\un x}_{0}, {\un x}_{1}; {\un x}_{2}) \, 
\Gamma^{gen}_{10, 21} (z',z_{\text{pol}})   \left[S_{20} (z') \, S_{21} (z') - S_{10} (z') \right] . \notag 
\end{align}
The last (red) term in \eq{eqn:Nc10} is separated from the rest because of its resemblance to the rcBK equation, written in the integral form as
\begin{align}
S_{10} (z_{\min}) = S^{(0)}_{10} + \frac{N_c}{2\pi^2} \int\limits_{\Lambda^2/s}^{z_{\min}}\frac{dz'}{z'} \int d^2 x_2 \,K_{\text{rcBK}} ({\un x}_{0}, {\un x}_{1}; {\un x}_{2}) \, \left[S_{20} (z') \, S_{21} (z') - S_{10} (z') \right] .
\label{eqn:Nc14}
\end{align}
In the last (red) term in \eq{eqn:Nc10} we have also put $\theta\left(x^2_{10}z_{\min}-x^2_{21}z'\right) =1$, since $z' \ll z$ and this term does not contain a logarithmic transverse integral (and, hence, this is a SLA$_L$ term, and not a DLA term, and does not need a lfetime-ordering theta-function).

It now appears tempting to cancel $S_{10} ( z_{\min} )$ on the left-hand side of \eq{eqn:Nc10} and in SLA$_T$ terms on its right-hand side. However, at first glance, such a cancellation cannot apply to the DLA+SLA$_L$ terms and to the inhomogenous term. As shown in Appendix~\ref{sec:GlargeNc}, this cancellation is, in fact, justified for the DLA+SLA$_L$ terms and for the inhomogenous term as well, if we simultaneously drop the last (red) term in \eq{eqn:Nc10}. The resulting large-$N_c$ evolution equation reads
\begin{align}\label{eqn:Nc17}
&G_{10} (z_{\min},z_{\text{pol}}) = G^{(0)}_{10} (z_{\text{pol}}) + \frac{N_c}{\pi^2} \int\limits_{\Lambda^2/s}^{z_{\min}}\frac{dz'}{z'} \int\limits_{1/(z's)}d^2 x_2 \\
&\;\;\;\;\times \left(\frac{\as(1/x^2_{21})}{x^2_{21}}\;\theta\left(x^2_{10}z_{\min}-x^2_{21}z'\right) - \as (\min\{1/x^2_{21},1/x^2_{20}\})\;\frac{\xx_{21}\cdot\xx_{20}}{x^2_{21}x^2_{20}}\;\theta\left(x^2_{10}z_{\min}-\max\left\{x^2_{21},x^2_{20}\right\}z'\right)\right) \notag \\
&\;\;\;\;\times \left[G_{21} (z',z' ) \, S_{20} (z' )  + \Gamma^{gen}_{20,21} (z',z') \, S_{21} (z' ) \right]  \notag \\
&+ \frac{N_c}{2\pi^2} \int\limits_{\Lambda^2/s}^{z_{\min}}\frac{dz'}{z'} \int\limits_{1/z's} d^2 x_2 \, K_{\text{rcBK}} ({\un x}_{0}, {\un x}_{1}; {\un x}_{2}) \, \theta\left(x^2_{10}z_{\min}-x^2_{21}z'\right)   \left[G_{21} (z',z_{\text{pol}} ) \, S_{20} (z')  - \Gamma^{gen}_{10, 21} (z',z_{\text{pol}})  \right] \notag \\
&\color{blue} - \frac{N_c}{\pi^2}\int\limits_0^{z_{\text{pol}}}\frac{dz'}{z_{\text{pol}}}\int\limits_{\frac{z_{\text{pol}}}{z'  (z_{\text{pol}} - z') s}} \frac{d^2 x_{32}}{x_{32}^2} \;\as\left(\frac{1}{x^2_{32}}\right) \theta (x_{10}^2 z_{\min} z_{\text{pol}}  - x_{32}^2 z' (z_{\text{pol}} - z'))  \notag \\
&\color{blue} \;\;\;\;\;\;\;\;\;\;\;\;\times \left[G_{\xx_{1}+\left(1-\frac{z'}{z_{\text{pol}}} \right) \xx_{32}, \, \xx_{1}-\frac{z'}{z_{\text{pol}}} \xx_{32}} ( z_{\min}, z' ) \, S_{10} (z_{\min}) + \Gamma_{10,32} ( z_{\min}, z' ) \right] \notag \\
&\color{blue} + \frac{N_c}{2\pi^2}\int\limits_0^{z_{\text{pol}}}\frac{dz'}{z_{\text{pol}}} \left(2-\frac{z'}{z_{\text{pol}}}+\frac{z'^2}{z_{\text{pol}}^2}\right) \int\limits_{\frac{z_{\text{pol}}}{z'  (z_{\text{pol}} - z') s}} \frac{d^2 x_{32}}{x_{32}^2} \;\as\left(\frac{1}{x^2_{32}}\right) \theta (x_{10}^2 z_{\min} z_{\text{pol}}  - x_{32}^2 z' (z_{\text{pol}} - z')) \notag \\
&\color{blue} \;\;\;\;\times \Gamma_{10,32} ( z_{\min}, z_{\text{pol}} ).  \notag 
\end{align}

Similarly, the evolution equation for the neighbor dipole amplitude is
\begin{align}\label{eqn:Nc18}
&\Gamma_{10,32} (z_{\min},z_{\text{pol}}) = G^{(0)}_{10} (z_{\text{pol}}) + \frac{N_c}{\pi^2} \int\limits_{\Lambda^2/s}^{z_{\min}}\frac{dz'}{z'} \int\limits_{1/(z's)}d^2 x_4 \\
&\;\;\;\;\times \left(\frac{\as(1/x^2_{41})}{x^2_{41}}\;\theta\left(x^2_{32}z_{\min}-x^2_{41}z'\right) - \as (\min\{1/x^2_{41},1/x^2_{40}\})\;\frac{\xx_{41}\cdot\xx_{40}}{x^2_{41}x^2_{40}}\;\theta\left(x^2_{32}z_{\min}-\max\left\{x^2_{41},x^2_{40}\right\}z'\right)\right) \notag \\
&\;\;\;\;\times \left[G_{41} (z',z' ) \, S_{40} (z' )  + \Gamma^{gen}_{40,41} (z',z') \, S_{41} (z' ) \right]  \notag \\
&+ \frac{N_c}{2\pi^2} \int\limits_{\Lambda^2/s}^{z_{\min}}\frac{dz'}{z'} \int\limits_{1/z's} d^2 x_4 \, K_{\text{rcBK}} ({\un x}_{0}, {\un x}_{1}; {\un x}_{4}) \, \theta\left(x^2_{32}z_{\min}-x^2_{41}z'\right)   \left[G_{41} (z',z_{\text{pol}} ) \, S_{40} (z')  - \Gamma^{gen}_{10, 41} (z',z_{\text{pol}})  \right] \notag \\
&\color{blue} - \frac{N_c}{\pi^2}\int\limits_0^{z_{\text{pol}}}\frac{dz'}{z_{\text{pol}}}\int\limits_{\frac{z_{\text{pol}}}{z'  (z_{\text{pol}} - z') s}} \frac{d^2 x_{54}}{x_{54}^2} \;\as\left(\frac{1}{x^2_{54}}\right) \theta (x_{32}^2 z_{\min} z_{\text{pol}}  - x_{54}^2 z' (z_{\text{pol}} - z'))  \notag \\
&\color{blue} \;\;\;\;\;\;\;\;\;\;\;\;\times \left[G_{\xx_{1}+\left(1-\frac{z'}{z_{\text{pol}}} \right) \xx_{54}, \, \xx_{1}-\frac{z'}{z_{\text{pol}}} \xx_{54}} ( z_{\min}, z' ) \, S_{10} (z_{\min}) + \Gamma_{10,54} ( z_{\min}, z' ) \right] \notag \\
&\color{blue} + \frac{N_c}{2\pi^2}\int\limits_0^{z_{\text{pol}}}\frac{dz'}{z_{\text{pol}}} \left(2-\frac{z'}{z_{\text{pol}}}+\frac{z'^2}{z_{\text{pol}}^2}\right) \int\limits_{\frac{z_{\text{pol}}}{z'  (z_{\text{pol}} - z') s}} \frac{d^2 x_{54}}{x_{54}^2} \;\as\left(\frac{1}{x^2_{54}}\right) \theta (x_{32}^2 z_{\min} z_{\text{pol}}  - x_{54}^2 z' (z_{\text{pol}} - z')) \notag \\
&\color{blue} \;\;\;\;\times \Gamma_{10,54} (z_{\min}, z_{\text{pol}} ).  \notag 
\end{align}
Notice the difference in the arguments of the $\theta$-functions between \eqref{eqn:Nc17} and \eqref{eqn:Nc18}. They reflect the fact mentioned above that the lifetime of emissions in the dipole described by the amplitude $\Gamma$ is constrained by its second transverse argument, the smaller size of the neighbor dipole, and not by the first transverse argument, which is its transverse size.

Equations~\eqref{eqn:Nc17} and \eqref{eqn:Nc18} are our main results at large $N_c$: they constitute a closed set of integral equations, generating small-$x$ evolution of the polarized dipole amplitude $G$ in the DLA+SLA approximation and including the running coupling corrections. The unpolarized dipole amplitude $S_{10} (z)$ has to be found separately from \eq{eqn:Nc14}, which includes saturation corrections. 

Outside the saturation region we can linearize Eqs.~\eqref{eqn:Nc17} and \eqref{eqn:Nc18} by putting $S=1$ in them. It also appears easier (though not necessary) to work with the impact-parameter integrated equation. To this end we define the dipole amplitudes integrated over all impact parameters,
\begin{subequations}
\begin{align}
S(x^2_{10},z_{\min}) & \equiv \int d^2 b \ S_{10} (z_{\min}), \\
G\left(x^2_{10},z_{\min},z_{\text{pol}}\right) & \equiv \int d^2 b \ G_{10} (z_{\min},z_{\text{pol}}), \\
\Gamma \left(x^2_{20},x^2_{21},z',z'\right) & \equiv \int d^2 b \ \Gamma_{20,21} (z',z'), 
\end{align}
\end{subequations}
where one can define the impact parameter as ${\un b} = (1/2) ({\un x}_0 + {\un x}_1)$ or simply as ${\un b} = {\un x}_1$ for instance. What is important is that the dipole transverse sizes $x_{10}$, $x_{20}$ and $x_{21}$ are kept fixed during the impact parameter integration. 

Linearizing Eqs.~\eqref{eqn:Nc17} and \eqref{eqn:Nc18} and integrating over the impact parameters we obtain 
\begin{align}\label{eqn:Nc15}
&G\left(x^2_{10},z_{\min},z_{\text{pol}}\right) = G^{(0)}\left(x^2_{10}, z_{\text{pol}}\right) + \frac{N_c}{\pi^2}\int\limits_{\Lambda^2/s}^{z_{\min}}\frac{dz'}{z'}\int\limits_{1/z's} d^2 x_2\\
&\;\;\;\;\times \left(\frac{\as(1/x^2_{21})}{x^2_{21}}\;\theta\left(x^2_{10}z_{\min}-x^2_{21}z'\right) - \as (\min\{1/x^2_{21},1/x^2_{20}\})\;\frac{\xx_{21}\cdot\xx_{20}}{x^2_{21}x^2_{20}}\;\theta\left(x^2_{10}z_{\min}-\max\left\{x^2_{21},x^2_{20}\right\}z'\right)\right) \notag \\
&\;\;\;\;\times \left[G\left(x^2_{21},z',z'\right) + \Gamma_{gen}\left(x^2_{20},x^2_{21},z',z'\right) \right]  \notag \\
&+ \frac{N_c}{2\pi^2}\int\limits_{\Lambda^2/s}^{z_{\min}}\frac{dz'}{z'} \int\limits_{1/z's}d^2 x_2 \, K_{\text{rcBK}} ({\un x}_{0}, {\un x}_{1}; {\un x}_{2}) \, \theta\left(x^2_{10}z_{\min}-x^2_{21}z'\right)    \left[G\left(x^2_{21},z',z_{\text{pol}}\right)  - \Gamma_{gen}(x^2_{10},x^2_{21},z',z_{\text{pol}}) \right]  \notag \\
&\color{blue} - \frac{N_c}{\pi^2}\int\limits_0^{z_{\text{pol}}}\frac{dz'}{z_{\text{pol}}}\int\limits_{1/z' s} \frac{d^2 x_{32}}{x_{32}^2} \;\as\left(\frac{1}{x^2_{32}}\right) \theta (x_{10}^2 z_{\min}   - x_{32}^2 z' ) \left[G\left(x^2_{32},z_{\min},z'\right) + \Gamma\left(x_{10}^2,x^2_{32},z_{\min},z'\right)  \right]  \notag \\
&\color{blue} + \frac{N_c}{2\pi^2}\int\limits_0^{z_{\text{pol}}}\frac{dz'}{z_{\text{pol}}} \left(2-\frac{z'}{z_{\text{pol}}}+\frac{z'^2}{z_{\text{pol}}^2}\right) \int\limits_{1/z' s} \frac{d^2 x_{32}}{x_{32}^2} \;\as\left(\frac{1}{x^2_{32}}\right) \theta (x_{10}^2 z_{\min}  - x_{32}^2 z' ) \,  \Gamma\left(x^2_{10},x^2_{32},z_{\min},z_{\text{pol}}\right)   \notag
\end{align}
and
\begin{align}\label{eqn:Nc16}
&\Gamma\left(x^2_{10},x^2_{32},z_{\min},z_{\text{pol}}\right) = G^{(0)}\left(x^2_{10}, z_{\text{pol}}\right) + \frac{N_c}{\pi^2}\int\limits_{\Lambda^2/s}^{z_{\min}}\frac{dz'}{z'}\int\limits_{1/z's}d^2 x_4 \\
&\;\;\;\;\times \left(\frac{\as(1/x^2_{41})}{x^2_{41}}\;\theta\left(x^2_{32}z_{\min}-x^2_{41}z'\right) - \as (\min\{1/x^2_{41},1/x^2_{40}\})\;\frac{\xx_{41}\cdot\xx_{40}}{x^2_{41}x^2_{40}}\;\theta\left(x^2_{32}z_{\min}-\max\left\{x^2_{41},x^2_{40}\right\}z'\right)\right) \notag \\
&\;\;\;\;\times \left[G\left(x^2_{41},z',z'\right) + \Gamma_{gen}\left(x^2_{40},x^2_{41},z',z'\right) \right] \notag \\
&+ \frac{N_c}{2\pi^2}\int\limits_{\Lambda^2/s}^{z_{\min}}\frac{dz'}{z'}\int\limits_{1/z's} d^2 x_4 \, K_{\text{rcBK}} ({\un x}_{0}, {\un x}_{1}; {\un x}_{4}) \, \theta\left(x^2_{32}z_{\min}-x^2_{41}z'\right) \left[G\left(x^2_{41},z',z_{\text{pol}}\right)  - \Gamma_{gen}(x^2_{10},x^2_{41},z',z_{\text{pol}}) \right] \notag \\
&\color{blue} - \frac{N_c}{\pi^2}\int\limits_0^{z_{\text{pol}}}\frac{dz'}{z_{\text{pol}}}\int\limits_{1/z' s} \frac{d^2 x_{54}}{x_{54}^2} \;\as\left(\frac{1}{x^2_{54}}\right) \theta (x_{32}^2 z_{\min}  - x_{54}^2 z' ) \left[G\left(x^2_{54},z_{\min},z'\right) + \Gamma\left(x_{10}^2,x^2_{54},z_{\min},z'\right) \right] \notag \\
&\color{blue} + \frac{N_c}{2\pi^2}\int\limits_0^{z_{\text{pol}}}\frac{dz'}{z_{\text{pol}}} \left(2-\frac{z'}{z_{\text{pol}}}+\frac{z'^2}{z_{\text{pol}}^2}\right) \int\limits_{1/z' s} \frac{d^2 x_{54}}{x_{54}^2} \;\as\left(\frac{1}{x^2_{54}}\right) \theta (x_{32}^2 z_{\min}  - x_{54}^2 z') \,  \Gamma\left(x^2_{10},x^2_{54},z_{\min},z_{\text{pol}}\right)   . \notag
\end{align}
In arriving at Eqs.~\eqref{eqn:Nc15} and \eqref{eqn:Nc16} we have further employed \eq{eqn:Nc11} to simplify the arguments of the theta-functions and the lower limits of the transverse integrals in the SLA$_T$ terms. Such simplifications only affect the constant under the logarithm.

Equations~\eqref{eqn:Nc15} and \eqref{eqn:Nc16} are the helicity small-$x$ evolution equations at DLA+SLA and at large $N_c$, linearized outside of the saturation region.


\subsection{Large-$N_c \& N_f$ Limit}
\label{sec:ncnf}

Now let us study the large-$N_c \& N_f$ limit of Eqs.~\eqref{evol_rc} and \eqref{Gevol_rc}. We will take the Veneziano limit \cite{Veneziano:1976wm}, in which $N_c$ and $N_f$ are both very large, while their ratio $N_f/N_c$ and $\as \, N_c$ are constant. In addition, we assume that $\as \, N_c \ll 1$ to be able to use QCD perturbation theory.  

In order to apply this limit, we first note that the polarized adjoint Wilson line can be written as (cf. \eq{eq:Wpol4}, see also Eqs.~(74) and (83) in \cite{Kovchegov:2018znm})
\begin{align}\label{briefU}
(U^{\text{pol}}_{\un x})^{ab} =  4 \, \tr \left[ W^{\text{pol} \, \dagger}_{\ul x} t^a V_{\ul x} t^b \right] + 4 \,  \tr \left[ V^\dagger_{\ul x} t^a W_{\ul x}^{\text{pol}} t^b \right] 
\end{align}
with
\begin{align}\label{Wdef}
W_{\ul x}^{\text{pol}} \equiv V_{\ul x}^{\text{pol} \, g} - \frac{g^2 \, p_1^+}{4 s} \, \int\limits_{-\infty}^\infty d x_1^- \, \int\limits_{x_1^-}^\infty d x_2^- \,  V_{\un x} [+ \infty, x_2^-] \, \psi (x_2^-, {\un x})_\alpha \, \left( \frac{1}{2} \, \gamma^+ \, \gamma_5 \right)_{\beta\alpha} \, {\bar \psi} (x_1^-, {\un x})_\beta \, V_{\un x} [x_1^-, - \infty] .
\end{align}
Defining another polarized dipole amplitude $G^{adj}_{10}$ by (cf. Eq.~(82) in \cite{Kovchegov:2018znm})
\begin{align}\label{UU}
\frac{1}{2 (N_c^2 -1)} \, \mbox{Re} \, \llangle \mbox{T} \, \mbox{Tr} \left[ U_{\ul 0} \, U_{{\un 1}}^{\text{pol} \, \dagger} \right] + \mbox{T} \, \mbox{Tr} \left[ U_{{\un 1}}^{\text{pol}} \, U_{\ul 0}^\dagger \right] \rrangle (z_{\min},z_{\text{pol}}) \equiv G^{adj}_{10} (z_{\min},z_{\text{pol}}) \, S_{10} (z_{\min})
\end{align}
we see that, at large $N_c$, dropping the Re and time-ordering signs for brevity, one has (cf. Eqs.~\eqref{Qdef10} and \eqref{Ggdef})
\begin{align}\label{adj_dip}
G^{adj}_{10} (z_{\min},z_{\text{pol}}) = \frac{2}{N_c} \, \llangle \tr \left[ V_{\ul 0} \, W_{{\un 1}}^{\text{pol} \, \dagger} \right] + \tr \left[ W_{{\un 1}}^{\text{pol}} \, V_{\ul 0}^\dagger \right] \rrangle (z_{\min},z_{\text{pol}}) .
\end{align}
Equations \eqref{UU} and \eqref{adj_dip} are related to each other in the same way as Eqs.~\eqref{eqn:Nc5} and \eqref{Ggdef}.

The similarity of \eq{briefU} to \eq{eq:Wpol4} allows us to write at large $N_c$
\begin{subequations}\label{eqn:NcNf0}
\begin{align}\label{eqn:NcNf1}
\frac{1}{N_c^2 -1} \llangle\text{Tr}\left[T^bU_{\underline{0}}T^aU_{\underline{1}}^{\dagger}\right] \left( U_{\underline{2}}^{\text{pol}} \right)^{ba}\rrangle\left(z_{\min},z_{\text{pol}}\right) 
= \frac{N_c}{2} \, S_{10} \left(z_{\min}\right) \, \Big[ & \, S_{20} (z_{\min}) \, G^{adj}_{21} (z_{\min},z_{\text{pol}}) \\
&+ S_{21} (z_{\min}) \, \Gamma^{adj \, gen}_{20, 21} (z_{\min},z_{\text{pol}})\Big] , \notag  \\
\label{eqn:NcNf2}
\frac{1}{N_c^2 -1}  \llangle\text{Tr}\left[T^bU_{\underline{0}}T^a \left( U_{\underline{1}}^{\text{pol}} \right)^{\dagger}\right]U_{\underline{2}}^{ba}\rrangle \left(z_{\min},z_{\text{pol}}\right) = \frac{N_c}{2} \, S_{20} (z_{\min}) \, \Big[& \, S_{10} (z_{\min}) \, G^{adj}_{21} (z_{\min},z_{\text{pol}} ) \\
& + S_{21} (z_{\min}) \, \Gamma^{adj \, gen}_{10,21} (z_{\min},z_{\text{pol}} )\Big] , \notag
\end{align}
\end{subequations}
by analogy to Eqs.~\eqref{eqn:Nc6} and \eqref{eqn:Nc8} (cf. also Eqs.~(98) and (99) in \cite{Kovchegov:2018znm}, while keeping in mind that those equations neglect the SLA evolution by putting $S=1$). Here $\Gamma^{adj \, gen}$ is the generalized polarized dipole amplitude 
\begin{align}\label{eqn:NcNf3}
\Gamma^{adj \, gen}_{10, 32} (z_{\min},z_{\text{pol}}) = G^{adj}_{10} (z_{\min},z_{\text{pol}} ) \, \theta\left(x_{32}-x_{10}\right) + \Gamma^{adj}_{10,32} (z_{\min},z_{\text{pol}}) \, \theta (x_{10}-x_{32} ) 
\end{align}
defined by analogy to \eq{eqn:Nc7} with $\Gamma^{adj}_{10,32}$ the corresponding neighbor dipole amplitude, defined in the same way as $G^{adj}$, but with the lifetime of subsequent emissions depending on the size of another dipole. 

In addition, using Eq.~\eqref{briefU} along with \eq{eqn:Nc2}, and applying Fierz identity, we obtain
\begin{align}\label{eqn:NcNf4}
& \frac{1}{N_c} \llangle \mbox{tr} \left[ t^b \, V_{\ul{0}} \, t^a \, V_{\ul{1}}^{\dagger} \right] \, U^{\text{pol} \, ba}_{\ul{2}} \rrangle \left(z_{\min},z_{\text{pol}}\right) + c.c. \\ & = \frac{N_c}{2} \, S_{20} (z_{\min}) \, G^{adj}_{21} (z_{\min},z_{\text{pol}}) + \frac{N_c}{2} \, S_{21} (z_{\min}) \, \Gamma^{adj \, gen}_{20, 21} (z_{\min},z_{\text{pol}}) . \notag 
\end{align}
Employing this result, along with \eq{eqn:Nc2} and definitions \eqref{Qdef10} and \eqref{eqn:Nc4} in \eq{evol_rc}, and implementing the simplifications from Eqs.~\eqref{eqn:Nc11} and \eqref{32approx} we arrive at the following equation in the large-$N_c \& N_f$ limit
\begin{align}\label{Qevol}
  & Q_{10} (z_{\min}, z_{\text{pol}}) =
  Q_{10}^{(0)} (z_{\text{pol}}) +
  \frac{N_c}{4 \pi^2} \int\limits_{\Lambda^2/s}^{z_{\min}} \frac{d z'}{z'} 
  \int\limits_{1/(z' s)} d^2 x_{2} \\ & \;\;\;\;\times \left[ \left(
      \frac{\as (1/x_{21}^2)}{x_{21}^2} \, \theta (x_{10}^2 z_{\min} - x_{21}^2 z') - \as (\mbox{min} \{ 1/x_{21}^2, 1/x_{20}^2 \} ) \, 
      \frac{\ul{x}_{21} \cdot \ul{x}_{20}}{x_{21}^2 \, x_{20}^2} \,
      \theta (x_{10}^2 z_{\min} - \mbox{max} \{ x_{21}^2, x_{20}^2 \} z')
    \right) \right. \notag \\ & \;\;\;\;\;\;\;\;\;\;\;\;\times \, \left[ S_{20} (z') \, G^{adj}_{21} (z' ,  z') + S_{21} (z') \, \Gamma^{adj \, gen}_{20, 21} (z' ,  z') \right]  \left. +
    \frac{\as (1/x_{21}^2)}{x_{21}^2} \, \theta (x_{10}^2 z_{\min} - x_{21}^2 z') \, S_{10} (z') \, 
   Q_{21} (z', z') \right] \notag \\ &
  + \frac{N_c}{2 \pi^2} \int\limits_{\Lambda^2/s}^{z_{\min}} \frac{d z'}{z'} 
  \int\limits_{1/(z' s)} d^2 x_{2} \, K_{\text{rcBK}} ({\un x}_{0}, {\un x}_{1}; {\un x}_{2}) \, \theta (x_{10}^2 z_{\min} - x_{21}^2 z')  \, 
  \left[ S_{02} (z') \, Q_{12} (z', z_{\text{pol}})  - \overline{\Gamma}^{gen}_{10,21}  (z', z_{\text{pol}}) \right] \notag \\ 
    & \textcolor{blue}{ - \frac{N_c}{8 \pi^2} \int\limits_{0}^{z_{\text{pol}}} \frac{d z'}{z_{\text{pol}}} 
  \int\limits_{\frac{1}{z' s}} \frac{d^2 x_{32}}{x_{32}^2} \, \theta (x_{10}^2 z_{\min}  - x_{32}^2 z' )  \ \as \left( \frac{1}{x_{32}^2} \right) } \notag \\ & \textcolor{blue}{  \;\;\;\;\times  \, \left[ S_{10} (z_{\min}) \, G^{adj}_{\ul{x}_1 + \left( 1 - \frac{z'}{z_{\text{pol}}} \right) \ul{x}_{32}, \, \ul{x}_1 - \frac{z'}{z_{\text{pol}}} \ul{x}_{32}} (z_{\min},z') + \Gamma^{adj}_{10,32} (z_{\min},z')  \right. } \notag \\
    & \textcolor{blue}{ \;\;\;\;\;\;\;\;\left. +  \, 2 \, S_{10} ( z_{\min}) \, Q _{\ul{x}_1 + \left( 1 - \frac{z'}{z_{\text{pol}}} \right) \ul{x}_{32}, \, \ul{x}_1 - \frac{z'}{z_{\text{pol}}} \ul{x}_{32}} (z_{\min},z') \right] } \notag \\
    & \textcolor{blue}{+ \frac{N_c}{4 \pi^2} \int\limits_{0}^{z_{\text{pol}}}  \frac{d z'}{z_{\text{pol}}} \left( 1 + \frac{z'}{z_{\text{pol}}} \right) \int\limits_{\frac{1}{z'  s}}  \frac{d^2 x_{32}}{x_{32}^2} \, \theta (x_{10}^2 z_{\min}  - x_{32}^2 z' )  \ \as \left( \frac{1}{x_{32}^2} \right)  \overline{\Gamma}_{10,32} (z_{\min}, z_{\text{pol}}) , }  \notag 
\end{align}
where we defined the generalized quark polarized dipole amplitude
 \begin{align}
\overline{\Gamma}^{gen}_{10,32} \left( z_{\min}, z_{\text{pol}} \right) &= Q_{10} \left(z_{\min},z_{\text{pol}}\right) \, \theta\left(x_{32}-x_{10}\right) + \overline{\Gamma}_{10,32} \left(z_{\min},z_{\text{pol}}\right) \, \theta\left(x_{10}-x_{32}\right) 
\label{eqn:Ncf4}
\end{align}
with $\overline{\Gamma}_{10,32}$ being the neighbor quark polarized dipole amplitude, i.e., the amplitude in which the transverse size $x_{32}$ of the neighbor dipole controls the lifetime of the subsequent emissions. $\overline{\Gamma}_{10,32}$ is the quark counterpart of $\Gamma^{adj \, gen}_{10, 32}$ from \eq{eqn:NcNf3}. This amplitude obeys the evolution equation which is derived similarly to \eq{Qevol}, taking the external lifetime constraint into the account, 
\begin{align}\label{Gamma_bar_evol}
  & \overline{\Gamma}_{10,32} (z_{\min}, z_{\text{pol}}) =
  Q_{10}^{(0)} (z_{\text{pol}}) +
  \frac{N_c}{4 \pi^2} \int\limits_{\Lambda^2/s}^{z_{\min}} \frac{d z'}{z'} 
  \int\limits_{1/(z' s)} d^2 x_{4} \\ & \;\;\;\;\times \left[ \left(
      \frac{\as (1/x_{41}^2)}{x_{41}^2} \, \theta (x_{32}^2 z_{\min} - x_{41}^2 z') - \as (\mbox{min} \{ 1/x_{41}^2, 1/x_{40}^2 \} ) \, 
      \frac{\ul{x}_{41} \cdot \ul{x}_{40}}{x_{41}^2 \, x_{40}^2} \,
      \theta (x_{32}^2 z_{\min} - \mbox{max} \{ x_{41}^2, x_{40}^2 \} z')
    \right) \right. \notag \\ & \;\;\;\;\;\;\;\;\;\;\;\;\times \, \left[ S_{40} (z') \, G^{adj}_{41} (z' ,  z') + S_{21} (z') \, \Gamma^{adj \, gen}_{40, 41} (z' ,  z') \right]  \left. +
    \frac{\as (1/x_{41}^2)}{x_{41}^2} \, \theta (x_{32}^2 z_{\min} - x_{41}^2 z') \, S_{10} (z') \, 
   Q_{41} (z', z') \right] \notag \\ &
  + \frac{N_c}{2 \pi^2} \int\limits_{\Lambda^2/s}^{z_{\min}} \frac{d z'}{z'} 
  \int\limits_{1/(z' s)} d^2 x_{4} \, K_{\text{rcBK}} ({\un x}_{0}, {\un x}_{1}; {\un x}_{4}) \, \theta (x_{32}^2 z_{\min} - x_{41}^2 z')  \, 
  \left[ S_{04} (z') \, Q_{14} (z', z_{\text{pol}})  - \overline{\Gamma}^{gen}_{10,41}  (z', z_{\text{pol}}) \right] \notag \\ 
    & \textcolor{blue}{ - \frac{N_c}{8 \pi^2} \int\limits_{0}^{z_{\text{pol}}} \frac{d z'}{z_{\text{pol}}} 
  \int\limits_{\frac{1}{z' s}} \frac{d^2 x_{54}}{x_{54}^2} \, \theta (x_{32}^2 z_{\min}  - x_{54}^2 z' )  \ \as \left( \frac{1}{x_{54}^2} \right) } \notag \\ & \textcolor{blue}{  \;\;\;\;\times  \, \left[ S_{10} (z_{\min}) \, G^{adj}_{\ul{x}_1 + \left( 1 - \frac{z'}{z_{\text{pol}}} \right) \ul{x}_{54}, \, \ul{x}_1 - \frac{z'}{z_{\text{pol}}} \ul{x}_{54}} (z_{\min},z') + \Gamma^{adj}_{10,54} (z_{\min},z')  \right. } \notag \\
    & \textcolor{blue}{ \;\;\;\;\;\;\;\;\left. +  \, 2 \, S_{10} ( z_{\min}) \, Q _{\ul{x}_1 + \left( 1 - \frac{z'}{z_{\text{pol}}} \right) \ul{x}_{54}, \, \ul{x}_1 - \frac{z'}{z_{\text{pol}}} \ul{x}_{54}} (z_{\min},z') \right] } \notag \\
    & \textcolor{blue}{+ \frac{N_c}{4 \pi^2} \int\limits_{0}^{z_{\text{pol}}}  \frac{d z'}{z_{\text{pol}}} \left( 1 + \frac{z'}{z_{\text{pol}}} \right) \int\limits_{\frac{1}{z'  s}}  \frac{d^2 x_{54}}{x_{54}^2} \, \theta (x_{32}^2 z_{\min}  - x_{54}^2 z' )  \ \as \left( \frac{1}{x_{54}^2} \right)  \overline{\Gamma}_{10,54} (z_{\min}, z_{\text{pol}}) . }  \notag 
\end{align}

For the gluon sector, the derivation is analogous to the large-$N_c$ case from the previous Section. Employing Eqs.~\eqref{UU}, \eqref{eqn:NcNf0} and \eqref{eqn:NcNf4} in \eq{Gevol_rc}, while applying the approximations \eqref{eqn:Nc11} and \eqref{32approx}, we obtain
\begin{align}\label{Gevol_largeNcNf1}
  & G^{adj}_{10} (z_{\min},z_{\text{pol}}) \, S_{10 } (z_{\min}) = G^{adj \, (0)}_{10} (z_{\text{pol}}) \, S_{10}^{(0)} + \frac{N_c}{\pi^2} \int\limits_{\Lambda^2/s}^{z_{\min}}
  \frac{d z'}{z'} \int\limits_{1/(z' s)} d^2 x_{2}  \\ & \;\;\;\;\times
  \left[ \left( \frac{\as (1/x_{21}^2)}{x_{21}^2} \, \theta (x_{10}^2 z_{\min} - x_{21}^2
      z') - \as (\mbox{min} \{ 1/x_{21}^2, 1/x_{20}^2 \} ) \, \frac{\ul{x}_{21} \cdot \ul{x}_{20}}{x_{21}^2 \, x_{20}^2}
      \, \theta (x_{10}^2 z_{\min} - \mbox{max} \{ x_{21}^2, x_{20}^2 \} z')
    \right) \right. \notag \\ & \;\;\;\;\;\;\;\;\;\;\;\;\times S_{10} \left(z' \right) \, \Big[ S_{20} (z') \, G^{adj}_{21} (z', z') + S_{21} (z') \, \Gamma^{adj \, gen}_{20, 21} (z', z') \Big] 
   \notag \\ & \;\;\;\;\;\;\;\;\left. - \frac{\as (1/x_{21}^2)}{x_{21}^2} \, \theta (x_{10}^2 z_{\min} -
    x_{21}^2 z') \, \frac{N_f}{2 N_c} \, S_{10} \left(z' \right) \, \overline{\Gamma}_{20, 21}^{gen} (z', z')  \right] \notag \\ &
  + \frac{N_c}{2 \pi^2} \int\limits_{\Lambda^2/s}^{z_{\min}} \frac{d z'}{z'} \!
  \int\limits_{1/(z' s)} d^2 x_{2} \, K_{\text{rcBK}} ({\un x}_{0}, {\un x}_{1}; {\un x}_{2}) \, \theta (x_{10}^2 z_{\min} - x_{21}^2 z')  \notag \\ 
& \;\;\;\;\times S_{10} (z') \,  \left[ S_{20} (z') \, G^{adj}_{21} (z',z_{\text{pol}} )  -  \Gamma^{adj \, gen}_{10,21} (z', z_{\text{pol}} ) \right] \notag \\
      & \textcolor{blue}{  - \frac{N_c}{\pi^2} \int\limits_{0}^{z_{\text{pol}}}
  \frac{d z'}{z_{\text{pol}}} \int\limits_{\frac{1}{z'  s}} \frac{d^2 x_{32}}{x_{32}^2} \, \theta (x_{10}^2 z_{\min}  - x_{32}^2 z' )  \ \as \left( \frac{1}{x_{32}^2} \right) \, S_{10} \left(z_{\min}\right)  } \notag \\ 
  & \textcolor{blue}{ \;\;\;\;\times \, \Big[ S_{10} (z_{\min}) \, G^{adj}_{\ul{x}_1 + \left( 1 - \frac{z'}{z_{\text{pol}}} \right) \ul{x}_{32}, \, \ul{x}_1 - \frac{z'}{z_{\text{pol}}} \ul{x}_{32}} (z_{\min}, z') + \Gamma^{adj}_{10, 32} (z_{\min}, z')\Big]} \notag \\ 
  & \textcolor{blue}{  + \frac{N_f}{\pi^2} \int\limits_{0}^{z_{\text{pol}}}
  \frac{d z'}{z_{\text{pol}}} \, \int\limits_{\frac{1}{z'  s}} \frac{d^2 x_{32}}{x_{32}^2} \, \theta (x_{10}^2 z_{\min} - x_{32}^2 z' )  \ \as \left( \frac{1}{x_{32}^2} \right) \, S_{10} (z_{\min}) \, \overline{\Gamma}_{10, 32} (z_{\min}, z') } \notag \\ 
  & \textcolor{blue}{  + \frac{1}{2 \pi^2} \int\limits_{0}^{z_{\text{pol}}}
  \frac{d z'}{z_{\text{pol}}} \int\limits_{\frac{1}{z' s}} \frac{d^2 x_{32}}{x_{32}^2} \left[ N_c \left( 2 - \frac{z'}{z_{\text{pol}}} + \frac{z'^2}{z_{\text{pol}}^2} \right) - \frac{N_f}{2} \left( \frac{z'^2}{z_{\text{pol}}^2} + \left( 1 - \frac{z'}{z_{\text{pol}}} \right)^2 \right) \right]   } \notag \\ 
  & \textcolor{blue}{  \;\;\;\;\times  \, \theta \left(x_{10}^2 z_{\min} - x_{32}^2 z' \right)  \as \left( \frac{1}{x_{32}^2} \right) \, S_{10} (z_{\min}) \, \Gamma^{adj}_{10, 32} (z_{\min}, z_{\text{pol}}) } \notag   \\
   & \textcolor{purple}{ + \frac{N_c}{2 \pi^2} \int\limits_{\Lambda^2/s}^{z_{\min}} \frac{d z'}{z'} \!
  \int d^2 x_{2} \, K_{\text{rcBK}} ({\un x}_{0}, {\un x}_{1}; {\un x}_{2}) \, \Gamma^{adj \, gen}_{10,21} (z', z_{\text{pol}} )  \, \left[ S_{20} (z') \, S_{21} (z') - S_{10} (z') \right] } . \notag
\end{align}
Here, similar to \eq{eqn:Nc10}, we have separated the last (red) term which contains an rcBK iteration. Again, applying the calculation outlined in Appendix~\ref{sec:GlargeNc}, we reduce \eq{Gevol_largeNcNf1} to
\begin{align}\label{Gevol_largeNcNf}
  & G^{adj}_{10} (z_{\min},z_{\text{pol}}) = G^{adj \, (0)}_{10} (z_{\text{pol}}) + \frac{N_c}{\pi^2} \int\limits_{\Lambda^2/s}^{z_{\min}}
  \frac{d z'}{z'} \int\limits_{1/(z' s)} d^2 x_{2}  \\ & \;\;\;\;\times
  \left[ \left( \frac{\as (1/x_{21}^2)}{x_{21}^2} \, \theta (x_{10}^2 z_{\min} - x_{21}^2
      z') - \as (\mbox{min} \{ 1/x_{21}^2, 1/x_{20}^2 \} ) \, \frac{\ul{x}_{21} \cdot \ul{x}_{20}}{x_{21}^2 \, x_{20}^2}
      \, \theta (x_{10}^2 z_{\min} - \mbox{max} \{ x_{21}^2, x_{20}^2 \} z')
    \right) \right. \notag \\ & \;\;\;\;\;\;\;\;\;\;\;\;\times \Big[ S_{20} (z') \, G^{adj}_{21} (z', z') + S_{21} (z') \, \Gamma^{adj \, gen}_{20, 21} (z', z') \Big] 
  \left. - \frac{\as (1/x_{21}^2)}{x_{21}^2} \, \theta (x_{10}^2 z_{\min} -
    x_{21}^2 z') \, \frac{N_f}{2 N_c} \, \overline{\Gamma}_{20, 21}^{gen} (z', z')  \right] \notag \\ &
  + \frac{N_c}{2 \pi^2} \int\limits_{\Lambda^2/s}^{z_{\min}} \frac{d z'}{z'} \!
  \int\limits_{1/(z' s)} d^2 x_{2} \, K_{\text{rcBK}} ({\un x}_{0}, {\un x}_{1}; {\un x}_{2}) \, \theta (x_{10}^2 z_{\min} - x_{21}^2 z')   \left[ S_{20} (z') \, G^{adj}_{21} (z',z_{\text{pol}} )  -  \Gamma^{adj \, gen}_{10,21} (z', z_{\text{pol}} ) \right] \notag \\
      & \textcolor{blue}{  - \frac{N_c}{\pi^2} \int\limits_{0}^{z_{\text{pol}}}
  \frac{d z'}{z_{\text{pol}}} \int\limits_{\frac{1}{z'  s}} \frac{d^2 x_{32}}{x_{32}^2} \, \theta (x_{10}^2 z_{\min}  - x_{32}^2 z' )  \ \as \left( \frac{1}{x_{32}^2} \right)  } \notag \\ 
  & \textcolor{blue}{ \;\;\;\;\times \, \Big[ S_{10} (z_{\min}) \, G^{adj}_{\ul{x}_1 + \left( 1 - \frac{z'}{z_{\text{pol}}} \right) \ul{x}_{32}, \, \ul{x}_1 - \frac{z'}{z_{\text{pol}}} \ul{x}_{32}} (z_{\min}, z') + \Gamma^{adj}_{10, 32} (z_{\min}, z')\Big]} \notag \\ 
  & \textcolor{blue}{  + \frac{N_f}{\pi^2} \int\limits_{0}^{z_{\text{pol}}}
  \frac{d z'}{z_{\text{pol}}} \, \int\limits_{\frac{1}{z'  s}} \frac{d^2 x_{32}}{x_{32}^2} \, \theta (x_{10}^2 z_{\min} - x_{32}^2 z' )  \ \as \left( \frac{1}{x_{32}^2} \right) \, \overline{\Gamma}_{10, 32} (z_{\min}, z') } \notag \\ 
  & \textcolor{blue}{  + \frac{1}{2 \pi^2} \int\limits_{0}^{z_{\text{pol}}}
  \frac{d z'}{z_{\text{pol}}} \int\limits_{\frac{1}{z' s}} \frac{d^2 x_{32}}{x_{32}^2} \left[ N_c \left( 2 - \frac{z'}{z_{\text{pol}}} + \frac{z'^2}{z_{\text{pol}}^2} \right) - \frac{N_f}{2} \left( \frac{z'^2}{z_{\text{pol}}^2} + \left( 1 - \frac{z'}{z_{\text{pol}}} \right)^2 \right) \right]   } \notag \\ 
  & \textcolor{blue}{  \;\;\;\;\times  \, \theta \left(x_{10}^2 z_{\min} - x_{32}^2 z' \right)  \as \left( \frac{1}{x_{32}^2} \right) \, \Gamma^{adj}_{10, 32} (z_{\min}, z_{\text{pol}}) .} \notag 
\end{align}
The only difference between the derivation here and that in Section~\ref{sec:nc} is that we now keep all the terms with the $\frac{N_f}{N_c}$ factor, which is no longer assumed to be small. For the neighbor gluon polarized dipole amplitude $\Gamma^{adj}_{10, 32}$, the derivation similarly produces
\begin{align}\label{Gamma_evol_largeNcNf}
  & \Gamma^{adj}_{10, 32} (z_{\min},z_{\text{pol}}) = G^{adj \, (0)}_{10} (z_{\text{pol}}) + \frac{N_c}{\pi^2} \int\limits_{\Lambda^2/s}^{z_{\min}}
  \frac{d z'}{z'} \int\limits_{1/(z' s)} d^2 x_{4}  \\ & \;\;\;\;\times
  \left[ \left( \frac{\as (1/x_{41}^2)}{x_{41}^2} \, \theta (x_{32}^2 z_{\min} - x_{41}^2 z') - \as (\mbox{min} \{ 1/x_{41}^2, 1/x_{40}^2 \} ) \, 
      \frac{\ul{x}_{41} \cdot \ul{x}_{40}}{x_{41}^2 \, x_{40}^2} \,
      \theta (x_{32}^2 z_{\min} - \mbox{max} \{ x_{41}^2, x_{40}^2 \} z')
    \right) \right. \notag \\ & \;\;\;\;\;\;\;\;\;\;\;\;\times \Big[ S_{40} (z') \, G^{adj}_{41} (z', z') + S_{41} (z') \, \Gamma^{adj \, gen}_{40, 41} (z', z') \Big] 
  \left. - \frac{\as (1/x_{41}^2)}{x_{41}^2} \, \theta (x_{32}^2 z_{\min} -
    x_{41}^2 z') \, \frac{N_f}{2 N_c} \, \overline{\Gamma}_{40, 41}^{gen} (z', z')  \right] \notag \\ &
  + \frac{N_c}{2 \pi^2} \int\limits_{\Lambda^2/s}^{z_{\min}} \frac{d z'}{z'} \!
  \int\limits_{1/(z' s)} d^2 x_{4} \, K_{\text{rcBK}} ({\un x}_{0}, {\un x}_{1}; {\un x}_{4}) \, \theta (x_{32}^2 z_{\min} - x_{41}^2 z')   \left[ S_{40} (z') \, G^{adj}_{41} (z',z_{\text{pol}} )  -  \Gamma^{adj \, gen}_{10,41} (z', z_{\text{pol}} ) \right] \notag \\
      & \textcolor{blue}{  - \frac{N_c}{\pi^2} \int\limits_{0}^{z_{\text{pol}}}
  \frac{d z'}{z_{\text{pol}}} \int\limits_{\frac{1}{z'  s}} \frac{d^2 x_{54}}{x_{54}^2} \, \theta (x_{32}^2 z_{\min}  - x_{54}^2 z' )  \ \as \left( \frac{1}{x_{54}^2} \right)  } \notag \\ 
  & \textcolor{blue}{ \;\;\;\;\times \, \Big[ S_{10} (z_{\min}) \, G^{adj}_{\ul{x}_1 + \left( 1 - \frac{z'}{z_{\text{pol}}} \right) \ul{x}_{54}, \, \ul{x}_1 - \frac{z'}{z_{\text{pol}}} \ul{x}_{54}} (z_{\min}, z') + \Gamma^{adj}_{10, 54} (z_{\min}, z')\Big]} \notag \\ 
  & \textcolor{blue}{  + \frac{N_f}{\pi^2} \int\limits_{0}^{z_{\text{pol}}}
  \frac{d z'}{z_{\text{pol}}} \, \int\limits_{\frac{1}{z'  s}} \frac{d^2 x_{54}}{x_{54}^2} \, \theta (x_{32}^2 z_{\min} - x_{54}^2 z' )  \ \as \left( \frac{1}{x_{54}^2} \right) \, \overline{\Gamma}_{10, 54} (z_{\min}, z') } \notag \\ 
  & \textcolor{blue}{  + \frac{1}{2 \pi^2} \int\limits_{0}^{z_{\text{pol}}}
  \frac{d z'}{z_{\text{pol}}} \int\limits_{\frac{1}{z' s}} \frac{d^2 x_{54}}{x_{54}^2} \left[ N_c \left( 2 - \frac{z'}{z_{\text{pol}}} + \frac{z'^2}{z_{\text{pol}}^2} \right) - \frac{N_f}{2} \left( \frac{z'^2}{z_{\text{pol}}^2} + \left( 1 - \frac{z'}{z_{\text{pol}}} \right)^2 \right) \right]   } \notag \\ 
  & \textcolor{blue}{  \;\;\;\;\times  \, \theta \left(x_{32}^2 z_{\min} - x_{54}^2 z' \right)  \as \left( \frac{1}{x_{54}^2} \right) \, \Gamma^{adj}_{10, 54} (z_{\min}, z_{\text{pol}}) .} \notag 
\end{align}
Equations \eqref{Qevol}, \eqref{Gamma_bar_evol}, \eqref{Gevol_largeNcNf}, and \eqref{Gamma_evol_largeNcNf} are a closed set of helicity evolution equations at DLA+SLA in the large-$N_c \& N_f$ limit including running coupling corrections. Their solution would yield the most advanced theoretical knowledge of the small-$x$ asymptotics of helicity PDFs and TMDs to date.


\section{Conclusions and Outlook}
\label{sec:conclusion}

In this paper we have calculated the single-logarithmic corrections to the existing double-logarithmic helicity evolution at small $x$ derived originally in \cite{Kovchegov:2015pbl, Kovchegov:2016zex, Kovchegov:2016weo, Kovchegov:2017jxc, Kovchegov:2017lsr, Kovchegov:2018znm, Cougoulic:2019aja}. The main results are given in Eqs.~\eqref{evol_rc} and \eqref{Gevol_rc}, which include the DLA+SLA evolution kernels along with the running coupling corrections. Closed equations are derived at large $N_c$ (Eqs.~\eqref{eqn:Nc17} and \eqref{eqn:Nc18}) and, separately, at large $N_c \& N_f$ (Eqs.~\eqref{Qevol}, \eqref{Gamma_bar_evol}, \eqref{Gevol_largeNcNf}, and \eqref{Gamma_evol_largeNcNf}). At the single-logarithmic level, these equations mix helicity evolution with the unpolarized BK evolution, bringing in the effects of gluon saturation. Saturation effects are known to suppress the unpolarized small-$x$ evolution \cite{Gribov:1984tu,Iancu:2003xm,Weigert:2005us,JalilianMarian:2005jf,Gelis:2010nm,Albacete:2014fwa,Kovchegov:2012mbw} and the Reggeon evolution \cite{Itakura:2003jp} in the saturation region. The extent of the impact of saturation effects on helicity evolution will be explored in the future work.  

Before the work in \cite{Kovchegov:2015pbl, Kovchegov:2016zex, Kovchegov:2016weo, Kovchegov:2017jxc, Kovchegov:2017lsr, Kovchegov:2018znm, Cougoulic:2019aja}, the helicity evolution had been studied in \cite{Bartels:1995iu,Bartels:1996wc} in the double-logarithmic approximation, resumming powers of $\as \, \ln^2 (1/x)$. The small-$x$ asymptotics of helicity PDFs obtained in \cite{Kovchegov:2016weo, Kovchegov:2017jxc, Kovchegov:2017lsr} disagreed with that obtained in \cite{Bartels:1995iu,Bartels:1996wc}: while the origin of this discrepancy is still a largely open problem, possible reasons for this disagreement were outlined in \cite{Kovchegov:2016zex}. In addition, as we mentioned above, the equations from \cite{Kovchegov:2015pbl, Kovchegov:2016zex, Kovchegov:2016weo, Kovchegov:2017jxc, Kovchegov:2017lsr, Kovchegov:2018znm, Cougoulic:2019aja} were recently independently re-derived using the background field method \cite{Chirilli:2021lif} (see also \cite{CKTT}). 

Our present results represent a significant step forward in our understanding of parton helicity at small $x$. It is not clear how to obtain the SLA$_L$ terms using the infrared renormalization group equations employed in \cite{Bartels:1995iu,Bartels:1996wc}. The only known corrections to the results of \cite{Bartels:1995iu,Bartels:1996wc}, constructed in \cite{Ermolaev:2003zx}, are due to the running of the coupling. Above, for the first time ever, we have constructed both the running-coupling and SLA corrections to the 
small-$x$  
DLA helicity evolution, including both the SLA$_L$ and SLA$_T$ terms in the evolution kernel. 

The expressions relating the dipole amplitude $Q_{10}$ to the $g_1$ structure function of a nucleon or nucleus and to the flavor-singlet quark helicity PDF and TMD can be found in \cite{Kovchegov:2015pbl, Kovchegov:2016zex, Kovchegov:2016weo, Kovchegov:2017lsr, Kovchegov:2018znm, Kovchegov:2020hgb, Adamiak:2021ppq} and in \eq{Qtog1} above. Using those expressions (the ``impact factors"), perhaps improved to include single logarithms of $x$ in addition to the double logarithms they already contain, along with the evolution equations derived above, one can now build on the recent work in \cite{Adamiak:2021ppq} to develop 
high-precision 
small-$x$ helicity phenomenology 
within the approach based purely on evolution in $x$.


\section*{Acknowledgments}

\label{sec:acknowledgement}

We would like to thank Dr.~Florian~Cougoulic for discussions during the final stages of the calculation and for reading a draft version of this work. 

This material is based upon work supported by the U.S. Department of
Energy, Office of Science, Office of Nuclear Physics under Award
Number DE-SC0004286. The work is performed within the framework of the TMD Topical Collaboration.



\appendix


\section{Cancellations due to Unitarity for SLA$_T$ Splittings on Unpolarized Lines}
\label{sec:unitarity}

In this Appendix, we explicitly work out the cancellation of diagrams $E$, $F$ and $G$ from \fig{fig:qkdiagrams} in the SLA$_T$ part of the quark dipole evolution equations. Our conclusions also justify neglecting diagrams $\{N, O, P\}$ and $\{Q,R,S\}$ from \fig{fig:gldiagrams} in the  SLA$_T$ terms of the adjoint dipole evolution. The procedure used in this Appendix to compute the contribution from each of the diagrams is similar to that used to compute the remaining SLA$_T$ terms in \eqref{eqn:qknorc2}.

\begin{figure}[h]
\includegraphics[width= 0.4 \textwidth]{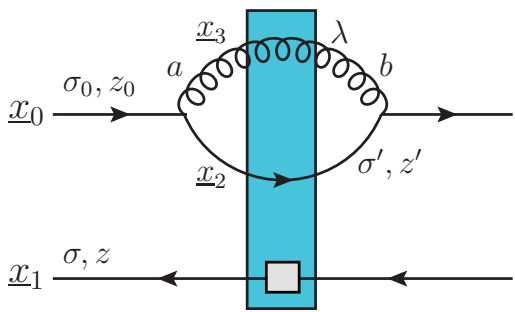} 
\caption{A more detailed rendering of the diagram E.}
\label{fig:diagramE}
\end{figure}

First, consider diagram $E$ shown in Fig \ref{fig:diagramE}. Its contribution to the polarized quark dipole evolution equations is
\begin{align}\label{eqn:qkE1}
\tilde{E} &= \int\limits_{\Lambda^2/s}^{z_0}\frac{dz'}{4\pi z'(1-z'/z_0)}\int\limits_{1/z's}d^2 x_{32}\;\theta\left(z_{\min} x^2_{10} - z' x^2_{32}\right)\frac{1}{2}\sum_{\sigma_0,\sigma',\lambda} \\ 
&\;\;\;\;\;\times \frac{1}{N_c}\llangle\text{tr}\left[\psi^{q\to qG*}_{b\sigma_0\sigma'\lambda}\left(\xx_{32},\frac{z'}{z_0}\right)V_{\underline{2}}\psi^{q\to qG}_{a\sigma_0\sigma'\lambda}\left(\xx_{32},\frac{z'}{z_0}\right)V_{\underline{1}}^{\text{pol }\dagger}\right]U_{\underline{3}}^{ba}\rrangle (\min\{z ,z'\},z)  \notag \\
&=\frac{\alpha_s}{8\pi^2} \int\limits_{\Lambda^2/s}^{z_0}\frac{dz'}{z_0(1-z'/z_0)}\int\limits_{1/z's}\frac{d^2 x_{32}}{x^2_{32}}\;\theta\left(z_{\min} x^2_{10} - z' x^2_{32}\right)\sum_{\sigma_0,\sigma',\lambda}\delta_{\sigma_0\sigma'} \left[\left(1+\frac{z'^2}{z_0^2}\right)+\sigma_0\lambda\left(1-\frac{z'^2}{z^2_0}\right)\right] \notag \\
&\;\;\;\;\;\times \frac{1}{N_c}\llangle\text{tr}\left[t^bV_{\underline{2}}t^aV_{\underline{1}}^{\text{pol }\dagger}\right]U_{\underline{3}}^{ba}\rrangle (\min\{z,z'\},z) \notag \\
&=\frac{\alpha_s}{2\pi^2} \int\limits_{\Lambda^2/s}^{z_0}\frac{dz'}{z_0(1-z'/z_0)}\left(1+\frac{z'^2}{z_0^2}\right) \int\limits_{1/z's}\frac{d^2 x_{32}}{x^2_{32}}\;\theta\left(z_{\min} x^2_{10} - z' x^2_{32}\right)  \frac{1}{N_c}\llangle\text{tr}\left[t^bV_{\underline{2}}t^aV_{\underline{1}}^{\text{pol }\dagger}\right]U_{\underline{3}}^{ba}\rrangle (\min\{z,z'\},z) , \notag 
\end{align}
where $\psi^{q\to qG}_{a\sigma\sigma'\lambda}(\underline{x},z)$ is given by \eq{eqn:psiqqG3} and $z_0 \xx_0 = z' \xx_2 + (z_0 - z') \xx_3$, as follows from the delta-function in \eq{eqn:psiqqG2}. Here we have also defined $z_{\min} = \min \{ z, z_0\}$. In the limit of the soft quark at $\xx_2$, when $z'\ll z_0$, \eq{eqn:qkE1} becomes
\begin{align}
\tilde{E}\big|_{z'\ll z_0} &= \frac{\alpha_s}{2\pi^2} \int\limits_{\Lambda^2/s}^{z_0}\frac{dz'}{z_0} \int\limits_{1/z's}\frac{d^2 x_{32}}{x^2_{32}}\;\theta\left(z_{\min} x^2_{10} - z' x^2_{32}\right)  \frac{1}{N_c}\llangle\text{tr}\left[t^bV_{\underline{2}}t^aV_{\underline{1}}^{\text{pol }\dagger}\right]U_{\underline{3}}^{ba}\rrangle (\min\{z,z'\},z) ,
\label{eqn:qkE2}
\end{align}
which is a part of SLA$_T$, since the longitudinal ($z'$) integral is not logarithmic now. However, in the limit where the gluon is soft, $z_0-z'\ll z_0$, \eqref{eqn:qkE1} is a part of the DLA kernel:
\begin{align}
\tilde{E}\big|_{z_0-z'\ll z_0} &= \frac{\alpha_s}{\pi^2} \int\limits_{\Lambda^2/s}^{z_0}\frac{dz'}{z_0-z'} \int\limits_{1/z's}\frac{d^2 x_{32}}{x^2_{32}}\;\theta\left(z_{\min} x^2_{10} - z' x^2_{32}\right)  \frac{1}{N_c}\llangle\text{tr}\left[t^bV_{\underline{2}}t^aV_{\underline{1}}^{\text{pol }\dagger}\right]U_{\underline{3}}^{ba}\rrangle (\min\{z,z'\},z) .
\label{eqn:qkE3}
\end{align}
Then, to obtain only the SLA$_T$ contribution of the diagram $E$, we subtract \eqref{eqn:qkE3} from \eqref{eqn:qkE1} to get
\begin{align}\label{eqn:qkE4}
E = \tilde{E} - \tilde{E}\big|_{z_0-z'\ll z_0} & = -\frac{\alpha_s}{2\pi^2} \int\limits_{\Lambda^2/s}^{z_0}\frac{dz'}{z_0}\left(1+\frac{z'}{z_0}\right) \int\limits_{1/z's}\frac{d^2 x_{32}}{x^2_{32}}\;\theta\left(z_{\min} x^2_{10} - z' x^2_{32}\right)  \\
& \times  \frac{1}{N_c}\llangle\text{tr}\left[t^bV_{\underline{x}_0-\left(1-\frac{z'}{z_0}\right)\xx_{32}}t^aV_{\underline{1}}^{\text{pol }\dagger}\right]U_{\underline{x}_0+\frac{z'}{z_0}\xx_{32}}^{ba}\rrangle (\min\{z,z'\},z) . \notag
\end{align}

\begin{figure}[hb]
\includegraphics[width= 0.8 \textwidth]{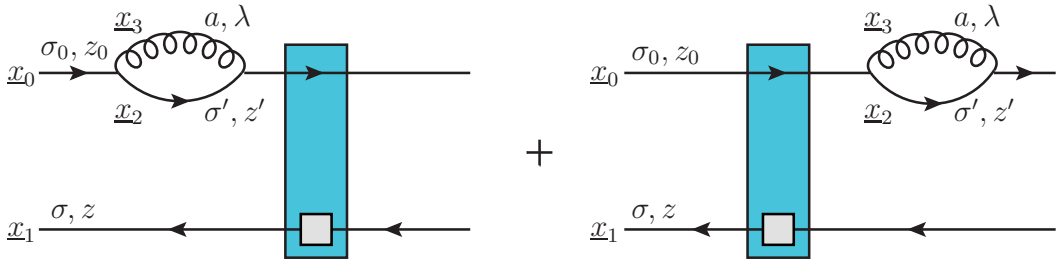} 
\caption{A detailed representation of the diagrams F and G.}
\label{fig:diagramFG}
\end{figure}

Now, consider diagrams $F$ and $G$ shown in \fig{fig:diagramFG}. These diagrams yield (for the polarized dipole evolution)
\begin{align}\label{eqn:qkF1}
\tilde{F}+\tilde{G} &= - \int\limits_{\Lambda^2/s}^{z_0}\frac{dz'}{4\pi z'(1-z'/z_0)}\int\limits_{1/z's}d^2 x_{32}\;\theta\left(z_{\min} x^2_{10} - z' x^2_{32}\right)\frac{1}{2}\sum_{\sigma_0,\sigma',\lambda} \\
&\;\;\;\;\;\times \frac{1}{N_c}\llangle\text{tr}\left[\left|\psi^{q\to qG}_{a\sigma_0\sigma'\lambda}\left(\xx_{32},\frac{z'}{z_0}\right)\right|^2V_{\underline{0}}V_{\underline{1}}^{\text{pol }\dagger}\right]\rrangle (\min\{z,z' \},z) \notag \\
&=-\frac{\alpha_s}{8\pi^2} \int\limits_{\Lambda^2/s}^{z_0}\frac{dz'}{z_0(1-z'/z_0)}\int\limits_{1/z's}\frac{d^2 x_{32}}{x^2_{32}}\;\theta\left(z_{\min} x^2_{10} - z' x^2_{32}\right)\sum_{\sigma_0,\sigma',\lambda}\delta_{\sigma_0\sigma'} \left[\left(1+\frac{z'^2}{z_0^2}\right)+\sigma_0\lambda\left(1-\frac{z'^2}{z^2_0}\right)\right] \notag \\
&\;\;\;\;\;\times \frac{1}{N_c}\llangle\text{tr}\left[t^at^aV_{\underline{0}}V_{\underline{1}}^{\text{pol }\dagger}\right]\rrangle (\min\{z,z' \},z) \notag \\
&= - \frac{\alpha_sC_F}{2\pi^2} \int\limits_{\Lambda^2/s}^{z_0}\frac{dz'}{z_0(1-z'/z_0)}\left(1+\frac{z'^2}{z_0^2}\right) \int\limits_{1/z's}\frac{d^2 x_{32}}{x^2_{32}}\;\theta\left(z_{\min} x^2_{10} - z' x^2_{32}\right)  \frac{1}{N_c}\llangle\text{tr}\left[V_{\underline{0}}V_{\underline{1}}^{\text{pol }\dagger}\right]\rrangle (\min\{z,z' \},z)  . \notag
\end{align}
The result \eqref{eqn:qkF1} is the same as \eqref{eqn:qkE1} except for the factor of $-C_F$ and the Wilson lines in the double angle brackets. Subtracting the soft-gluon (DLA) contribution with $z_0-z'\ll z_0$ out of \eq{eqn:qkF1} gives us the SLA$_T$ contribution of the diagrams $F$ and $G$,
\begin{align}\label{eqn:qkF4}
& F+G  = \tilde{F}+\tilde{G} - [\tilde{F}+\tilde{G}]\big|_{z_0-z'\ll z_0} \\ & = \frac{\alpha_sC_F}{2\pi^2} \int\limits_{\Lambda^2/s}^{z_0}\frac{dz'}{z_0}\left(1+\frac{z'}{z_0}\right) \int\limits_{1/z's}\frac{d^2 x_{32}}{x^2_{32}}\;\theta\left(z_{\min}  x^2_{10} - z' x^2_{32}\right)  \frac{1}{N_c}\llangle\text{tr}\left[V_{\underline{0}}V_{\underline{1}}^{\text{pol }\dagger}\right]\rrangle (\min\{z,z' \},z) . \notag
\end{align}
Combining \eqref{eqn:qkE4} and \eqref{eqn:qkF4}, we see that the sum $E+F+G$ is proportional to
\begin{align}\label{eqn:qkEFG1}
& E+F+G \propto \llangle\text{tr}\left[t^bV_{\underline{x}_0-\left(1-\frac{z'}{z_0}\right)\xx_{32}}t^aV_{\underline{1}}^{\text{pol }\dagger}\right]U_{\underline{x}_0+\frac{z'}{z_0}\xx_{32}}^{ba}\rrangle (\min\{z,z'\},z) -  C_F\llangle\text{tr}\left[V_{\underline{0}}V_{\underline{1}}^{\text{pol }\dagger}\right]\rrangle (\min\{z,z' \} ,z)  \\
\!\!\! = \, & 2 \, \llangle\text{tr}\left[t^bV_{\underline{x}_0-\left(1-\frac{z'}{z_0}\right)\xx_{32}}t^aV_{\underline{1}}^{\text{pol }\dagger}\right]\text{tr}\left[t^bV_{\underline{x}_0+\frac{z'}{z_0}\xx_{32}}t^aV_{\underline{x}_0+\frac{z'}{z_0}\xx_{32}}^{\dagger}\right]\rrangle (\min\{z,z'\},z) -  C_F\llangle\text{tr}\left[V_{\underline{0}}V_{\underline{1}}^{\text{pol }\dagger}\right]\rrangle (\min\{z,z' \},z) \notag \\
& =  \frac{1}{2}\llangle\text{tr}\left[V_{\underline{x}_0-\left(1-\frac{z'}{z_0}\right)\xx_{32}}V_{\underline{x}_0+\frac{z'}{z_0}\xx_{32}}^{\dagger}\right]\text{tr}\left[V_{\underline{x}_0+\frac{z'}{z_0}\xx_{32}}V_{\underline{1}}^{\text{pol }\dagger}\right]\rrangle (\min\{z,z'\},z) \notag \\
&\;\;\;\;\;- \frac{1}{2N_c}\llangle\text{tr}\left[V_{\underline{x}_0-\left(1-\frac{z'}{z_0}\right)\xx_{32}}V_{\underline{1}}^{\text{pol }\dagger}\right]\rrangle (\min\{z,z'\},z) -  C_F\llangle\text{tr}\left[V_{\underline{0}}V_{\underline{1}}^{\text{pol }\dagger}\right]\rrangle (\min\{z,z' \} ,z), \notag
\end{align}
where we have used the Fierz identity and \eq{eqn:Nc2}. If we, further, apply the $x_{32}\ll x_{10}$ condition, which is appropriate for the SLA$_T$ terms and which also follows from the theta-function $\theta\left(z_{\min}  x^2_{10} - z' x^2_{32}\right)$ since $z_{\min} \lesssim z'$ due to $z' \sim z_0$, we see that the expression in \eqref{eqn:qkEFG1} approximates to
\begin{align}\label{eqn:qkEFG2}
 E+F+G \propto & \, \frac{1}{2}\llangle\text{tr}\left[V_{\underline{0}}V_{\underline{0}}^{\dagger}\right]\text{tr}\left[V_{\underline{0}}V_{\underline{1}}^{\text{pol }\dagger}\right]\rrangle (\min\{z,z' \},z) \\
& - \frac{1}{2N_c}\llangle\text{tr}\left[V_{\underline{0}}V_{\underline{1}}^{\text{pol }\dagger}\right]\rrangle (\min\{z,z' \},z) -  C_F\llangle\text{tr}\left[V_{\underline{0}}V_{\underline{1}}^{\text{pol }\dagger}\right]\rrangle (\min\{z,z' \},z) = 0 \notag
\end{align}
because $\text{tr}\left[V_{\underline{0}}V_{\underline{0}}^{\dagger}\right] = N_c$. Hence, \fbox{$E+F+G=0$}, and the complete SLA$_T$ terms in \eq{evol1}, and, consequently, \eq{eqn:qknorc2} only come from the diagrams $A+B+C+D$ in \fig{fig:qkdiagrams}.

Performing similar computations and approximations allows us to draw the conclusion that $N+O+P = Q+R+S = 0$ for the diagrams $N$ through $S$ in \fig{fig:gldiagrams}. As a result, only diagrams $H$ through $M$ contribute to the SLA$_T$ part of the polarized gluon dipole evolution equation \eqref{eqn:glnorc1}.


\section{Constants under the Logarithms}
\label{sec:logconst}

Consider unpolarized evolution in the SLA$_L$ approximation, 
\begin{align}
S_{10} (z_{\min}) = S^{(0)}_{10} + \frac{N_c}{2\pi^2} \int\limits_{\Lambda^2/s}^{z_{\min}}\frac{dz'}{z'} \int d^2 x_2 \,K_{\text{rcBK}} ({\un x}_{0}, {\un x}_{1}; {\un x}_{2}) \, \left[S_{20} (z') \, S_{21} (z') - S_{10} (z') \right] .
\label{BK0}
\end{align}
One may worry that, if a more detailed calculation would result in some constant multiplying one of the limits of the $z'$ integral, such corrections would affect the evolution at the next-to-leading logarithmic (NLL) level. Really, an equation similar to \eq{BK0}, but with a constant $C$ in the lower limit of integration, reads
\begin{align}
S_{10} (z_{\min}) = S^{(0)}_{10} + \frac{N_c}{2\pi^2} \int\limits_{C \Lambda^2/s}^{z_{\min}}\frac{dz'}{z'} \int d^2 x_2 \,K_{\text{rcBK}} ({\un x}_{0}, {\un x}_{1}; {\un x}_{2}) \, \left[S_{20} (z') \, S_{21} (z') - S_{10} (z') \right] ,
\label{BK10}
\end{align}
and differs from \eq{BK0} at NLL. 

However, for the unpolarized evolution this problem can be easily avoided if it is cast in the differential form as
\begin{align}
z_{\min} \frac{\pd}{\pd z_{\min}} S_{10} (z_{\min}) = \frac{N_c}{2\pi^2} \int d^2 x_2 \,K_{\text{rcBK}} ({\un x}_{0}, {\un x}_{1}; {\un x}_{2}) \, \left[S_{20} (z_{\min}) \, S_{21} (z_{\min}) - S_{10} (z_{\min}) \right] ,
\label{BK1}
\end{align}
or, defining rapidity $\eta = \ln (z s /\Lambda^2)$ as
\begin{align}
\frac{\pd}{\pd \eta} S_{10} (\eta) = \frac{N_c}{2\pi^2} \int d^2 x_2 \,K_{\text{rcBK}} ({\un x}_{0}, {\un x}_{1}; {\un x}_{2}) \, \left[S_{20} (\eta) \, S_{21} (\eta) - S_{10} (\eta) \right] .
\label{BK2}
\end{align}
Where did the problem of the constant under the logarithm from \eq{BK0} disappear in \eq{BK2}? The problem became associated with defining the rapidity value for setting the initial conditions. Namely, \eq{BK2} is equivalent to \eq{BK0} if we set the initial conditions by requiring
\begin{align}
S_{10} (\eta = \eta_0 = 0) = S^{(0)}_{10}.
\end{align}
At the same time, \eq{BK2} is equivalent to \eq{BK10} if we set the initial conditions by requiring
\begin{align}
S_{10} (\eta = \eta'_0 = \ln (C)) = S^{(0)}_{10}.
\end{align}
The starting rapidities $\eta_0 =0$ and $\eta'_0 = \ln (C)$ are different in the two cases. In the practical phenomenological applications $\eta_0$ is adjusted by hand to better describe the data. Theoretically one can say that the evolution parameter is $\eta - \eta_0$, and the question of the constant $C$ is just a question of proper definition of the evolution parameter, not affecting the evolution or the intercept. Moreover, if we choose to start our evolution at some $\eta' \in [\eta_0, \eta]$, then the evolution in $\eta - \eta'$ will be independent of $C$. 

The same logic applies to helicity evolution. Start with the DLA large-$N_c$ equations with $x_{10} < 1/\Lambda$ \cite{Kovchegov:2015pbl},
\begin{subequations}\label{GNc}
\begin{align}
& G_{10} (z) = G_{10}^{(0)} (z) + \frac{\alpha_s \, N_c}{2 \pi} \int\limits_{\frac{1}{x_{10}^2 s}}^{z}
\frac{dz'}{z'} \int\limits^{x_{10}^2}_\frac{1}{z' s} \frac{d x_{21}^2}{x_{21}^2} \: \left[ \Gamma_{10,21} (z') + 3 \, G_{21} (z')  \right], \label{GNc1} \\
& \Gamma_{10,21} (z') = \Gamma_{10,21}^{(0)} (z') + \frac{\alpha_s \, N_c}{2 \pi} \int\limits_{ \frac{1}{x_{10}^2 s }}^{z'}
\frac{dz''}{z''} \int\limits^{\min \{ x_{10}^2, x_{21}^2 z'/z'' \} }_\frac{1}{z'' s} \frac{d x_{32}^2}{x_{32}^2} \: \left[ \Gamma_{10,32} (z'') + 3 \, G_{32} (z'')  \right]. \label{GNc2}
\end{align}
\end{subequations}
Rewrite these equations in terms of the variables
\begin{align}
    \eta = \ln \frac{z s}{\Lambda^2}, \ \ \ \ s_{10} = \ln \frac{1}{x_{10}^2 \Lambda^2},
\end{align}
obtaining \cite{Kovchegov:2016weo,Kovchegov:2020hgb,Adamiak:2021ppq}\footnote{Note that in references \cite{Kovchegov:2016weo,Kovchegov:2020hgb,Adamiak:2021ppq} the definitions of $\eta$ and $s_{10}$ include $\sqrt{\frac{\alpha_s N_c}{2\pi}}$ on the right-hand side, which, in turn, removes the factor of $\frac{\alpha_s N_c}{2\pi}$ on the right of the version of Eqs.~\eqref{e:HelEv2_f} obtained there.} 
\begin{subequations} \label{e:HelEv2_f}
\begin{align}
    &G (s_{10}, \eta) = G^{(0)} (s_{10}, \eta) 
    + \frac{\alpha_s N_c}{2\pi} \int\limits_{s_{10}}^{\eta} d\eta' 
      \int\limits_{s_{10}}^{\eta'} d s_{21} \: 
      \big[ \Gamma (s_{10} , s_{21} , \eta') + 3\, G (s_{21}, \eta') \big] , \\
    &\Gamma (s_{10} , s_{21} , \eta') = G^{(0)}(s_{10} , \eta') 
    + \frac{\alpha_s N_c}{2\pi} \int\limits_{s_{10}}^{\eta'} d\eta''
     \int\limits_{\max\!\left\{\!s_{10} \, , \, s_{21} - \eta' + \eta''\right\}}^{\eta''} ds_{32} \, 
     \big[ \Gamma (s_{10} , s_{32} , \eta'') + 3\, G (s_{32}, \eta'') \big]. 
\end{align}
\end{subequations}
Consider these equations giving us evolution in the rapidity parameter
\begin{align}
y = \eta - s_{10} =  \ln (z s x_{10}^2). 
\end{align}
Let us generalize the evolution such that it would start at $y = y_0$ instead of $y=0$, as Eqs.~\eqref{GNc} do. Here $y_0$ is some arbitrary initial rapidity. One gets (for $y = \eta - s_{10} > y_0$) \cite{Adamiak:2021ppq}
\begin{subequations} \label{e:HelEv3_f}
\begin{align}
    &G (s_{10}, \eta) = G^{(0)}(s_{10}, \eta)  + 
    \frac{\alpha_s N_c}{2\pi} \int\limits_{s_{10}+y_0}^{\eta} d\eta' 
    \int\limits_{s_{10}}^{\eta' - y_0} d s_{21} \: 
    \big[ \Gamma (s_{10} , s_{21} , \eta') + 3\, G (s_{21}, \eta') \big] , \\
    &\Gamma (s_{10} , s_{21} , \eta') = G^{(0)}(s_{10} , \eta')  + \frac{\alpha_s N_c}{2\pi} \int\limits_{s_{10}+y_0}^{\eta'} d\eta''
    \int\limits_{\max\!\left\{\!s_{10} \, , \, s_{21} - \eta' + \eta''\right\}}^{\eta'' - y_0} ds_{32} \, 
    \big[ \Gamma (s_{10} , s_{32} , \eta'') + 3\, G (s_{32}, \eta'') \big] . 
\end{align}
\end{subequations} 
Reverting Eqs.~\eqref{e:HelEv3_f} back to the original coordinates yields
\begin{subequations}\label{GNc_y}
\begin{align}
& G_{10} (z) = G_{10}^{(0)} (z) + \frac{\alpha_s \, N_c}{2 \pi} \int\limits_{\frac{e^{y_0}}{x_{10}^2 s}}^{z}
\frac{dz'}{z'} \int\limits^{x_{10}^2}_\frac{e^{y_0}}{z' s} \frac{d x_{21}^2}{x_{21}^2} \: \left[ \Gamma_{10,21} (z') + 3 \, G_{21} (z')  \right], \label{GNc1_y} \\
& \Gamma_{10,21} (z') = \Gamma_{10,21}^{(0)} (z') + \frac{\alpha_s \, N_c}{2 \pi} \int\limits_{ \frac{e^{y_0}}{x_{10}^2 s }}^{z'}
\frac{dz''}{z''} \int\limits^{\min \{ x_{10}^2, x_{21}^2 z'/z'' \} }_\frac{e^{y_0}}{z'' s} \frac{d x_{32}^2}{x_{32}^2} \: \left[ \Gamma_{10,32} (z'') + 3 \, G_{32} (z'')  \right]. \label{GNc2_y}
\end{align}
\end{subequations}
We see that the constants under both logarithms are indeed related to the starting rapidity $y_0$. 

Note that the constant should indeed be the same under both logarithms: the lower limit of the $x_{21}^2$-integral in \eq{GNc1_y} results from the lifetime ordering condition requiring that the emitted gluon lifetime is longer than the shock wave width, $z' x_{21}^2 > 1/s$, which indeed may bring in the constant denoted $e^{y_0}$ above. The upper limit of the $x_{21}^2$-integral in \eq{GNc1_y} is given by the boundary of integration in the DLA region: beyond it, for $x_{21} > x_{10}$, the equation becomes SLA$_L$, and the problem of the constant under the log is solved in the same way as for the unpolarized evolution \eqref{BK0} above. The lower limit of the $z'$ integral in \eq{GNc1_y} results from requiring that the upper limit of the $x_{21}^2$-integral is larger than the lower limit: hence, one gets the same $e^{y_0}$ constant factor. The upper limit of the $z'$ integral is dictated by light-cone minus momentum conservation for most diagrams, and is, hence, exact. 

The discussion of cutoffs in \eq{GNc2_y} is almost the same, with one difference: the $x_{32}^2$-integral may also be bounded by $x_{21}^2 z'/z''$ from above. This is another lifetime ordering limit, and the constant under the log here gets passed on to the next iteration of the evolution equation as a lower limit in its transverse integral. Indeed, the integration limits imply that $x_{21}^2 z' > x_{32}^2 z'' > e^{y_0}/s$, which, for the next iteration, implies that $x_{21}^2 > e^{y_0}/(z' s)$. It is, therefore, consistent to leave the upper limit of the $x_{32}^2$-integral to include $x_{21}^2 z'/z''$ without multiplying it by any additional constant. 

Finally, beyond the large-$N_c$ limit, some terms in the DLA evolution equations come in with the integration kernel
\begin{align}\label{K2}
\int\limits_{\frac{\Lambda^2}{s}}^{z} \frac{dz'}{z'} \int\limits^{x_{10}^2 z/z'}_\frac{1}{z' s} \frac{d x_{21}^2}{x_{21}^2}.
\end{align}
Imposing the same rapidity constraint as above, $y = \eta - s_{10} > y_0$, transforms this kernel into
\begin{align}\label{K3}
\int\limits_{\frac{\Lambda^2}{s}}^{z} \frac{dz'}{z'} \int\limits^{x_{10}^2 z/z'}_\frac{e^{y_0}}{z' s} \frac{d x_{21}^2}{x_{21}^2},
\end{align}
similar to the kernels in Eqs.~\eqref{GNc_y}. The difference here is that the lower limit of the $z'$ integral is not changed in \eq{K3}. However, any change in that lower limit of the $z'$ integral due to other factors would only result in the re-definition of the IR cutoff $\Lambda$, and, therefore, affects the transverse coordinate dependence of the solution in a trivial scaling way. 

If $\Lambda$ is taken to be the scale characterizing the target, instead of an IR cutoff, then one can use variables $y$ and $\eta$ to write
\begin{align}\label{K4}
\int\limits_{\frac{\Lambda^2}{s}}^{z} \frac{dz'}{z'} \int\limits^{x_{10}^2 z/z'}_\frac{1}{z' s} \frac{d x_{21}^2}{x_{21}^2} = \int\limits_0^\eta d \eta' \int\limits_0^y d y' \to \int\limits_{y_0}^\eta d \eta' \int\limits_{y_0}^y d y' = \int\limits_{\frac{e^{y_0} \, \Lambda^2}{s}}^{z} \frac{dz'}{z'} \int\limits^{x_{10}^2 z/z'}_\frac{e^{y_0}}{z' s} \frac{d x_{21}^2}{x_{21}^2}.
\end{align}
Target-projectile symmetry leads to the same cutoff $y_0$ applied to the projectile and target rapidities $y$ and $\eta$ resulting in the same factor $e^{y_0}$ in the $z'$ and $x_{21}^2$ integrals, analogous to Eqs.~\eqref{GNc_y}. Again, the constants under the logarithms are related to the starting rapidity $y_0$.


 \section{Running Coupling in DGLAP vs Small-$x$ Evolution}

 \label{A}

The goal of this Appendix is twofold. On the one hand, we want to justify the running coupling prescriptions of the last two bullets of Sec.~\ref{sec:rc_terms} by comparing them to explicit diagrammatic calculations using the BLM scheme \cite{Brodsky:1983gc}. On the other hand, in the process of that calculation, we will demonstrate how including the running coupling in the SLA$_T$ part of the helicity evolution kernel using the simple prescription of Sec.~\ref{sec:rc_terms} generates the running coupling corrections to the diagrams behind the DLA and SLA$_L$ terms in the kernel, with these corrections being similar in their structure to the ``triumvirates" found in \cite{Balitsky:2006wa,Kovchegov:2006vj} for the unpolarized small-$x$ evolution. Hence, the evolution equations which include DGLAP-type of terms with running coupling in the kernel, as do the DLA+SLA evolution equations we derived in the main text, by construction also generate the more complex running coupling corrections, like the ``triumvirates" previously observed in the equations with small-$x$ evolution only \cite{Balitsky:2006wa,Kovchegov:2006vj}. 

\begin{figure}[ht]
\begin{center}
\includegraphics[width= 0.6 \textwidth]{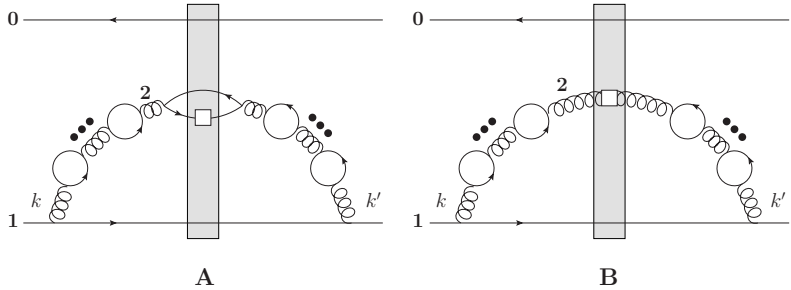} 
\caption{Quark-loop corrections for the polarized gluon emission diagram from \eq{pol1}. The diagram contributes to the DLA+SLA$_L$ evolution kernel. The ellipsis indicate a sum over all possible numbers of quark bubble corrections, from zero to infinity.}
\label{fig:bubbles_evol}
\end{center}
\end{figure}

Following the BLM prescription, we ``dress" the diagram in \eq{pol1} by quark loops. We arrive at the two types of diagrams labeled A and B in \fig{fig:bubbles_evol}, with and without a quark loop crossing the shock wave (cf. \cite{Balitsky:2006wa,Kovchegov:2006vj}). First we note that the quark-loop sum in the diagram B leads to \cite{Balitsky:2006wa,Kovchegov:2006vj}
\begin{align}\label{Bdiag}
B \propto \frac{\amu}{\left[ 1 + \amu \, \beta_2 \, \ln \frac{k_\perp^2}{\mu^2} \right] \, \left[ 1 + \amu \, \beta_2 \, \ln \frac{k_\perp^{\prime \, 2}}{\mu^2} \right]},
\end{align}
where $\mu$ is the renormalization scale and, following BLM, we complete $N_f \to - 6 \pi \beta_2$ to obtain the full one-loop QCD beta-function. The overall factor of $\amu$ is due to the gluon emissions. The interaction with the shock-wave is not considered in \eq{Bdiag}, in order to concentrate on the running coupling corrections to the evolution kernel. The running coupling corrections for the background fields in the shock wave (e.g., quark loops on $t$-channel gluons) are known to factorize from the evolution corrections (quark loops on $s$-channel gluons) \cite{Kovchegov:2007vf}.

Diagram A in \fig{fig:bubbles_evol} is different from the diagram B by a quark loop going through the shock wave. In principle, the loop can be inside or outside the shock wave. The calculation for the loop inside the shock wave is rather involved (see \cite{Balitsky:2006wa} for a similar calculation in the unpolarized case). Instead of performing this calculation, we will argue that the quark loop going through the shock wave in diagram A should give us a logarithm of $1/\mu^2$ with the $\amu \, \beta_2$ in the prefactor (after the $N_f \to - 6 \pi \beta_2$ completion). That is, the quark loop should give
\begin{align}\label{bubble1}
\amu \, \beta_2 \, \ln \frac{\perp^2}{\mu^2} + \ldots , 
\end{align}
with $\perp$ denoting some transverse momentum scale. Indeed, without such term the sum of A+B would not be expressible in terms of the physical $\mu^2$-independent running coupling $\as$ from \eq{rc1}. The ellipsis in \eq{bubble1} denote the remaining terms, which are logarithmic in energy. By that we mean either the DLA or SLA$_T$ terms which give us a logarithm of energy coming from the transverse integral. We write the entire contribution of the quark loop crossing the shock wave in A as
\begin{align}\label{bubble2}
\amu \, \beta_2 \, \ln \frac{\perp^2}{\mu^2} + \amu \, C_2 \, \ln \frac{zs}{\perp^{2}} , 
\end{align}
where $z$ is the longitudinal momentum fraction carried by the gluon in \fig{fig:bubbles_evol}, while $C_2$ denotes both a term $\sim \ln (zs)$ coming from the DLA contribution of a quark or gluon loop and a term without a logarithm of energy coming from the SLA$_T$ part of the quark or gluon loop. 
  
Now, if we only include the contribution coming from the quark loop crossing the shock wave in diagram A with the $G \to q \bar q$ splitting and $q {\bar q} \to G$ merger being outside (on different sides of) the shock wave, then all the UV divergences would be regulated by $z s$ coming from the (inverse) width of the shock wave. This corresponds to replacing $\mu^2 \to z s$ in \eq{bubble2}. We see that
\begin{align}\label{bubble3}
\mbox{quark} \ \mbox{loop} \ \mbox{begins/ends} \ \mbox{on} \ \mbox{different} \ \mbox{sides} \ \mbox{of} \ \mbox{shock} \ \mbox{wave} \sim \amu \, (C_2 - \beta_2) \, \ln \frac{zs}{\perp^{2}} .
\end{align}
Comparing Eqs.~\eqref{bubble2} and \eqref{bubble3} we conclude that the quark loop contribution from \eq{bubble2} can be rewritten as
\begin{align}\label{bubble4}
\amu \, \beta_2 \, \ln \frac{zs}{\mu^2} + \amu \, (C_2 - \beta_2) \, \ln \frac{zs}{\perp^{2}} , 
\end{align}
where the first term must originate from the contribution of the quark loop inside of the shock wave, while the second term comes from the $G \to q \bar q$ splitting and $q {\bar q} \to G$ merger being outside the shock wave, as per \eq{bubble3}.

We thus arrive at the contribution of diagram A from \fig{fig:bubbles_evol} being proportional to\footnote{In our discussion here we have avoided the question on how exactly the UV divergent part of the quark loop crossing the shock wave in diagram A is extracted, which is an important question for the running coupling corrections to the unpolarized evolution kernel \cite{Balitsky:2006wa,Kovchegov:2006vj,Albacete:2007yr}, closely related to the determination of the scale $\perp$ in \eq{bubble4}: in the helicity evolution case the separation of the UV-divergent part is dictated by the DLA and SLA evolution kernels, which keep the transverse position of the parent parton fixed, similar to what was done in \cite{Kovchegov:2006vj} for the unpolarized evolution.} 
\begin{align}\label{Adiag}
A \propto \frac{\amu^2 \, \left[ \beta_2 \, \ln \frac{zs}{\mu^2} + (C_2 - \beta_2) \, \ln \frac{zs}{\perp^{2}} \right]}{\left[ 1 + \amu \, \beta_2 \, \ln \frac{k_\perp^2}{\mu^2} \right] \, \left[ 1 + \amu \, \beta_2 \, \ln \frac{k_\perp^{\prime \, 2}}{\mu^2} \right]}.
\end{align}
Adding up equations \eqref{Adiag} and \eqref{Bdiag} we arrive at
\begin{align}\label{diagA+B}
A + B \propto \frac{\as (k_\perp^2) \, \as (k_\perp^{\prime \, 2})}{\as (z s)} + \as (k_\perp^2) \, \as (k_\perp^{\prime \, 2}) \, (C_2 - \beta_2) \, \ln \frac{zs}{\perp^{2}} .
\end{align}
Note the ``triumvirate" structure first observed in \cite{Kovchegov:2006vj,Balitsky:2006wa}. Fourier-transforming \eq{diagA+B} into transverse coordinate space, we obtain for the diagrams in \fig{fig:bubbles_evol},
\begin{align}\label{diagA+B_coord}
A + B \propto \frac{\left[ \as \left( \frac{1}{x_{21}^2} \right) \right]^2 }{\as (z s)} + \left[ \as \left( \frac{1}{x_{21}^2} \right) \right]^2 \, (C_2 - \beta_2) \, \ln \frac{zs}{\perp^{2}} .
\end{align}
Note that the second term contains $(C_2 - \beta_2) \, \ln \frac{zs}{\perp^{2}}$ which is one iteration of the real part of the SLA$_T$ kernel (excluding the coupling constant). 

Now that we have established how the running coupling corrections enter the diagrams A and B in \fig{fig:bubbles_evol} let us see whether the evolution in Eqs.~\eqref{evol_rc} and \eqref{Gevol_rc}, or, equivalently, the prescriptions of Sec.~\ref{sec:rc_terms}, reproduce the result \eqref{diagA+B_coord} of our calculation. 

We begin with the diagram B in \fig{fig:bubbles_evol}. It looks like all-order iteration of the virtual part of the SLA$_T$ kernel due to quark and gluon loops on the gluon line. Indeed, taking the last two terms in \eq{KGGGG} and integrating over $z$ (while discarding the helicity dependence, whose iterations are already taken into account) yields
\begin{align}\label{LLATrc}
\beta_2 \, \int \frac{d x_{32}^2}{x_{32}^2} \, \as \left( \frac{1}{x_{32}^2} \right) ,
\end{align}
where we have also inserted the running coupling following the prescription suggested in \eq{SLATloop}. 
Iterating this kernel to all orders yields
\begin{align}\label{Biter}
& \sum_{n=0}^\infty (\beta_2)^n \int\limits_{1/zs}^{x_{21}^2} \frac{d r_1^2}{r_1^2} \, \as \left( \frac{1}{r_1^2} \right) \, \int\limits_{1/zs}^{r_{1}^2} \frac{d r_2^2}{r_2^2} \, \as \left( \frac{1}{r_2^2} \right) \, \ldots \, \int\limits_{1/zs}^{r_{n-1}^2} \frac{d r_n^2}{r_n^2} \, \as \left( \frac{1}{r_n^2} \right) \notag \\ & = \exp \left\{ \int\limits_{1/zs}^{x_{21}^2} \frac{d r_\perp^2}{r_\perp^2} \, \as \left( \frac{1}{r_\perp^2} \right) \, \beta_2 \right\} = \frac{\as \left( \frac{1}{x_{21}^2} \right) }{\as (z s)} .
\end{align}
This result would be the same as the first term in \eq{diagA+B_coord} if we multiply it by another factor of $\as \left( \tfrac{1}{x_{21}^2} \right)$ coming from emitting the gluon in diagrams A and B. This is exactly what the prescription in \eq{pol1} tells us to do. We see that the application of prescriptions given in Eqs.~\eqref{pol1} and \eqref{SLATloop} gives us the first term on the right of \eq{diagA+B_coord}, that is, it reproduces the ``triumvirate" coupling structure as a result of DLA+SLA$_T$ evolution.\footnote{Note that the first term in \eq{diagA+B_coord} results from the diagram B and the part of diagram A with the quark loop inside the shock wave: thus the all-order iteration of the DLA+SLA$_T$ kernels gives us not just the diagram B, but a part of A as well. This is natural, since the diagram B by itself is not $\mu$-independent, and we need to add a part of A to make it $\mu$-independent; our prescriptions for the running of the coupling in the evolution kernels given in Sec.~\ref{sec:rc_terms} are all $\mu$-independent, and, hence, cannot reproduce diagram B alone. There is no double-counting here, since the part of the diagram A with the quark loop inside the shock wave is not included anywhere else in the calculation.}

We can further test this conclusion by reproducing the remainder of diagram A, that is, the second term in \eq{diagA+B_coord}. It seems to arise from one iteration of the real part of the full kernel coming from a quark or a gluon loop along with the all-order iterations of the virtual part of the SLA$_T$ kernel \eqref{LLATrc}, multiplied by $\as \left( \tfrac{1}{x_{21}^2} \right)$ accounting for the gluon emission \eqref{pol1}. We write
\begin{align}\label{Aiter}
& \as \left( \frac{1}{x_{21}^2} \right)  \, \sum_{n=1}^\infty (\beta_2)^{n-1} \, (C_2 - \beta_2) \, \int\limits_{1/zs}^{x_{21}^2} \frac{d r_1^2}{r_1^2} \, \as \left( \frac{1}{r_1^2} \right) \, \int\limits_{1/zs}^{r_{1}^2} \frac{d r_2^2}{r_2^2} \, \as \left( \frac{1}{r_2^2} \right) \, \ldots \, \int\limits_{1/zs}^{r_{n-1}^2} \frac{d r_n^2}{r_n^2} \, \as \left( \frac{1}{r_n^2} \right) \notag \\ & = \as \left( \frac{1}{x_{21}^2} \right)  \, \frac{C_2 - \beta_2}{\beta_2} \, \left[ \exp \left\{ \int\limits_{1/zs}^{x_{21}^2} \frac{d r_\perp^2}{r_\perp^2} \, \as \left( \frac{1}{r_\perp^2} \right) \, \beta_2 \right\} -1 \right] = \left[ \as \left( \frac{1}{x_{21}^2} \right) \right]^2 \, (C_2 - \beta_2) \, \ln (zs x_{21}^{2})  ,
\end{align}
in agreement with the second term on the right of \eq{diagA+B_coord}. We have thus shown that the prescription in Eqs.~\eqref{pol1} and \eqref{SLATloop} agrees with the direct diagrammatic calculations.

\begin{figure}[ht]
\begin{center}
\includegraphics[width= 0.6 \textwidth]{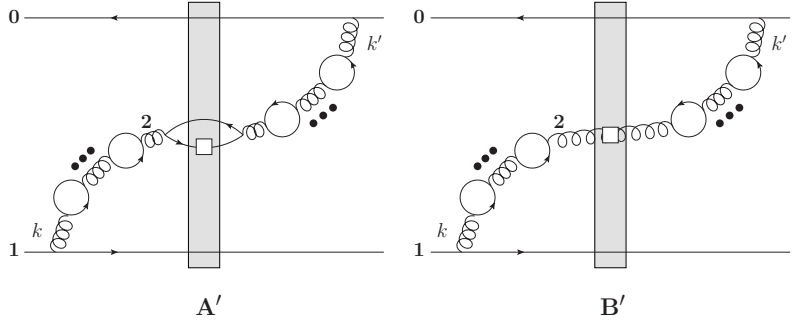} 
\caption{Quark-loop corrections for the polarized gluon emission diagram from \eq{pol2}. Again, the ellipsis indicate a sum over all possible numbers of quark loop corrections, summed from zero to infinity.}
\label{fig:bubbles_evol2}
\end{center}
\end{figure}

We are only left with the prescription for the diagram in \eq{pol2} to consider. Again we start by considering all-order quark loop corrections to the diagram, as depicted in \fig{fig:bubbles_evol2}. Note that the above calculation resulting in \eq{diagA+B} was not specific for the diagram in \eq{pol1}: it can be easily applied to the case of the diagram \eqref{pol2}. The only difference now is that in performing the Fourier transform we need to conjugate the momentum $k'_\perp$ to the distance $x_{20}$ (see \fig{fig:bubbles_evol2}), which yields 
\begin{align}\label{diagA+B_coord2}
A' + B' \propto \frac{\as \left( \frac{1}{x_{21}^2} \right) \,  \as \left( \frac{1}{x_{20}^2} \right)}{\as (z s)} + \as \left( \frac{1}{x_{21}^2} \right) \,  \as \left( \frac{1}{x_{20}^2} \right) \, (C_2 - \beta_2) \, \ln \frac{zs}{\perp^{2}} 
\end{align}
instead of \eq{diagA+B_coord}.

Now, the DLA+SLA evolution would again give \eq{Biter} after iterating the virtual SLA$_T$ loops to all orders, except for the $x_{21} \to \min \{ x_{21}, x_{20} \}$  replacement due to the shortest lifetime condition, giving
\begin{align}\label{Biter2}
\frac{\as \left( \max \left\{ \frac{1}{x_{21}^2} , \frac{1}{x_{20}^2} \right\} \right) }{\as (z s)} 
\end{align}
for the first term on the right-hand side of \eq{diagA+B_coord2}. To make the terms match one needs to multiply \eq{Biter2} by $\as \left( \min \left\{ \frac{1}{x_{21}^2} , \frac{1}{x_{20}^2} \right\} \right)$ due to the gluon emission in the dipole 10, which is the running coupling prescription we put in \eq{pol2}. Similarly, repeating the calculation in \eq{Aiter} with the $x_{21} \to \min \{ x_{21}, x_{20} \}$  replacement in the upper limit of the $r$-integrals and with the
\begin{align}
\as \left( \frac{1}{x_{21}^2} \right)  \to \as \left( \min \left\{ \frac{1}{x_{21}^2} , \frac{1}{x_{20}^2} \right\} \right)
\end{align}
replacement for the coupling factor in the front leads to the second term in the \eq{diagA+B_coord2}. We have thus justified the running coupling prescription in \eq{pol2}. 

The analysis of the diagram in \eq{pol3} is identical to the analysis of the diagram in \eq{pol1} above. Therefore, we have confirmed that the running coupling prescriptions outlined in Sec.~\ref{sec:rc_terms} and, therefore, in Eqs.~\eqref{evol_rc} and \eqref{Gevol_rc} are consistent with diagrammatic calculations.



\section{Extraction of a Closed Equation at Large $N_c$ and at Large $N_c\& N_f$ from the Adjoint Polarized Dipole Evolution}
\label{sec:GlargeNc}

In this Appendix we clarify the steps needed to obtain equations \eqref{eqn:Nc17} and \eqref{Gevol_largeNcNf} starting, respectively, from the evolution equations \eqref{eqn:Nc10} and \eqref{Gevol_largeNcNf1} for the adjoint polarized dipole amplitude in the large-$N_c$ and large-$N_c\& N_f$ limits.

As discussed in Section~\ref{sec:nc}, in the limit where $N_c$ is large, a polarized gluon dipole can be regarded as two fundamental dipoles of the same transverse separation, one polarized and one unpolarized. In particular, \eq{eqn:Nc5} reads (cf. also \eq{UU} for the large-$N_c \& N_f$ limit)
\begin{align}
\frac{1}{N_c^2-1}\llangle\text{Tr}\left[U_{\underline{0}} \, \left( U_{\underline{1}}^{\text{pol} \, g} \right)^{\dagger}\right]\rrangle \left(z_{\min},z_{\text{pol}}\right) = 4 \, G_{10} (z_{\min},z_{\text{pol}}) \, S_{10} (z_{\min}).
\label{eqn:GS1}
\end{align}
The unpolarized fundamental dipole amplitude evolves under (rc)BK equation, which is of the SLA$_L$ type. For brevity, we write (cf. \eq{eqn:Nc14})
\begin{align}\label{eqn:GS2}
S_{10} (z_{\min}) =  S^{(0)}_{10} + \left(\mathcal{K}\otimes S\right) = S^{(0)}_{10} + \int\limits^{z_{\min}}_{\Lambda^2/s}dz'\int d^2 x_2 \; K\left(\xx_0,\xx_1,\xx_2\right) \, \theta\left(x^2_{10}z_{\min}-x^2_{21}z'\right) \, \tilde{S}_{012} \left(z'\right), 
\end{align}
where 
\begin{align}
\tilde{S}_{012} \left(z'\right) = S_{20} (z') \, S_{21} (z') - S_{10} (z').
\label{eqn:GS3}
\end{align}

Our goal here is to construct the evolution equation for the polarized fundamental dipole amplitude $G_{10}$. At the moment we do not know what it is. We assume that the polarized fundamental dipole amplitude $G_{10}$ evolves under some (helicity) evolution equation, whose kernel we denote by $\tilde{K}$. The kernel $\tilde{K}$ will act on the quantity we denote $\tilde{G}_{012} \left(z'\right)$, which represents an unknown combination of $G_{10}$, $Q_{10}$ and the neighbor dipole amplitudes on the right-hand side of the evolution equation for $G_{10}$. The amplitude $G_{10}$ represents here either the amplitude defined in \eq{Ggdef} for the large-$N_c$ evolution or the amplitude $G_{10}^{adj}$ defined in \eq{adj_dip} for the large-$N_c \& N_f$ evolution. The kernel $\tilde{K}$ acting on $\tilde{G}_{012} \left(z'\right)$ will ultimately represent the right-hand sides of Eqs.~\eqref{eqn:Nc17} and \eqref{Gevol_largeNcNf} in the large-$N_c$ and large-$N_c\& N_f$ limits, respectively. However, let us stress that at this point we do not know the structure of $\tilde{K}$ and $\tilde{G}_{012} \left(z'\right)$: rather, we are trying to construct them. 

With this in mind, concentrating on the DLA+SLA$_L$ terms, we anticipate the evolution for $G_{10}$ to be
\begin{align}\label{eqn:GS4}
G_{10} \left(z_{\min},z_{\text{pol}}\right) & = G_{10}^{(0)}\left(z_{\text{pol}}\right) + \left(\tilde{\mathcal{K}}\otimes \tilde{G} \right) \\ & = G^{(0)}\left(z_{\text{pol}}\right) + \int\limits^{z_{\min}}_{\Lambda^2/s}dz'\int d^2 x_2\;\tilde{K}\left(\xx_0,\xx_1,\xx_2\right)\theta\left(x^2_{10}z_{\min}-x^2_{21}z'\right)\tilde{G}_{012} \left(z'\right) . \notag
\end{align}
Equation \eqref{eqn:GS4} implies the following equation for the generalized neighbor fundamental dipole amplitude (defined by, e.g., \eq{eqn:Nc7} above): 
\begin{align}\label{eqn:GS5}
\Gamma^{gen}_{10,32} &\left(z_{\min},z_{\text{pol}}\right) = G_{10}^{(0)}\left(z_{\text{pol}}\right) \\
&+ \int\limits^{z_{\min}}_{\Lambda^2/s}dz' \int d^2 x_4\;\tilde{K}\left(\xx_0,\xx_1,\xx_4\right)\theta\left(\min\left\{x^2_{10},x^2_{32}\right\}z_{\min}-x^2_{41}z'\right)\tilde{G}_{014} \left(z'\right). \notag
\end{align}

Now let us work with the evolution equations \eqref{eqn:Nc10} or \eqref{Gevol_largeNcNf1} for the adjoined polarized dipole amplitude, which are really our starting points in the large-$N_c$ and large-$N_c\& N_f$ calculations, respectively. Employing the notation of Eqs.~\eqref{eqn:GS2}--\eqref{eqn:GS5}, we re-write the DLA+SLA$_L$ parts of \eqref{eqn:Nc10} or \eqref{Gevol_largeNcNf1} as
\begin{align}\label{eqn:GS6}
G_{10} &\left(z_{\min},z_{\text{pol}}\right) \, S_{10} \left(z_{\min}\right) = G_{10}^{(0)}\left(z_{\text{pol}}\right) \, S_{10}^{(0)} + \int\limits^{z_{\min}}_{\Lambda^2/s}dz'\int d^2 x_2\;\tilde{K}\left(\xx_0,\xx_1,\xx_2\right)\theta\left(x^2_{10}z_{\min}-x^2_{21}z'\right) \tilde{G}_{012} \left(z'\right) \, S_{10} \left(z'\right) \notag \\
&+ \int\limits^{z_{\min}}_{\Lambda^2/s}dz'' \int d^2 x_3\;K\left(\xx_0,\xx_1,\xx_3\right)\theta\left(x^2_{10}z_{\min}-x^2_{31}z'' \right)\Gamma^{gen}_{10,31} \left(z'',z_{\text{pol}}\right) \, \tilde{S}_{013} \left(z''\right). 
\end{align}
Note that \eq{eqn:GS6} can serve as the definition of $\tilde{K}$ and $\tilde{G}_{012}$, with \eq{eqn:GS4} being our ansatz for the  $G_{10}$ evolution following from \eq{eqn:GS6}.  

Employing Eqs.~\eqref{eqn:GS2} and \eqref{eqn:GS5} in \eq{eqn:GS6} we arrive at
\begin{align}\label{eqn:GS7}
G_{10} &\left(z_{\min},z_{\text{pol}}\right) \, S_{10} \left(z_{\min}\right) = G_{10}^{(0)}\left(z_{\text{pol}}\right) \, S_{10}^{(0)}   \\
&+ \int\limits^{z_{\min}}_{\Lambda^2/s}dz'\int d^2 x_2\;\tilde{K}\left(\xx_0,\xx_1,\xx_2\right)\theta\left(x^2_{10}z_{\min}-x^2_{21}z'\right) \tilde{G}_{012} \left(z'\right) \, S^{(0)}_{10} \notag \\
&+ \int\limits^{z_{\min}}_{\Lambda^2/s}dz'\int d^2 x_2\;K\left(\xx_0,\xx_1,\xx_2\right)\theta\left(x^2_{10}z_{\min}-x^2_{21}z'\right) \, \tilde{S}_{012} \left(z'\right) \, G^{(0)}_{10} \left(z_{\text{pol}}\right) \notag \\
&+ \int\limits^{z_{\min}}_{\Lambda^2/s}dz'\int\limits_{\Lambda^2/s}^{z'}dz''\int d^2 x_2 \int d^2 x_3\;\theta\left(x^2_{10}z_{\min}-x^2_{21}z'\right)\theta\left(x^2_{21}z'-x^2_{31}z''\right) \notag \\
&\;\;\;\;\;\times K\left(\xx_0,\xx_1,\xx_3\right)\tilde{K}\left(\xx_0,\xx_1,\xx_2\right) \, \tilde{G}_{012} \left(z'\right)  \, \tilde{S}_{013} \left(z''\right) \notag \\
&+ \int\limits^{z_{\min}}_{\Lambda^2/s}dz''\int\limits_{\Lambda^2/s}^{z''}dz'\int d^2 x_3\int d^2 x_2\;\theta\left(x^2_{10}z_{\min}-x^2_{31}z''\right)\theta\left( x^2_{31} z''-x^2_{21}z'\right) \notag \\
&\;\;\;\;\;\times K\left(\xx_0,\xx_1,\xx_3\right)\tilde{K}\left(\xx_0,\xx_1,\xx_2\right) \, \tilde{G}_{012} \left(z'\right) \, \tilde{S}_{013} \left(z''\right). \notag
\end{align}
In the second to last term of \eq{eqn:GS7}, the lifetime $x^2_{21}z'$ in the latter $\theta$-function follows from the fact that the subsequent splittings are limited by the lifetime of the previous splitting, thus imposing a proper lifetime ordering \cite{Kovchegov:2016zex,Cougoulic:2019aja}. This leads to $z''\ll z'\ll z_{\min}$ and $x^2_{31}z''\ll x^2_{21}z'\ll x^2_{10}z_{\min}$ in that term. Similar lifetime ordering argument leads to the second $\theta$-function in the last term of \eq{eqn:GS7} containing the lifetime $x^2_{31} z''$. In contrast to the second to last term, the last term of \eq{eqn:GS7} comes in with the $z'\ll z''\ll z_{\min}$ and $x^2_{21}z'\ll x^2_{31}z''\ll x^2_{10}z_{\min}$ orderings. Together, these two terms complete the phase space of $\left(\tilde{\mathcal{K}}\otimes \tilde{G} \right)\left(\mathcal{K}\otimes S\right)$ at DLA+SLA$_L$ allowing us to write \eqref{eqn:GS7} as
\begin{align}\label{eqn:GS8}
G_{10} \left(z_{\min},z_{\text{pol}}\right) \, S_{10} \left(z_{\min}\right) & = G_{10}^{(0)}\left(z_{\text{pol}}\right) \, S_{10}^{(0)} + \left(\tilde{\mathcal{K}}\otimes \tilde{G} \right) \, S_{10}^{(0)} + G_{10}^{(0)} \, \left(z_{\text{pol}}\right)\left(\mathcal{K}\otimes S\right)+ \left(\tilde{\mathcal{K}}\otimes \tilde{G} \right) \left(\mathcal{K}\otimes S\right) \\
= \left[G_{10}^{(0)}\left(z_{\text{pol}}\right) + \left(\tilde{\mathcal{K}}\otimes \tilde{G} \right)\right] &\left[S_{10}^{(0)} + \left(\mathcal{K}\otimes S\right)\right] = \left[G_{10}^{(0)}\left(z_{\text{pol}}\right) + \left(\tilde{\mathcal{K}}\otimes \tilde{G} \right)\right] \, S_{10} (z_{\min}), \notag
\end{align}
where in the last step we have employed \eq{eqn:GS2}. After canceling the overall factor of $S_{10} \left(z_{\min}\right)$ on both sides of \eq{eqn:GS8} we arrive at \eq{eqn:GS4}, as desired, thus proving that the ansatz \eqref{eqn:GS4} follows from our starting point \eqref{eqn:GS6}. 

Comparing equations \eqref{eqn:GS6} and \eqref{eqn:GS8}, we see that the effective prescription following from our calculation for obtaining the latter out of the former is to drop the last (BK-looking) term in \eq{eqn:GS6} while simultaneously replacing $S_{10}^{(0)} \to S_{10} \left(z_{\min}\right)$ and $S_{10} (z') \to S_{10} \left(z_{\min}\right)$ in the other two terms. One can also write down \eq{eqn:GS8} from the outset, instead of starting with \eq{eqn:GS6}, if one treats the evolved unpolarized dipole amplitude $S_{10} \left(z_{\min}\right)$ as a ``background field" for helicity evolution: this philosophy was adopted in the Appendix~A of \cite{Kovchegov:2016zex}. 

Our discussion above concerned the DLA and SLA$_L$ terms only. Note that every SLA$_T$ term in the evolution of a polarized adjoint dipole at large $N_c$ (and at large $N_c\& N_f$) contains a factor of $S_{10} \left(z_{\min}\right)$ which directly cancels the unpolarized dipole amplitude on the left-hand side of the equation and requires no subtle considerations presented here.

We have thus shown how \eq{eqn:Nc17} results from \eq{eqn:Nc10} in the large-$N_c$ limit, while \eq{Gevol_largeNcNf} follows from \eq{Gevol_largeNcNf1} in the large-$N_c\& N_f$ limit.


\bibliographystyle{JHEP}
\bibliography{references}

\end{document}